\documentclass[pre,superscriptaddress,twocolumn,amssymb,amsmath,nofootinbib]{revtex4-1}
\usepackage{graphicx}
\usepackage{latexsym}
\usepackage{color}

\definecolor{Blue}{rgb}{0.00, 0.00, 1.00}
\definecolor{Red}{rgb}{1.00, 0.00, 0.00}

\newcommand{\red}{\color{Red}}
\newcommand{\rmd}{\mathrm{d}}

\newcommand{\tr}{\mathrm{tr}}
\newcommand{\nn}{\nonumber}

\def\be{\begin{equation}}
\def\ee{\end{equation}}

\def\bea{\begin{eqnarray}}
\def\eea{\end{eqnarray}}

\newcommand{\1}{1\hspace*{-.48ex}{\rm l}}

\usepackage{feynmp}
\graphicspath{{./}{./figures/}}

\newcommand{\Fig}[1]{\includegraphics[width=8.6cm]{#1}}

\begin{document}

\title{Collective excitations in a large-$d$ model for graphene}

\author{Francisco Guinea}
\address{Instituto de Ciencia de Materiales de Madrid (CSIC), Sor Juana In\'es de la Cruz 3, 28049 Madrid, Spain}
\author{Pierre Le Doussal}
\address{CNRS-Laboratoire de Physique Th\'eorique de l'Ecole
Normale Sup\'erieure, 24 rue Lhomond, 75005 Paris, France.}
\author{Kay J\"org Wiese}
\address{CNRS-Laboratoire de Physique Th\'eorique de l'Ecole
Normale Sup\'erieure, 24 rue Lhomond, 75005 Paris, France.}

\begin{abstract}
We consider a model of Dirac fermions coupled to flexural phonons to describe
a graphene sheet fluctuating in dimension $2+d$. We derive the self-consistent screening equations
for the quantum problem, exact in the limit of large $d$. We first treat the membrane alone, and
work out the quantum to classical, and harmonic to anharmonic crossover. For the coupled electron-membrane
problem we calculate the dressed two-particle propagators of the elastic and electron interactions and find that it exhibits a collective mode
which becomes unstable at some wave-vector $q_{\rm c}$ for large enough coupling $g$.
The saddle point analysis, exact at large $d$, indicates that this instability corresponds to spontaneous and simultaneous
appearance of gaussian curvature and electron puddles. The relevance to ripples in graphene is
discussed.

\end{abstract}

\maketitle




\section{introduction}

Graphene is a one atom thick membrane  \cite{Netal04,Netal05,NGPNG09} with a high bulk and Young elastic modulii,  which can withstand large strains before fracture  \cite{LWKH08}. Both suspended samples and samples on substrates show corrugations on a variety of scales. In some cases these corrugations are due to inhomogeneities in the substrate  \cite{Setal08,Getal09} (see also  \cite{KCNG11}), or to the mismatch between the graphene and the substrate lattice parameters  \cite{Vetal08}. Freely suspended samples also show ripples, whose origin is still undetermined  \cite{Metal07} (see also  \cite{FLK07}). The scale of the observed corrugations can lead to significant modifications in the electronic band structure of graphene \cite{HorovitzDoussal2002,GHL08,GHL09}).

The flexural modes of graphene are coupled to the in plane phonons, leading to anharmonic effects even in the zero temperature, quantum limit \cite{MO08}. Flexural modes couple to the electrons in graphene, and change the electrical conductivity \cite{MO08,Cetal10,MO10}. Ripples might arise from the coupling between the lattice deformations and the electrons \cite{G09,SGG11}. Structural corrugations induce a shift in the electronic chemical potential, which lead to the formation of charge puddles \cite{GTGKP12}. Instabilities at finite momenta in models where the electrons are described as a perfect metal, and the formation of charge inhomogeneities is only prevented by the Coulomb interaction \cite{G09}. On the other hand, the low density of states in graphene leads to a small quantum capacitance, although, again, a sufficiently large coupling between electrons and lattice deformations can induce instabilities \cite{SGG11}.

We study here the nature of the instabilities due to the combination of anharmonic effects \cite{NP87} and electron-phonon coupling at zero and at finite temperature. We extend the model used in \cite{SGG11} by considering a membrane fluctuating in $d$ transverse dimensions and coupled to $N_{\rm f} d$ fermion species. This extension allows for an exact solution at large $d$. We derive the large $d$ equations which provide a generalization of the $1/d$ expansion  \cite{AL88,DavidGuitter1988} and of the Self Consistent Screening Approximation (SCSA) \cite{LR92} for the classical membrane to (i) the quantum membrane problem, (ii) the coupled electron-membrane quantum problem. Given the  success of the SCSA to describe both classical anharmonic effects in the elastic problem \cite{ZRFK10}, and
interaction effects in the electron problem alone (see e.g. \cite{GGV94}, and confirming experiments in \cite{Eetal11}), it is indeed tempting to apply it  to the coupled problem. Here we solve mainly the $d=\infty$ limit, and discuss some of the $1/d$ corrections, leaving the full study of the SCSA equations to the future. We find that as the electron-phonon coupling increases, the
charge excitations become strongly hybridized with flexural phonons, and the frequencies of these excitations decrease,
until a threshold is reached where an instability occurs. A saddle-point analysis, exact at large $d$,
indicates that this instability corresponds to the spontaneous and simultaneous
appearance of gaussian curvature and electron puddles, a hallmark of the ripples. Note that our mechanism is consistent, although different in details from  \cite{SGG11}, since the instability occurs already for $d=\infty$ hence does not require the renormalization of the
bending rigidity of the membrane. While consideration of these additional $1/d$ effects may lead to quantitative changes,
it is not expected to radically alter the picture proposed here.

In addition to the coupled problem, the SCSA equations for the quantum membrane alone lead to a new ``phase diagram" where we identify regions in the temperature/wave-vector plane where quantum to classical as well as harmonic to anharmonic elasticity crossovers occur, and which should be
useful in analyzing experiments.

This article is organized as follows:

In section \ref{s:Model}, we introduce our model, and compare it to previous studies.

Section \ref{s:SCSA} introduces the equations to be solved in the self-consistent-screening method.

In Section \ref{s:AnaResults} we first analyze the membrane alone, and study the crossover quantum/classical and harmonic/anharmonic for the flexural modes.
Then we study the coupled membrane-electron problem and present our results for the instability in section \ref{s:AnaResults}.

In section \ref{sec:sp}, we analyze further the instability by deriving the exact effective action in the large $d$-limit.

Our conclusions are presented in section \ref{s:Conclusion}.

Several technical, but important details are presented in the appendices: In appendix \ref{a:in-plane-phonons} we discuss how to integrate over the in-plane phonons. In appendices \ref{a:phonon-bubble} and \ref{a:fermion-bubble}  we evaluate the most important diagrams, the phonon (flexural) and fermion bubbles.

\section{Model}
\label{s:Model}

\subsection{Hamiltonian of flexural phonons coupled to Dirac electrons}

To consider a model with a tractable limit, we extend the model for a graphene sheet to a membrane
in embedding dimension $d$, interacting with $N_{\rm f} d$ copies of a free Dirac fermion.
Here $N_{\rm f}$ is the number of flavors (valleys plus spin).
The physical case is recovered by setting $d=1$ and $N_{\rm f}=4$. The parameter $d$ is convenient to consider in the solvable
limit $d \to \infty$. The deformation of the sheet with respect to the perfect flat crystal is parameterized by 2 in-plane phonon displacement fields
$u_i$, $i=1,2$, and $d$ out-of-plane phonon modes $h_a$, $a=1,...,d$ (flexural modes). The deformation energy is the sum of
 curvature and elastic energy,
\bea  \label{elasten}
H_{\rm ph} &=& H_{\rm kin} + H_{\rm elas}\\
H_{\rm elas}&=& \frac{1}{2} \int \rmd^2 x \, \Big[  \kappa (\nabla^2 h_a)^2 +  \lambda u_{ii}^2 + 2 \mu u_{ij}^2 \Big].
\eea
It is given
in terms of the Lam\'e coefficients $\lambda,\mu$ and the strain field $u_{ij}:= \frac{1}{2} (\partial_i u_j + \partial_j u_i + \partial_i h_a \partial_j h_a)$.
Adding the kinetic energy $H_{\rm kin}$ leads to the quantum action which describes the membrane dynamics (in real time $t$, and with mass density $\rho$),
\be
{\cal S}_{\rm ph} =  \int \rmd t\, \Big\{  \rmd^2 x\, \frac{\rho}{2} [ (\partial_t u_i)^2 + (\partial_t h_a)^2 ]  - H_{\rm elas} \Big\}.
\ee
We now couple the long-wavelength modes of the membrane to Dirac fermions. Following previous work, we define a scalar potential, which describes a global shift of the chemical potential, and a gauge field, which describes the hopping between the two sublattices which make up graphene \cite{VKG10}. The scalar potential modifies the local chemical potential, and induces charge fluctuations. The change in the electronic energy associated to charge fluctuations is described, in second order perturbation theory, by the charge susceptibility, $\chi_\rho (q, \omega ) = \sum_n  \left| \left< 0 | \hat{\rho}_{q} ) | n \right> \right|^2 \delta ( \omega - \epsilon_n + \epsilon_0 )$ where $\hat{\rho}_{q}$ creates an electron-hole pair of momentum $q$, $\left| 0 \right>$ and $\left| n \right>$ are the ground and excited states, and $\epsilon_0$ and $\epsilon_n$ their energies. The gauge potential, on the other hand, couples to the current operator, and it induces current fluctuations. The term which describes the effect of these fluctuations on the total energy is given by the current susceptibility, $\chi_j ( q , \omega )$, which is related to the charge susceptibility by charge conservation $\chi_j ( q, \omega ) = \omega^2 / ( v_F^2 q^2 ) \times \chi_\rho ( q , \omega )$. As, for flexural modes, $\omega_{\rm fl} ( q ) \propto q^2 \ll v_F q$ over the entire Brillouin Zone, and we can neglect the contribution of the gauge potential as $q \rightarrow 0$ \cite{SGG11}.

In this article we consider the coupling to a scalar potential, modeled by
\bea  \label{coupl}
   H_{\rm e\text-ph}&=& - g_0 \int \rmd^2 x\, \delta \rho(x) u_{ii}(x) \\
 \delta \rho(x) &=& \rho(x) - \rho_0 = \frac{1}{d} \sum_{\gamma=1}^{N_{\rm f} d} \bar \Psi_\gamma \1 \Psi_\gamma - \rho_0\quad
\eea
which is the standard form of the long-wavelength coupling assuming (i) rotational invariance, i.e.\ no substrate, (ii) no membrane tension (arising from e.g.\  clamping)-- it can be added later. Here $\rho_0$ is the equilibrium carrier density counted from the neutrality point. 
Estimations for the value of $g_0$ vary over one order of magnitude  \cite{OS66,SA02,CSS10}, $g_0 \approx 4 - 50$ eV. 

In previous work  \cite{G09,SGG11} the strategy was to first integrate over fermions (within some approximation, see below) and only in a second stage
sum over in-plane modes, to obtain an effective (approximate) theory for the flexural modes only. Our present strategy is different. We first integrate
over in-plane phonons leading to a coupled theory of flexural modes and electrons. Since the action is quadratic in these modes, the integration
can be performed exactly. The calculation is performed in details in  Appendix \ref{a:in-plane-phonons}. Because of the frequency dependence of the in-plane phonon propagator we obtain a more complicated expression than in the standard (i.e.\ classical) case. It contains new, frequency dependent, terms. Since in this article we focus on frequencies of the order of the
Debye frequencies of flexural modes, which are much lower than the one for in-plane phonons,
this new frequency dependence can  safely be  neglected. Hence we arrive at our starting
(effective) Hamiltonian for the flexural modes coupled to the free Dirac electrons (we set $\hbar=1$):
\begin{eqnarray} \label{model0}
 H =  H_{\rm kin} &+& H_\rho + \int \rmd^2 x \bigg\{ \sum_{a=1}^d \frac{\kappa}{2} (\nabla^2 h_a)^2
\nn \\
&& + \frac{K_0}{2 d} \bigg[ \frac{1}{2} {\rm P}^{\rm T}_{ij}(\partial)  \sum_{a=1}^d \partial_i h_a  \partial_j h_a \bigg]^2 \nn
\\
&& - \frac{g}{d} \sum_{\gamma=1}^{N_{\rm f} d} \bar \Psi_\gamma \1  \Psi_\gamma \bigg[ \frac{1}{2} {\rm P}^{\rm T}_{ij}(\partial)  \sum_{a=1}^d \partial_i h_a \partial_j h_a \bigg]
\nn\\
&& +  \sum_{\gamma=1}^{N_{\rm f} d}  \bar \Psi_\gamma \big[ - v_{\rm F} { {\boldsymbol{\sigma}}} \cdot (- i \nabla)\big]  \Psi_\gamma \bigg\}.
\end{eqnarray}
Here $K_0= 4 \mu(\mu+\lambda)/(2 \mu+\lambda) d$ is the bare Young modulus, to which should be added
the kinetic energy. Note that the resulting coupling becomes
\be
g= \frac{2 \mu}{2 \mu + \lambda} g_0 .
\ee
In graphene, $\lambda / \mu \approx 1/6$, so that $g \approx g_0$.

In a second stage (see below) we will add to this model the electron-electron interaction.
The energy for the charge fluctuations then take the form:
\be
H_{\rm ee}= \frac{1}{2}
 \int \frac{\rmd^2 q}{(2 \pi)^2}   V_0(q) |\rho(q)|^2.\ee
We consider below the Coulomb interaction $V_0(r)=e^2/(\epsilon_0 r)$, i.e.\ in Fourier $V_0(q) = \frac{2 \pi e^2}{\epsilon_0 q}$, where
$\epsilon_0$ is the dielectric constant of the environment.
This term $H_{ee}$ will be added to (\ref{model0}). 

After integration over the in-plane
 phonons the interaction becomes
 \be \label{newV} 
 V(q) = V_0(q) - \frac{g_0^2}{\lambda + 2 \mu} .
 \ee
i.e.\ it acquires a short-ranged attractive part, as shown in  Appendix \ref{a:in-plane-phonons}. By power counting
that part is formally irrelevant and can be neglected at small $q$ compared to the Coulomb repulsion \footnote{Note that even for free Dirac fermions it does not lead to superconducting instability at the neutrality point, since that would require a non vanishing density of states.} 
. At higher $q$ however, and especially if a ripple instability develops, it does play a role and may not be neglected.
This will be discussed below and in Section \ref{ss:mem+ele}. 

%

Finally, note that in the elastic interaction the uniform mode is excluded, i.e.\ everywhere in this article the composite field ${\rm P}^{\rm T}_{ij}(\partial)  \sum_{a=1}^d \partial_i h_a \partial_j h_a$
 is evaluated only for Fourier components $q \neq 0$  \cite{JerusalemWinterSchool1989,LR92,WieseHabil}. This field, which plays an important role below, has a nice geometrical interpretation, i.e.\ it is equal (say for $d=1$), in Fourier, to ${\cal K}(q)/q^2$ where ${\cal K}(x)$ is the Gaussian curvature of the membrane.

\subsection{Comparison with previous work}

Let us contrast again our approach with previous work  \cite{G09,SGG11}. There one first integrated the coupling term (\ref{coupl}) over the electrons using a Gaussian approximation. There the degree of freedom are the charge fluctuations $\delta \rho$, and one replaces the electronic part of the Hamiltonian with:
\bea  \label{susc_el}
   H_{\rm \rho}&=& \frac{1}{2}  \int \rmd^2 q\, |\delta \rho(q)|^2 \left[ \frac{1}{\chi_\rho (q)} + V_0 (q ) \right]  
\eea
As the term in eq.(\ref{susc_el}) is quadratic, it can be combined with eq.(\ref{coupl}) and the charge fluctuations can be integrated out leading to an additional term in the elastic
energy which could be interpreted as a $q$-dependent shift in the Lame coefficient:
\be
\lambda \to \lambda(q)= \lambda - g_0^2 \langle \delta \rho(-q) \delta \rho(q) \rangle
.
\ee
In this calculation, the electron-density correlation was estimated either from a fluid model for the interacting electrons  \cite{G09} (a finite-$T$ classical calculation using $H_\rho$ without the first term), or from the susceptibility $\chi_\rho (q)$
of non-interacting Dirac fermions  \cite{SGG11} (a $T=0$ quantum calculation, using $H_\rho$ including the first term). 
\footnote{Note that the dependence of $\chi_\rho (q, \omega )$ can be calculated analytically \cite{WSSG06} for any homogeneous charge $\rho_0$. For simplicity, we study here the case $\rho_0 = 0$. The difference between the two expressions is only significant at small momenta, $q  \sim k_F = \sqrt{\pi \rho_0}$. As discussed below, the effect of the electronic degrees of freedom vanishes as $q \rightarrow 0$, so that the approximation is justified if $\rho_0$ is sufficiently small.}

In a second stage one
integrated over the in-plane modes, resulting in the usual membrane action, but with a modified, wave-vector dependent,
Young's modulus $K_0(q) = \frac{4 \mu [\mu+\lambda(q)]}{2 \mu + \lambda(q)}$.
In the classical fluid estimate  \cite{G09}, one finds $\lambda \to \lambda - g^2 q/(2 \pi e^2)$
and $K_0(q)$ changes sign in some (relatively narrow) region of momenta
$\frac{2 \pi e^2}{g^2} (2 \mu + \lambda) > q > \frac{2 \pi e^2}{g^2} (\mu + \lambda)$. Using the standard SCSA method for classical membranes
to treat the effect of $K_0(q)$, this was then argued to lead to a maximum in the normal-normal correlation of the membrane,
interpreted as ripples. In  \cite{SGG11} the renormalization of the wave-vector dependent bending rigidity
$\kappa(q)$, resulting from this dispersive Young modulus was estimated in the quantum $T=0$ limit, and argued to lead to two different
regimes. In one regime $\kappa(q)$ softens near a finite $q$, which was argued to lead to ripples at that
wave-vector. Note that other proposals, based on buckling, also exist in the literature  \cite{GHL08,BP08,Gazit2009}.

While it is tempting to first integrate over the Dirac fermions, it is in practice difficult to do it accurately, beyond the classical-fluid approximation.
Even for non-interacting Dirac fermions, an exact calculation leads to a functional determinant and higher non-linearities in $u_{ii}$. In addition,
as we will see, it may obscure one piece of the physics, which is that the instability that we seek to describe is a {\it combined instability in flexural modes and electron density}. Furthermore, interactions are easily seen to stabilize the system, hence we need to include them for any realistic theory, which makes  integration over fermions in a first step even more problematic.

Hence  we choose a different route and first integrate over in-plane modes, a step which is well controlled.
The resulting theory (\ref{model0}) is quartic in both flexural modes and electrons, and quite non-trivial. We then solve this theory in the large-$d$ limit.
The instability occurs in a different manner as in  previous work, namely as a pole in the combined 2-particle propagators for phonons and electrons.
In particular, we do not need to consider the renormalization of $\kappa$ to obtain a transition. Although the bending rigidity is corrected to higher orders in our $1/d$ expansion, it may change  estimates for the numbers, but not the general scenario, which is a
phase transition. Note that we can recover in our model
$K_0(q)$ obtained by the previous methods (see Section \ref{ss:mem+ele}); it does not seem to play  an important role.

\subsection{The parameters of the model}

Before we analyze the model, let us recall the dimensions of the parameters (in units of length $L$ and energy $E$) and provide some estimates.
The model is described by the parameters $K_0 \equiv E L^{-2} , \kappa \equiv E, \rho^{-1}=E L^4, v_{\rm F} \equiv E L , e^2 \equiv E L , g\equiv E  $,
and the momentum cutoff $\Lambda\equiv L^{-1}$ (and time $t \sim E^{-1}$). By multiplying the parameters which describe the interactions, $K_0 , e^2$ and $g$, with the susceptibilities, three dimensionless coupling constants can be defined
\bea \label{11}
&& \lambda_{\rm anh} :=\frac{K_0}{\kappa^{3/2} \rho^{1/2}} \quad , \quad \alpha_{\rm e} := \frac{e^2}{v_{\rm F}}  \\
&& \lambda_{\rm e\text-fl} := \frac{g}{\sqrt{\kappa^{3/2} \rho^{1/2}  v_{\rm F} / \Lambda }} .\nn
\eea
The parameter $\alpha_{\rm e}$ is the fine structure constant of graphene and characterizes the strength of the electron-electron interaction.
The coupling $\lambda_{\rm anh}$ characterizes the strength of the anharmonic elasticity. The dependence of $\lambda_{\rm e\text-fl}$ on the cutoff $\Lambda$ shows that the electron-flexural phonon coupling is irrelevant at large distance,
 while the two other couplings are marginal. This analysis applies to the quantum, low temperature, regime. In the classical regime (higher temperatures) there is a single coupling constant (which does not contain $\rho$), given by
 \be
 \lambda_{\rm cl} = \frac{\lambda_{\rm e\text-fl}}{\sqrt{\lambda_{\rm anh}}} = \frac{g}{\sqrt{K_0 v_{\rm F}/\Lambda}}.
 \ee
It measures the strength of the coupling and is again irrelevant at large scale.

Numerical estimates of the parameters appearing in Eq.~(\ref{11}) are  \cite{NGPNG09}
\bea \nn
&& a=1.4 {\rm\AA}  , \qquad a^2 K_0 \approx 20  {\rm eV}  , \qquad \kappa \approx 1 {\rm eV} \\
&&\Lambda_{{\rm c},h} \sim \pi/a   , \qquad a^4 \rho =M_C a^2 =1/E_C\approx 1/(10^{-3}  {\rm eV}) \nn \\
&& \omega_{{\rm c},h} \approx \sqrt{\frac{\kappa}{\rho}} \Lambda_{\rm c,h}^2 \approx \sqrt{10^{-3}}  {\rm eV}  , \nn \\
&& \frac{v_{\rm F}}a \approx 5 {\rm eV}  , \qquad  \alpha_{\rm e}=\frac{e^2}{\epsilon_0 v_{\rm F}} \approx 2 \label{param} 
\eea
where $\epsilon_0$ is the dielectric constant of the environment. The value above is obtained for suspended samples, $\epsilon_0 = 1$.
The parameter $\Lambda_{{\rm c},h}$ gives the UV-cutoff \footnote{A more accurate value is $\Lambda_{{\rm c},h} = \frac{2\pi}{\sqrt3 a}$.} for the $h$ field, and $\omega_{{\rm c,h}}$ is the corresponding frequency.

With these values of the parameters, the dimensionless couplings defined above are all of order unity
$\lambda_{\rm anh} \approx 0.6 , \alpha_{\rm e} \approx 2$ and $\lambda_{\rm e\text-fl} \approx 0.6 - 7$.
 The value of the last parameter is subject to a significant uncertainty, since, as mentioned above,
 estimates for $g_0 \approx g$ can vary by one order of magnitude. The ensuing probable range for the
 classical coupling constant is $ \lambda_{\rm cl} \approx g/5.6 \approx 0.7 - 8.8$.

 At finite temperature the flexural-phonon propagator is modified by the inclusion of the Bose-Einstein distribution, which tends to the Boltzmann distribution at temperatures higher than the phonon frequencies. Fluctuations are enhanced as the temperature increases, making the anharmonic effects discussed here more important. A detailed analysis is carried out in Section \ref{membrane0}.

\section{self-consistent screening method}
\label{s:SCSA}
We now give the complete SCSA equations in the Matsubara equilibrium setting. They are much more
general than what we will be able to achieve below, but we hope they can stimulate further studies.

\subsection{Matsubara partition sum}

The equilibrium Matsubara partition sum is  $Z= \int {\cal D}[h] {\cal D}[\Psi] {\cal D}[\bar \Psi] e^{- S}$ in terms of the imaginary-time Matsubara action
$S =  S_0 + S_{\rm int}$ with \begin{widetext}
\begin{eqnarray} \label{mats}
S_0 &=&  \int \rmd^2 x  \int_0^\beta \rmd \tau\,  \sum_{a=1}^d \bigg[\frac{\rho}{2} (\partial_\tau h_a)^2 + \frac{\kappa}{2} (\nabla^2 h_a)^2  \bigg] +  \frac{1}{\beta} \sum_{\omega'_n} \int_{q} \sum_{\gamma=1}^{N_{\rm f} d}  \bar \Psi_\gamma(-q,-\omega'_n) \bigg[ - v_{\rm F} \boldsymbol{\sigma} \cdot (- i \nabla) - ( i \omega'_n + \mu ) \1 \bigg]  \Psi_\gamma(q,\omega'_n)  \nn\\
 S_{\rm int} &=& \frac{1}{d} \frac{1}{\beta} \sum_{\omega_n} \int_{q}  \Bigg\{  \frac{1}{2} \bigg[ \frac{1}{2} {\rm P}^{\rm T}_{ij}(\partial)  \sum_{a=1}^d \partial_i h_a  \partial_j h_a \bigg]_{q,\omega_n} K_0(q,\omega_n) \bigg[ \frac{1}{2} {\rm P}^{\rm T}_{ij}(\partial)  \sum_{a=1}^d \partial_i h_a  \partial_j h_a \bigg]_{-q,-\omega_n}  \nn
\\
&& -  \sum_{\gamma=1}^{N_{\rm f} d} [\bar \Psi_\gamma \1  \Psi_\gamma]_{-q,-\omega_n} g(q,\omega_n) \left[ \frac{1}{2} {\rm P}^{\rm T}_{ij}(\partial)  \sum_{a=1}^d \partial_i h_a  \partial_j h_a \right]_{q,\omega_n} + \frac{1}{2} \sum_{\gamma=1}^{N_{\rm f} d} \left[\bar \Psi_\gamma \1  \Psi_\gamma\right]_{q,\omega_n} V(q,\omega_n) \sum_{\gamma=1}^{N_{\rm f} d} \left[\bar \Psi_\gamma \1  \Psi_\gamma\right]_{-q,-\omega_n}
\Bigg\}.\nn\\ 
\end{eqnarray}\end{widetext}
We have enlarged the model to frequency and momentum dependent couplings for future convenience. The bare couplings are $K_0(q,\omega)=K_0$,
$g(q,\omega)=g$. The bare electron-electron interaction is  $V(q,\omega)=V(q)$. We denote by $\tau$ the imaginary time, and by $\omega_n: = 2 \pi n/\beta$
 the bosonic Matsubara frequencies. The fermionic ones are
$\omega'_n:= 2 \pi (n+\frac{1}{2})/\beta$; we will need them only rarely, since the composite fields $\bar \Psi_a \1 \Psi_a$,
as well as the polarization bubble (denoted $J$ below), contain only bosonic frequencies. We have added a chemical potential $\mu$
for the electrons, but we will set it to zero in the following. We denote $\int_{q}:=\int_{|q|<\Lambda}
\frac{\rmd^2q}{(2 \pi)^2}$ with an implicit UV cutoff $\Lambda$. The Pauli matrices are denoted in bold face,  $\boldsymbol{\sigma}_x:=\left({0\atop1} {1\atop 0}\right)$, $\boldsymbol{\sigma}_y:=\left({0\atop-i}
{i\atop 0}\right)$, to not confuse them with the auxiliary field $\sigma$ to be introduced later. By $q \cdot \boldsymbol{\sigma}$ we denote the matrix $q_x  \boldsymbol{\sigma}_x+q_y \boldsymbol{\sigma}_y$.

\subsection{Bare propagators: Quantum and classical limits}

In the absence of interactions, the bare propagators of the flexural phonon and of
the free fermions  are obtained from $S_0$ as
\bea
&&  \langle h_a(-q,-\omega_n) h_b(q,\omega_n) \rangle_0 = \delta_{ab} G(q,\omega_n) ,
\\
&& G(q,\omega_n)  = \frac{1}{\kappa q^4 + \rho \omega_n^2}  \\
&& \langle \bar \Psi_\gamma(-q,-\omega'_n) \Psi_\beta(q,\omega'_n) \rangle =  F_{\gamma \beta}(q,\omega'_n)  \qquad \\
&& F(q,\omega'_n) = (i \omega'_n \1 + v_{\rm F} q\cdot{\boldsymbol{\sigma}})^{-1} \nn\\
&&\qquad ~~~~~= - \frac{1}{v_{\rm F}^2 q^2 + { \omega'_n }^2} (i \omega'_n \1- v_{\rm F} q \cdot \boldsymbol{\sigma}).
\eea
We recall that real-time equilibrium response functions are recovered from these propagators via the analytical continuation
$i \omega_n \to \omega + i \delta$ and $\delta =0^+$. For instance the equal-time equilibrium correlation function  in real
time is obtained from the fluctuation-dissipation theorem (FDT) as (restoring the $\hbar$ factors)
\bea
   C(q) &:=& \frac{1}{d} \langle h^a_{q,t} h^a_{q,t} \rangle
\nn\\ &=& \int \frac{\rmd\, \omega}{2 \pi} \hbar \coth\!\left(\frac{\beta \hbar \omega}{2}\right) {\rm Im\!}\left( \frac{1}{\kappa q^4 - \rho \omega^2 - i \delta \omega} \right) \nn\\
&=& \frac{\hbar \coth\!\big(\frac{\beta \hbar \omega_{\rm fl}(q)}{2}\big) }{2 \rho \omega_{\rm fl}(q)}. \label{Cq}
\eea
Here $\omega_{\rm fl}(q) = q^2 \sqrt{\kappa/\rho} $ is the frequency of the flexural phonons. Eq. (\ref{Cq}) interpolates between $C(q)=\frac{\hbar}{2 \rho \omega_{\rm fl}(q)}$ in the
quantum (i.e.\ zero-temperature) limit, to $C(q) = T/\kappa q^4$ in the classical (i.e.\ high-temperature) limit.
The same result is obtained from the imaginary-time, equal-time
average by performing the Matsubara summation\footnote{We use that
$\frac{1}{\beta} \sum_{\omega_n} h(i \omega_n) = \zeta \sum_k {\rm Res}[ h(z) \frac{1}{e^{\beta z} + \zeta}]\big|_{z=z_k}$
with $\zeta=-1$ for bosons and $\zeta=+1$ for fermions, if $h(z)$ has isolated poles at $z_k$.}
\be
C(q) = \frac{1}{\beta} \sum_{\omega_n}  G(q,\omega_n) \stackrel{\beta \to 0} \longrightarrow \frac{1}{\beta} G(q,\omega_n=0) =  \frac{T}{\kappa q^4} \label{Cq}\ .
\ee
In the high-temperature limit, $\beta \to 0,$ one can  replace $G(q,\omega_n) \to \delta_{n0} \frac{1}{\kappa q^4}$,
hence only the mode $\omega_n=0$ contributes, and (\ref{Cq}) reproduces the classical result. This remark
will be important to recover the classical SCSA equations from the quantum ones below.

\subsection{SCSA equations}

As is well-known from the $O(N)$ model (here $N\equiv d$) for $d=\infty,$ one can calculate exactly the dressed quartic
interactions as a geometric sum of the polarization bubbles. In these bubbles one  uses the bare propagators,  $G$ for the
phonons and $F$ for the fermions, which will be mainly what we achieve to do explicitly here. However, one can aim to go further and also calculate the corrections
to the self-energies to first order in $1/d$, leading to the {\em dressed} propagators denoted  $\tilde G$ and $\tilde F$. The SCSA
equations are the coupled Dyson equations which determine both the dressed propagators and the dressed interactions.
They provide a self-consistent approximation for any $d$. If one uses the bare propagators in these equations (as done below),
they give the dominant order at large $d$ for the interaction and the self-energy. Hence they are exact at large
$d$ to dominant order in $1/d$.

One starts by defining the (dressed) phonon bubble and (dressed) fermion loop as
\bea
 I(q,\omega) &=& \frac{1}{\beta} \sum_{\Omega_n} \int_{p} [p\cdot {\rm P}^{\rm T}(q) \cdot p]^2 \tilde G(p,\Omega_n) \nn\\
 && \qquad ~~~~~\times \tilde G(p+q,\omega+\Omega_n) \label{Idef}\\
 J(q,\omega) &=& - \frac{1}{\beta} \sum_{\Omega'_n}  \int_{p}\tr\Big( \tilde F(p,\Omega'_n) \tilde F(p+q,\omega+\Omega'_n)\Big) \label{Jdef}\qquad
\eea
They are given
in terms of the (dressed) propagators. Everywhere in this Section we work in Matsubara time, hence $\omega$ designates everywhere $\omega_n$ (or $\omega'_n$ for fermions), thus \( I(q,\omega)\) and \( J(q,\omega)\) {\em are not defined for all} $\omega$, but only the quantized, bosonic Matsubara frequencies.
The same equations also hold  in real time,  substituting $\omega \to - i  \omega + \delta$.

\subsubsection{Decoupled problem: Membrane}

In the absence of a coupling between electrons and phonons,  we can consider separately the membrane  and the electron problem.
The flexural propagator, including the correction to the self-energy of the flexural modes due to the quartic interaction, reads
\bea  \label{Gdressed}
\tilde G(q,\omega)^{-1} &=& \kappa q^4 + \rho \omega^2 \nn\\
&& + \frac{1}{d} \frac{1}{\beta} \sum_{\Omega_n}  \int_{p} [q\cdot {\rm P}^{\rm T}(p)\cdot q]^2 \tilde K_0(p,\Omega_n) \nn\\
&& \qquad~~~~~~~~\times \tilde G(p+q,\omega+\Omega_n).
\eea
It is given
in terms of the dressed interaction,
\be   \label{K0dressed}
\tilde K_0(q,\omega) = \frac{K_0}{1+ \frac{1}{2} K_0 I(q,\omega)}.
\ee
Equations (\ref{Idef}), (\ref{Gdressed}) and (\ref{K0dressed}) define the (quantum) SCSA equations for the membrane, i.e.\ for the phonon problem {\em alone}.

In the high-temperature limit $\beta \to 0$, as discussed above, $\tilde G(q,\omega_n)=\delta_{n0} \tilde G(q)$,
 $\tilde K_0(q,\omega_n) = \delta_{n0} \tilde K_0(q)$, $I(q,\omega_n) = \delta_{n0} I(q)$, and the above equations
reduce to the classical SCSA equations of Ref.\  \cite{LR92} (with the correspondence from here to there $K_0 \to 2 b$, $I \to 3 I$, $T \to 1$).
As is well known, the self-consistent solution of these equations at small $q$ leads to: (i) the softening of the elastic modulii $\tilde K_0(q) \sim q^{\eta_u}$ due to the screening of the elastic interactions by thermally excited out-of-plane modes, (ii) a stiffening of the bending rigidity $\tilde G(q) \sim q^{4-\eta}$ (equivalently $\kappa(q) \sim q^{-\eta}$)  with $\eta=4/(d+\sqrt{16-2 d+ d^2})$, $\eta_u=2-2\eta$, i.e.\ in $d=1$, $\eta \approx 0.82$, $\eta_u \approx 0.36$ in good agreement with first-principle numerical studies of graphene sheets \cite{ZRFK10}. While
the SCSA provides a reasonable approximation for any $d$, to obtain the direct-expansion result
$\eta=2/d + O(1/d^2)$ and $\eta_u=2 + O(1/d),$ it is sufficient to use  the bare propagator $\tilde G(q) \to G(q)$ in all  integrals of the SCSA equations. Note that more recently the SCSA has been extended to the next order in $1/d$ \cite{G09b}.

\subsubsection{Decoupled problem: Dirac electrons}

Consider now the SCSA equations for the electrons alone, in presence of a bare electron-electron interaction
$V(q)$. The correction to the self-energy of the
electrons due to the quartic electron-electron interaction reads\bea  \label{Fdressed}
\tilde F(q,\omega)^{-1} &=& v_{\rm F} q\cdot\boldsymbol{\sigma} + i \omega \nn\\
&& + \frac{1}{d \beta} \sum_{\Omega_n} \int_{p}  \tilde V(p,\Omega_n) \tilde F(p+q,\omega+\Omega_n). ~~~~~~~
\eea
$\tilde V$ is  the dressed interaction,
\be  \label{Vdressed}
\tilde V(q,\omega) = \frac{V(q)}{1 + N_{\rm f} V(q) J(q,\omega)}
.\ee
Equations  (\ref{Jdef}), (\ref{Fdressed}) and (\ref{Vdressed}) are the SCSA equations
for the electron problem alone. If the bare propagators are inserted, these equations are called GW and RPA
and have been studied \cite{GGV94} for the Coulomb interactions at $T=0$.
They exhibit a logarithmic divergence; thus the corresponding RG flow for $v_{\rm F}$ and the wave-function renormalization $Z$ was
obtained,  implicitly, to first order in $1/(d N_{\rm f})$. In the infrared $v_{\rm F}$ increases, leading to a downward flow  of
the dimensionless coupling $\alpha_{\rm e}$ (since $e^2$ is not renormalized), which is  marginally irrelevant \cite{GGV94}. This effect was observed
in experiments \cite{Eetal11}.

\subsubsection{Coupled problems}

Since the SCSA has been so successful to describe separately the membrane and the
electron problem, it is tempting to apply it to the coupled problem.

In the presence of an electron-phonon coupling, the bubbles $I$ and $J$ are still
defined by (\ref{Idef}) and (\ref{Jdef}), and the equations (\ref{Gdressed}) and (\ref{Fdressed})
are still valid. To express the dressed interactions however, we must now consider 2 by 2 matrices.
We define (for each $q$ and $\omega$, which are implicit),\bea
\tilde {\cal V} &:=& \left(\begin{array}{cc}
\tilde K_0 & - \tilde g  \\
- \tilde g  & \tilde V
\end{array} \right)    , \\
 {\cal V} &:=& \left(\begin{array}{cc}
K_0 & - g  \\
- g  & V
\end{array} \right)   , \\
 {\cal J} &:=& \left(\begin{array}{cc}
\frac{1}{2} I & 0  \\
0 &  N_{\rm f} J
\end{array} \right). \label{J}
\eea
The last SCSA equation expresses the dressed interactions as
\be
\tilde {\cal V} = {\cal V} (\1 + {\cal J}   {\cal V})^{-1} .
\ee
The matrix elements are
\bea \label{coupled1}
 \tilde K_0(q,\omega) &=& \frac{K_0 [1 + N_{\rm f} V(q) J(q,\omega) ] - g^2 N_{\rm f} J(q,\omega)}{D(q,\omega)} ,~~~~~~~\\
\tilde g(q,\omega) &=& \frac{g}{D(q,\omega)},  \\
\tilde V(q,\omega) &=& \frac{V(q) [ 1+ \frac{1}{2} K_0 I(q,\omega)] - \frac{1}{2} g^2 I(q,\omega) }{D(q,\omega)}.
\eea
We have defined the determinant
\bea
D = {\rm det} (\1 + {\cal J}   {\cal V}). \label{det1}
\eea
More precisely,\bea  \label{coupled2}
   D(q,\omega) &=& \Big[ 1+ \frac{1}{2} K_0 I(q,\omega)\Big]\Big[1
+ N_{\rm f} V(q) J(q,\omega)\Big] \nn\\
&& - \frac{N_{\rm f}}{2} g^2 I(q,\omega) J(q,\omega).
\eea
The closed set of equations (\ref{Idef}), (\ref{Jdef}), (\ref{Gdressed}), (\ref{Fdressed}), and
(\ref{coupled1})--(\ref{coupled2}) are the SCSA equations for the coupled electron-phonon
problem. Again, they are exact at dominant order for $d \to \infty$ (in which case
we can use bare propagators in the integrals). Alternatively, using the dressed propagators, they provide an approximation for
any $d$. They also   contain the two special cases  of the uncoupled systems discussed above.

It is important to note the equivalent formulation in terms of  ``dressed bubbles", or dressed two-particle propagators, or susceptibilities as
\bea\label{36}
\tilde {\cal J} &=&  ( \1+ {\cal J}   {\cal V})^{-1} {\cal J}   = {\cal J}  (\1 +  {\cal V} {\cal J}  )^{-1}  ,\\ \tilde {\cal J} &=&  \left(\begin{array}{cc}
\frac{1}{2} \tilde I & \tilde \Pi  \\
~ \tilde \Pi   &  N_{\rm f} \tilde J
\end{array} \right).
\eea
It satisfies $\tilde {\cal V} = {\cal V} - {\cal V} \tilde {\cal J} {\cal V}$ and $\tilde {\cal J} = {\cal J} - {\cal J} {\cal V} \tilde {\cal J}$,
more specifically,
\bea
\tilde I(q,\omega) &=&  I(q,\omega) \Big[1 + N_{\rm f} V(q) J(q,\omega)\Big]/D(q,\omega)\qquad \\
 \tilde \Pi(q,\omega) &=&  \frac{1}{2} g N_{\rm f} I(q,\omega) J(q,\omega)/D(q,\omega)  \\
 \tilde J(q,\omega) &=& J(q,\omega) \Big[1+ \frac{1}{2} K_0 I(q,\omega)\Big]/D(q,\omega)
\label{40}\eea
The interest of this approach is that if one calls the composite fields,
\bea
&& \Phi(x) = \frac{1}{d} \sum_{a} \frac{1}{2} {\rm P}^{\rm T}_{ij}(\partial) \partial_i h^a(x) \partial_j h^a(x) \\
&& \rho(x) = \frac{1}{d}  \sum_{\gamma=1}^{N_{\rm f} d} \bar \Psi_\gamma(x) \1 \Psi_\gamma(x)   - \rho_0
\eea
Then
\be \label{susc1}
\tilde {\cal J} =d
\left(\begin{array}{cc}
\langle \Phi(-q,-\omega) \Phi(q,\omega)) \rangle & \langle \Phi(-q,-\omega) \rho(q,\omega) \rangle  \\
\langle \rho(-q,-\omega) \Phi(q,\omega)) \rangle & \langle \rho(-q,-\omega) \rho(q,\omega) \rangle
\end{array} \right).
\ee
We will not attempt  to solve here the self-consistent equations
(\ref{Idef}), (\ref{Jdef}), (\ref{Gdressed}), (\ref{Fdressed}), and
(\ref{coupled1})--(\ref{coupled2}). Instead we will use them by inserting
the bare propagators $\tilde G \to G$ and $\tilde F \to F$ in all the
integrals in the SCSA equations. Then to leading order in
$d \to \infty$ we need only (\ref{Idef}), (\ref{Jdef}) and
(\ref{coupled1})--(\ref{coupled2}). The additional equations
 (\ref{Gdressed}), (\ref{Fdressed}) then give the $O(1/d)$ corrections
to the propagators. These lead to the renormalizations of $\kappa,\rho,\alpha_e,v_{\rm F}$
which we will not study in detail here, as they are not needed to leading order at
large $d$.

\section{Analysis of the results}
\label{s:AnaResults}
We start by giving the explicit expression for the bubbles $I$ and $J$, calculated with the
bare propagators, hence denoted below $I_0$ and $J_0$. Then we analyze the consequences
first for the membrane alone, and then for the coupled system.

\subsection{Flexural bubble and membrane alone}  \label{membrane0}

A general expression for the flexural bubble $I_0$ at any temperature is given in  Appendix \ref{a:phonon-bubble}.
An explicit form is obtained in the quantum limit, $T=0,$ and in the classical limit.
The result there is given in Matsubara frequency.


\begin{figure*}
\includegraphics[width=0.68\columnwidth]{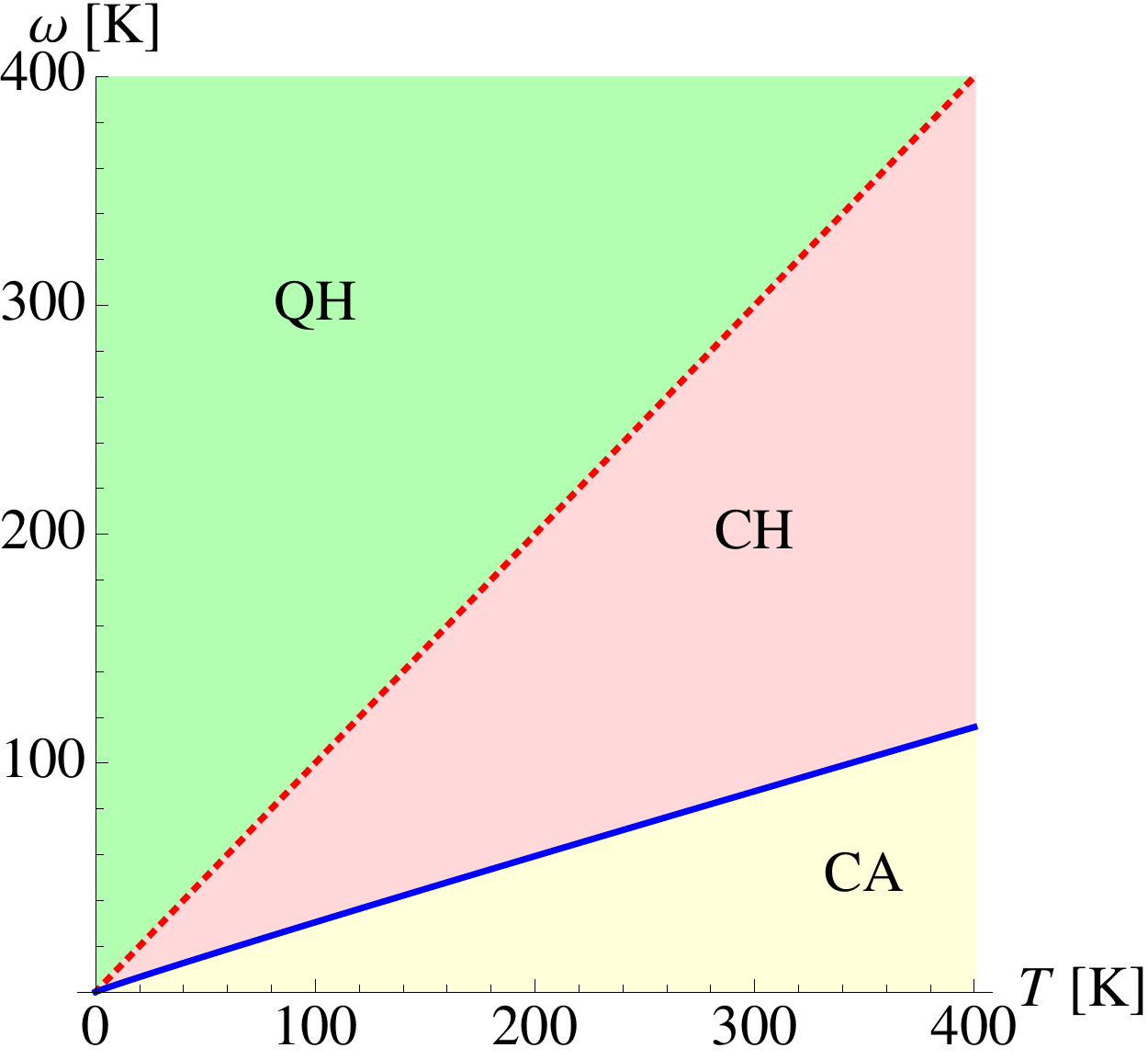}
\includegraphics[width=0.68\columnwidth]{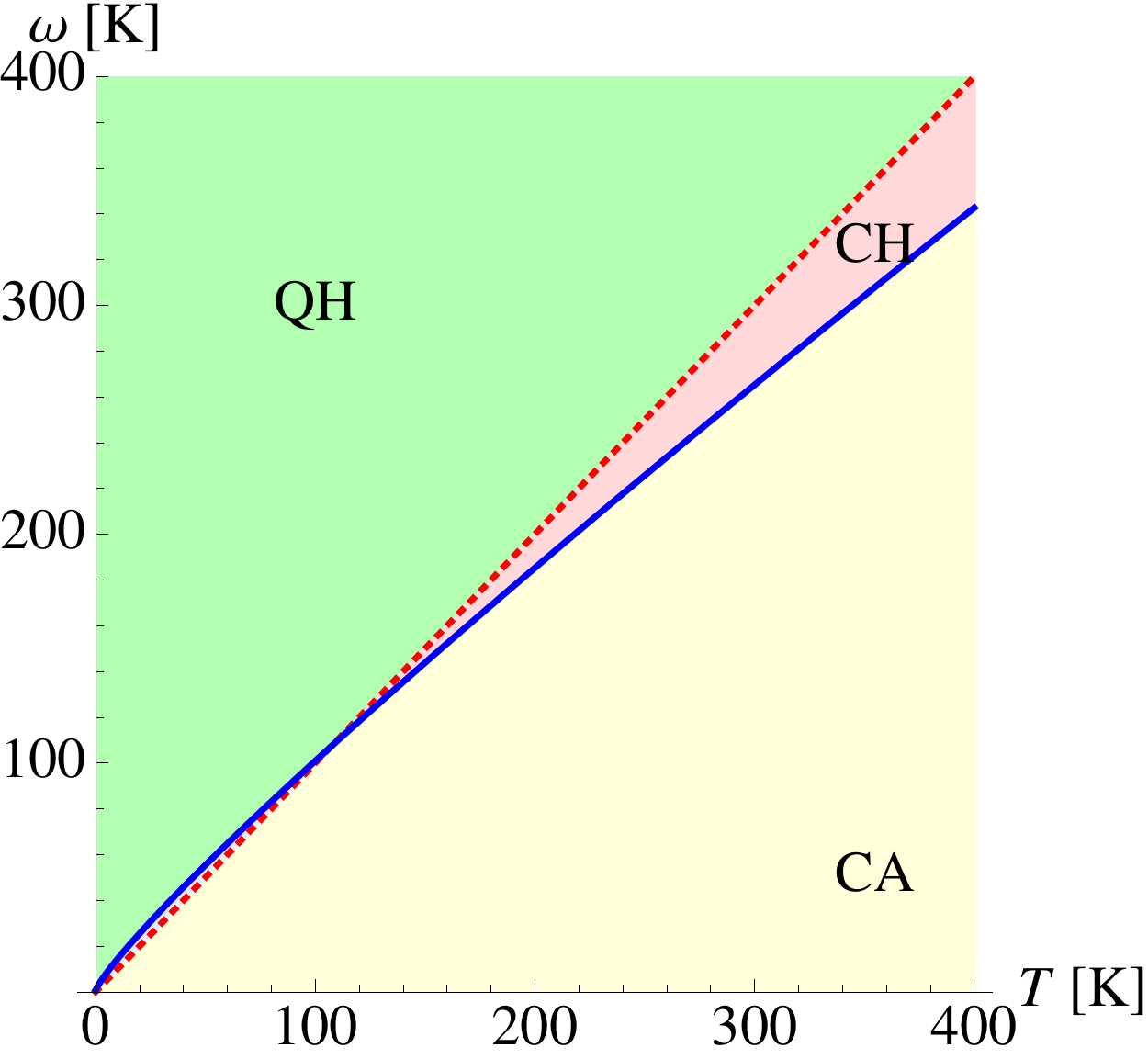}
\includegraphics[width=0.68\columnwidth]{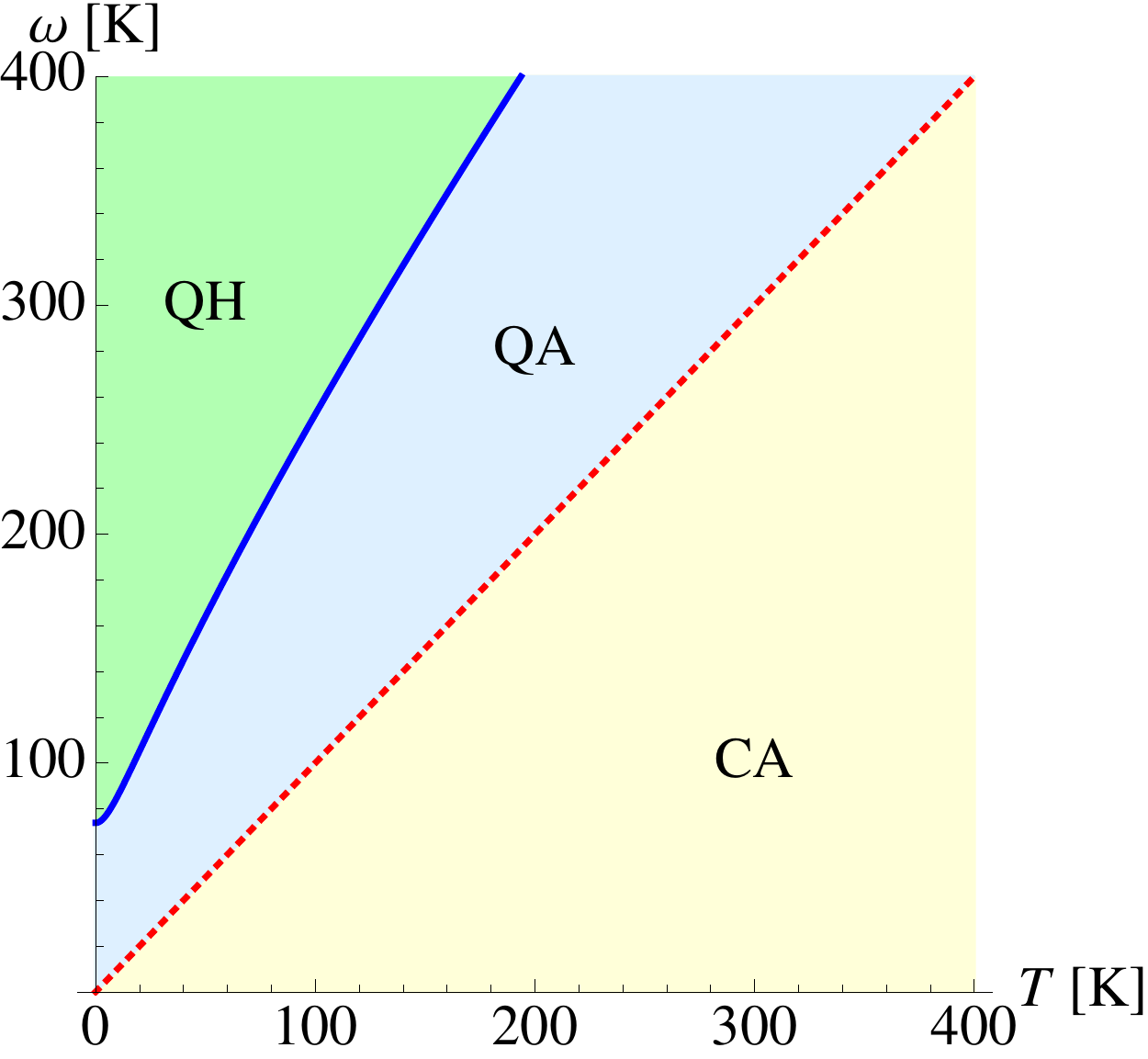}
\caption{Crossovers in scales: plot of $\omega_a(T)$ versus $T$ (both in Kelvin) for $\lambda_{\rm anh}=5,16,35$ (from top left to bottom) corresponding
to the three regimes (i-iii) described in the text. The diagonal $\omega_a=T$ divides quantum from classical region.
Four combined regions are possible, as represented (QH: quantum-harmonic,
QA: quantum-anharmonic,CH: classical-harmonic,CA: classical-anharmonic). The
vertical axis equivalently measures the wavector squared $q_{\rm anh}^2 = \frac{\omega_a}{100K} \times 0.14$ \AA$^{-2}$
For simplicity the approximation $g(x) \approx 1/(1+5 x)$ is used, which is found to be accurate. 
}\label{fig1}
\end{figure*}

\subsubsection{Zero frequency and zero temperature (quantum) limit:}

Let us start with the result at zero frequency (which is the same in real and imaginary time).
In the quantum case the momentum integral is UV divergent and depends on the UV cutoff $\Lambda$.
At $T=0$ it reads, in  dimensionless form,
\bea  \label{bubblezero}
   K_0 I_0(q,0) &=&  \frac{3}{64 \pi} \lambda_{\rm anh} f(s), \\
f(s) &=&  \frac{2}{3 (s+1)}+\frac{1}{2} \log
   \left(\frac{s+1}{16}\right)+\frac{3}{4},  \qquad  \\
   s &=& \frac{4 \Lambda^2}{q^2} .
\eea
The parameter $\lambda_{\rm anh}$ was defined in (\ref{11}).
The function $f(s)$ is of order unity, and we have given its explicit form for the circular cutoff used, see Appendix  \ref{a:phonon-bubble}. Let us stress that its details depend on the  chosen cutoff.
However, $I_0(q,0)$ is a log-divergent integral, and its dependence on $\ln \Lambda/(2 p)$ is universal, which can be summarized by a RG equation at $T=0$,
\begin{equation}\label{a19}
 -\frac{ q \partial}{\partial {q}} I_0(p,0) =  \frac{\Lambda\partial }{\partial \Lambda } I_0(q,0) =\frac{3}{64 \pi  \kappa ^{3/2} \sqrt{\rho }}, \mbox{ for } q \ll \Lambda.
\end{equation}
For $d=\infty$, inserting the form of $I_0(q,0)$ in (\ref{K0dressed}), we obtain the
effective Young modulus $\tilde K_0 := \tilde K_0(q,0)$ at momentum $q$. It satisfies the exact
RG equation (valid at all $T$) obtained from (\ref{K0dressed}),
\be
- q \partial_{q} \tilde K_0 (q,0)= - \frac{1}{2} \Big[ - q \partial_{q} I_0(q,0)\Big] \tilde K_0^2 (q,0).
\ee
Using (\ref{a19}) it yields the RG equation in the quantum limit $T=0$ as
\be
- \frac{q \partial} {\partial {q}} \tilde K_0 (q,0)
=  - \frac{3}{128 \pi  \kappa ^{3/2} \sqrt{\rho }} \tilde K_0^2   (q,0), \ T=0,\  q \ll \Lambda, \label{rgq}
\ee
recovering the result\footnote{Note however the
discrepancy of a factor of $2,$ due presumably to a misprint in  \cite{SGG11}.} of Ref.  \cite{SGG11}. 

To estimate the importance of the anharmonic effects, let us write schematically the relative
correction to the Young modulus at any $T$ as
\be  \label{add}
 - \frac{\delta K_0}{K_0} =  - \frac{\tilde K(q,0)- K_0}{K_0}  \approx \frac{1}{2} K_0 I_0(q,0),
\ee
and define the {\it anharmonic scale} by the wave-vector $q_{\rm anh}(T)$ such that
\bea \label{add2}
\left|\frac{\delta K_0}{K_0} \right| \approx 1/2 \quad , \quad K_0 I_0(q_{\rm anh},0) \approx 1\ .
\eea
It means that above this length scale, i.e.\ for $q<q_{\rm anh}(T)$, the anharmonic effects are important, while for
smaller length scales the corrections to the bare elastic energy due to the quartic $h$ vertex can be neglected, a regime
which we call ``harmonic".

Our result at $T=0$ is thus:
\bea
&& - \frac{\delta K_0}{K_0}  \approx \frac{3 \lambda_{\rm anh}}{128 \pi}  \ln\left(\frac{2 e^{3/4} \Lambda}{q}\right)  \\
&& q_{\rm anh}(T=0) \approx \Lambda e^{- \frac{64 \pi}{3 \lambda_{\rm anh}}} .
\eea
This means that the {\it quantum anharmonic scale} is very large (i.e.\ $q_{\rm anh}(T=0)$ is very small), unless
$\lambda_{\rm anh}$ is significant. In summary, the quantum anharmonic effects
are weak.

\subsubsection{Zero frequency: quantum-classical crossover at finite temperature}
\label{sec:crossoverT}

\begin{figure*}[t]
\includegraphics[height=6cm]{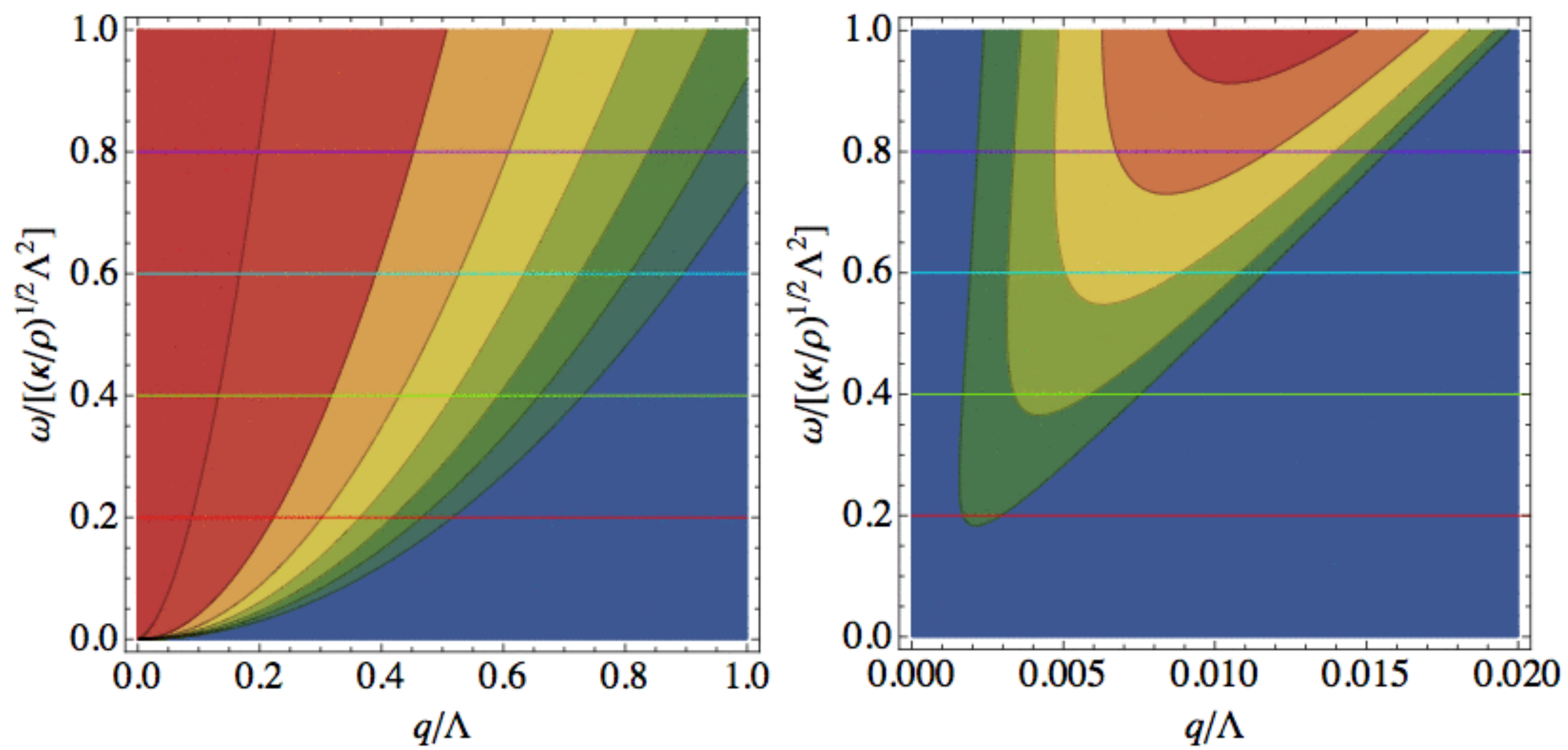}\\
\includegraphics[width=6cm]{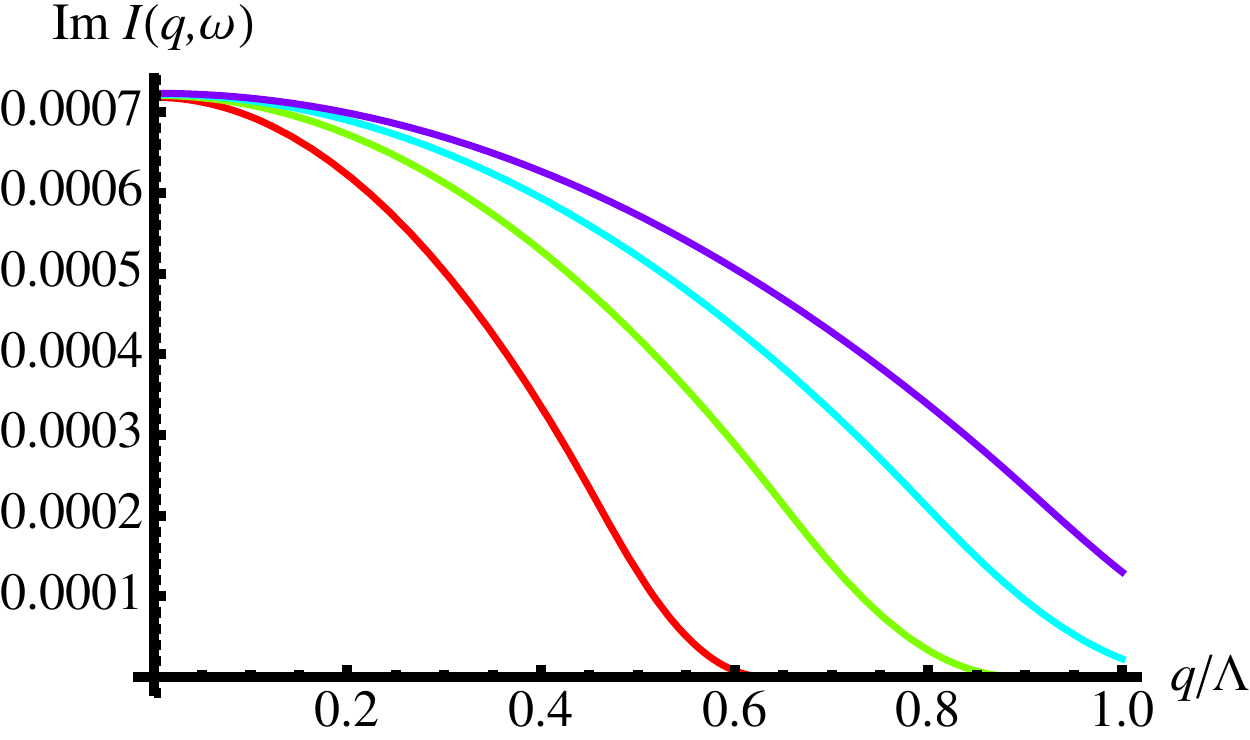}
\includegraphics[width=6cm]{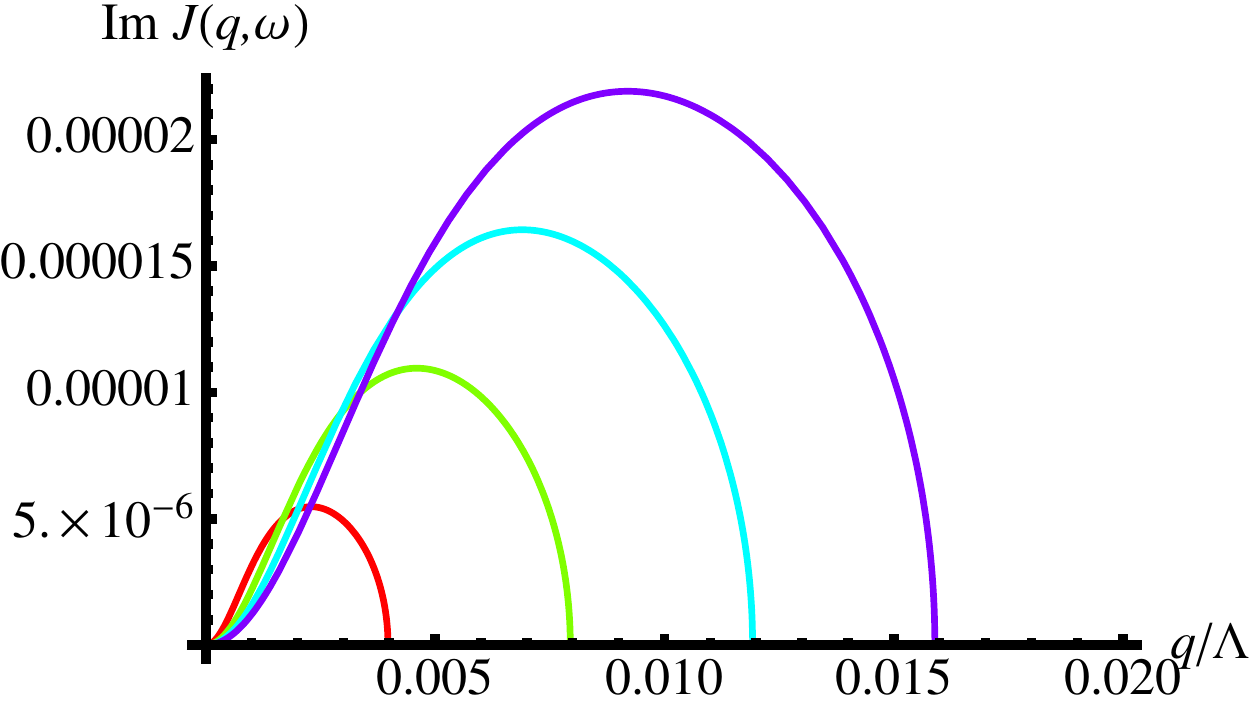}
\caption[fig]{(Color online). Top: Plot of the imaginary part of the propagators $\tilde I ( q , \omega )$ (left), and $\tilde J ( q , \omega )$ (right) for $\lambda_{\rm e\text-fl} = 0$ in the absence of the attractive part of the interaction. Bottom: Cuts of the  same functions for for $\omega = 0.2 \omega_{\rm c} ,
 0.4 \omega_{\rm c} , 0.6 \omega_{\rm c} {\rm and} \, 0.8 \omega_{\rm c}$. Colors on the lower plots correspond to colors of the slices in the upper plots. Note that for $\tilde J(q,\omega)$, we have restricted the range of $q/\Lambda$ to small values.}
\label{fig2}
\end{figure*}%

We now discuss the flexural bubble $I_0(q,0)$ at finite temperature $T>0$, which allows to describe the quantum to classical
crossover as a function of temperature. Evaluating  $\tilde K_0(q,0)$ then allows to ascertain the importance
of the anharmonic effects as a function of $T$ and wave vector $q$.

{\it (i) classical, high $T$ limit:} First we recall that in that limit
the flexural bubble is UV convergent and given by (\ref{B2}),
\be \label{class1}
I_0(q,0) = \frac{3}{16 \pi}  \frac{T}{\kappa^2 q^2},
\ee
a well-known expression. It results in the classical RG equation\be
- q \partial_{q} \tilde K_0 (q,0)=  - \frac{3}{16 \pi}  \frac{T}{\kappa^2 q^2}  \tilde K_0^2   (q,0), \quad {\rm classical},
\ee
which is exact in the $d=\infty$ limit. Comparing with (\ref{rgq}) we see that
in both cases the Young modulus is reduced at small $q$, but the classical, i.e.\ thermal, screening  is much
stronger than the quantum one. The {\it classical anharmonic scale}, defined from (\ref{add2}), is 
\be
q_{\rm anh}^{\rm clas}(T)^2  =  \frac{3}{16 \pi}  \frac{K_0 T}{\kappa^2}
,\ee
the well-known scale beyond which the standard SCSA predicts a softening of the elastic moduli of graphene.

\medskip

{\it (ii) arbitrary $T$: quantum to classical crossover:} In  Appendix \ref{a:phonon-bubble} we obtain that
\be
I_0(q,0) \simeq \frac{3 \log \left(
   \frac{\Lambda}{2 q} e^{3/4} \right)}{64 \pi  \kappa ^{3/2} \sqrt{\rho }}  +\frac{3}{16 \pi} \frac{T}{\kappa^2
q^2}\, g\!\left(     \frac{\omega_{\rm fl}(q)}{8 T} \right)  \label{crossoverT}
\ee
for $q \ll \Lambda$, and we recall the flexural phonon frequency $\omega_{\rm fl}(q)=q^2 \sqrt{\kappa/\rho}$.
The decreasing function $g(x)$, calculated in  Appendix  \ref{a:phonon-bubble},
thus describes the thermal crossover from the classical result (\ref{class1}) with $g(0)=1,$
to the quantum one (\ref{bubblezero}), with $g(\infty)=0$, as the temperature is decreased. The crossover scale
extracted from the function $g(x)$ occurs for $q \approx q_{\rm Q}(T)$ with \be q_{\rm Q}(T)^2 ={\red } T \sqrt{\frac{\rho}{\kappa}}.\ee
 i.e.\ , not surprisingly, the Debye scale. Smaller length scales show quantum behavior, while larger length scales behaves
 classically. Let us recall that the Debye temperature $T_\Lambda$ corresponds to $q_{\rm Q}(T_\Lambda)=\Lambda,$ beyond which all scales behave classically. For $T$ of the order or larger than $T_\Lambda$ the crossover behaves differently (see Appendix \ref{a:phonon-bubble})
 however this is not relevant for graphene, where $T_\Lambda \approx 3400$K.

The structure of our result (\ref{crossoverT}) is interesting: It can be interpreted as a sum of quantum and thermal fluctuations.
When $T \gg \omega_{\rm fl}(q)$, i.e.\ $q < q_{\rm Q}(T)$ one obtains the direct sum of (\ref{bubblezero}) and (\ref{class1}):
\bea
K_0 I_0(q,0) \simeq \frac{3 \lambda_{\rm anh}}{64 \pi}  \left[4  \frac{q_{\rm Q}(T)^2}{q^2} + \ln\left(\frac{\Lambda}{q}\right) \right].
\eea
From now on we approximate $e^{3/4}/2 \approx 1$.
Note that at any finite $T$ the integral is UV divergent, hence (\ref{class1}) is recovered only when
the thermal part overwhelms the quantum part, i.e.\ for $T > \frac{1}{4} \omega_{\rm fl}(q) \ln(\Lambda/q)$.

By contrast, at low $T$, the leading corrections are $O(T^2)$ and using the results of  Appendix \ref{a:phonon-bubble}, we find
\be \!K_0 I_0(q,0) \simeq \lambda_{\rm anh} \bigg[  \frac{3}{64 \pi}  \log \left(
   \frac{\Lambda e^{3/2}}{q} \right) +\frac{3 C}{2 \pi} \frac{T^2}{\omega_{\rm fl}(q)^2}  \bigg]
,\ee
with $C=0.205617..$.

\begin{figure*}
\includegraphics[height=6cm]{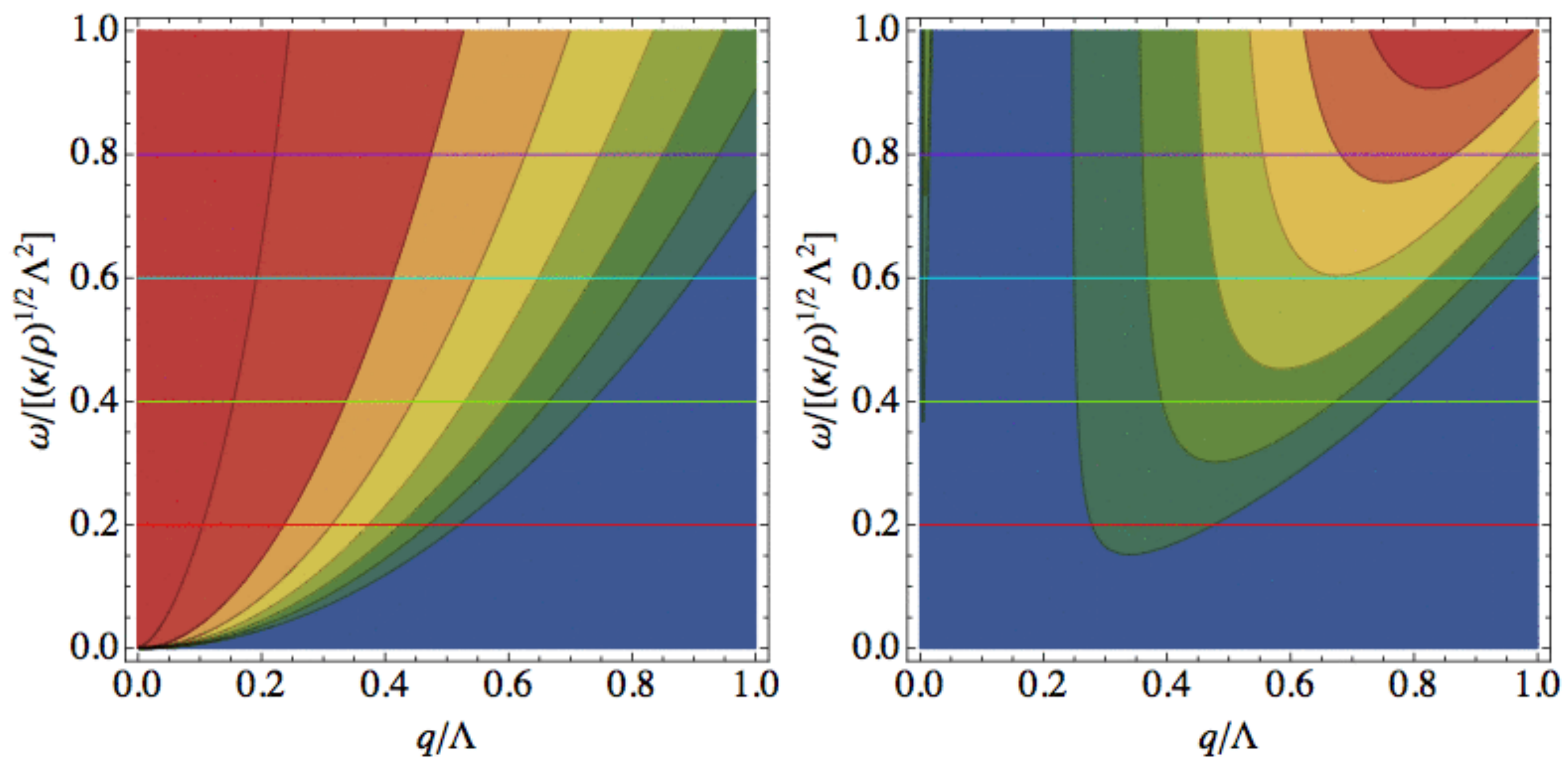}\\
\includegraphics[width=6cm]{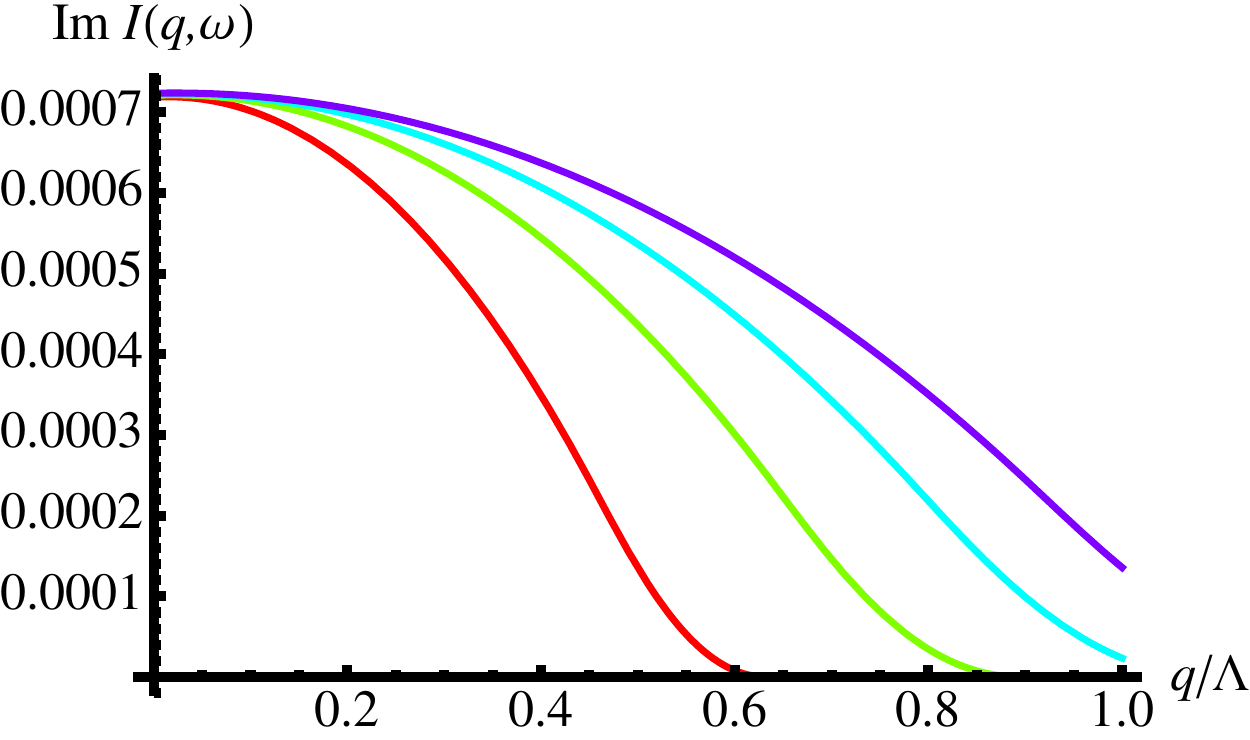}
\includegraphics[width=6cm]{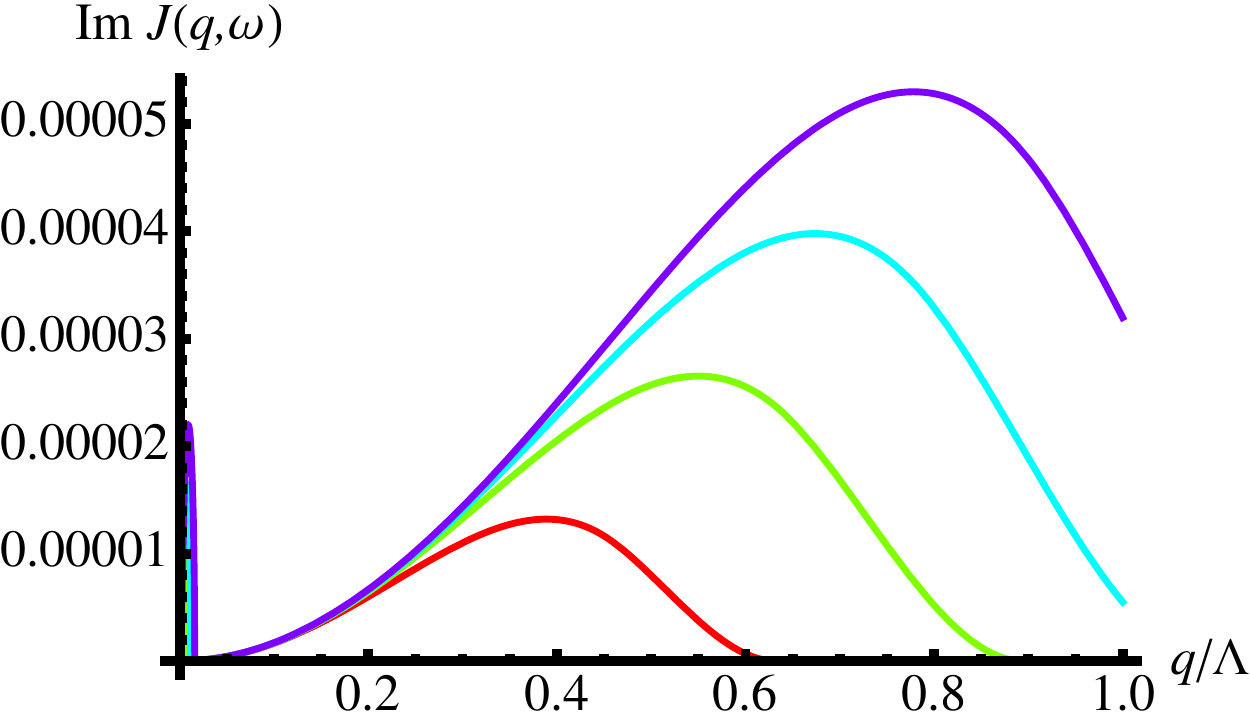}
\caption[fig]{(Color online). Top: Plot of the imaginary part of the propagators $\tilde I ( q , \omega )$ (left), and $\tilde J ( q , \omega )$ (right) for $\lambda_{\rm e\text-fl} = 10$ in the absence of the attractive part of the interaction. Bottom: Cuts of the  same functions for for $\omega = 0.2 \omega_{\rm c} ,
 0.4 \omega_{\rm c} , 0.6 \omega_{\rm c} {\rm and} \, 0.8 \omega_{\rm c}$. Colors on the lower plots correspond to colors of the slices in the upper plots.}
\label{fig4}
\end{figure*}%

We now obtain the ``phase diagram", delimiting anharmonic/harmonic and quantum/classical regions in the $T,q$ plane.
It is more conveniently expressed in terms of the frequency (i.e.\ energy) variable
\be
\omega_{a}(T) \leftrightarrow q_{\rm anh}(T)    , \quad \omega_{a} =  \omega_{\rm fl}(q_{\rm anh}) = q^2_{\rm anh} \sqrt{\kappa/\rho}
\ ,\ee
which is the root of the equation
\bea
\frac{1}{8} \ln \frac{T_\Lambda}{\omega_a} + \frac{T}{\omega_a} g\left(\frac{\omega_a}{8 T}\right) = \frac{16 \pi}{3 \lambda_{\rm anh}}
.\eea The curve $\omega_a(T)$ is represented in Fig \ref{fig1}. There the
vertical axis measures $\omega_a$ in Kelvin, equivalently wavector squared $q_{\rm anh}^2$, with the
correspondence $q_{\rm anh}^2 = \frac{\omega_a}{100K} \times 0.14$ \AA$^{-2}$. Below
this curve anharmonic effects are important. We have also plotted the diagonal line
$\omega_a=T$, which corresponds to the crossover $q^2=q^2_{Q}(T)$ from quantum (to the left)
to classical (to the right). Two important features are:

(i) the curve $\omega_a(T)$ crosses the diagonal $\omega_a=T$ only
if $\lambda_{\rm anh} < \lambda^*=\frac{16 \pi}{3 g(1/8)} \approx 24$.

(ii) the curve $\omega_a(T)$ is asymptotic at high $T$ to a straight line with a slope $z$,
solution of $\frac{1}{z} g(z/8)=\frac{16 \pi}{3 \lambda_{\rm anh}}$. Hence at high $T$,
$\omega_a(T) \simeq z T$ with
\bea
&& z \simeq \frac{3 \lambda_{\rm anh}}{16 \pi}  \approx \lambda_{\rm anh}/16.8   \ , \quad \lambda_{\rm anh} \ll \lambda^* \\
&& z \simeq \sqrt{\frac{24 C \lambda_{\rm anh}}{16 \pi}}
\approx \sqrt{\lambda_{\rm anh}/10.5} \  , \quad \lambda_{\rm anh} \gg \lambda^*
~~~~~~\eea
and $z=1$ for $\lambda=\lambda^*$.

Hence, as a function of $\lambda_{\rm anh}$, we can distinguish three regimes, represented in Fig \ref{fig1}:

(i) small $\lambda_{\rm anh} \ll \lambda^*$: The value of $q_{\rm anh}(T=0)$ is immeasurably small, hence
$\omega_a(T)$ is essentially a straight line lying well below the diagonal. There are
three regions QH, CH and CA (from left-up to right-down). Observing the region QA would require
gigantic length scales.

(ii) moderate $\lambda_{\rm anh} < \lambda^*$: The two curves now cross, hence there are now four regimes,
although the region QA remains quite limited.

(iii) $\lambda_{\rm anh} > \lambda^* $: $\omega_a(T)$ lies above the diagonal. There are three regimes
QH, QA and CA.

In summary, we have given here, for completeness, a general discussion of the crossover
as a function of the anharmonic coupling
$\lambda_{\rm anh}$. In graphene,
however, the situation seems to be (i), i.e.\ small coupling. Note however that there
is still some uncertainties on the value of the {\it bare} Young modulus since experiments
extract a renormalized one. Also, while the present scenario seems robust, the precise
values, e.g.\ of $\lambda^*$ will be affected by the renormalisation of $\kappa$, not
taken into account here.



\begin{figure*}
\includegraphics[height=6cm]{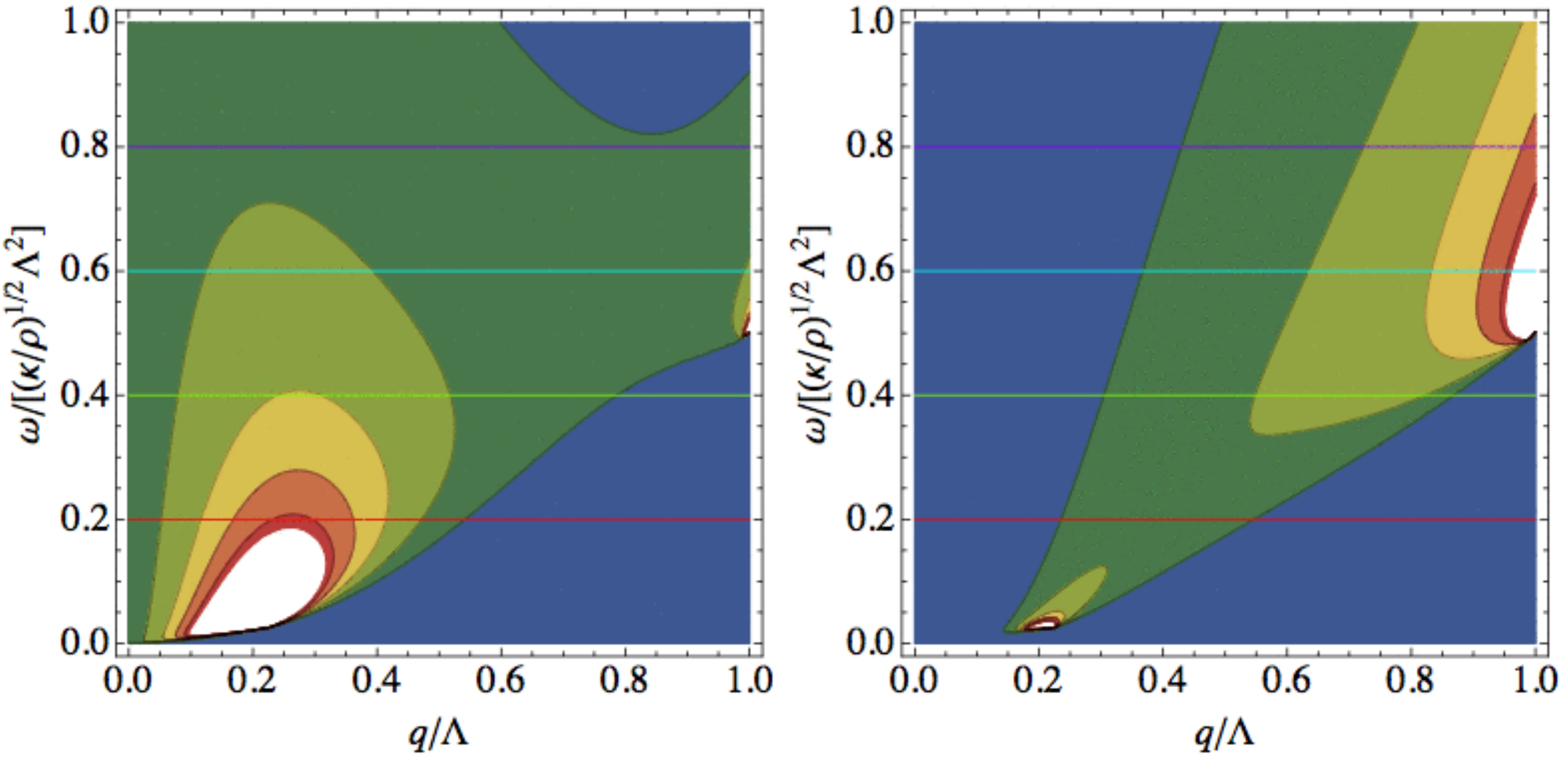}\\
\includegraphics[width=6cm]{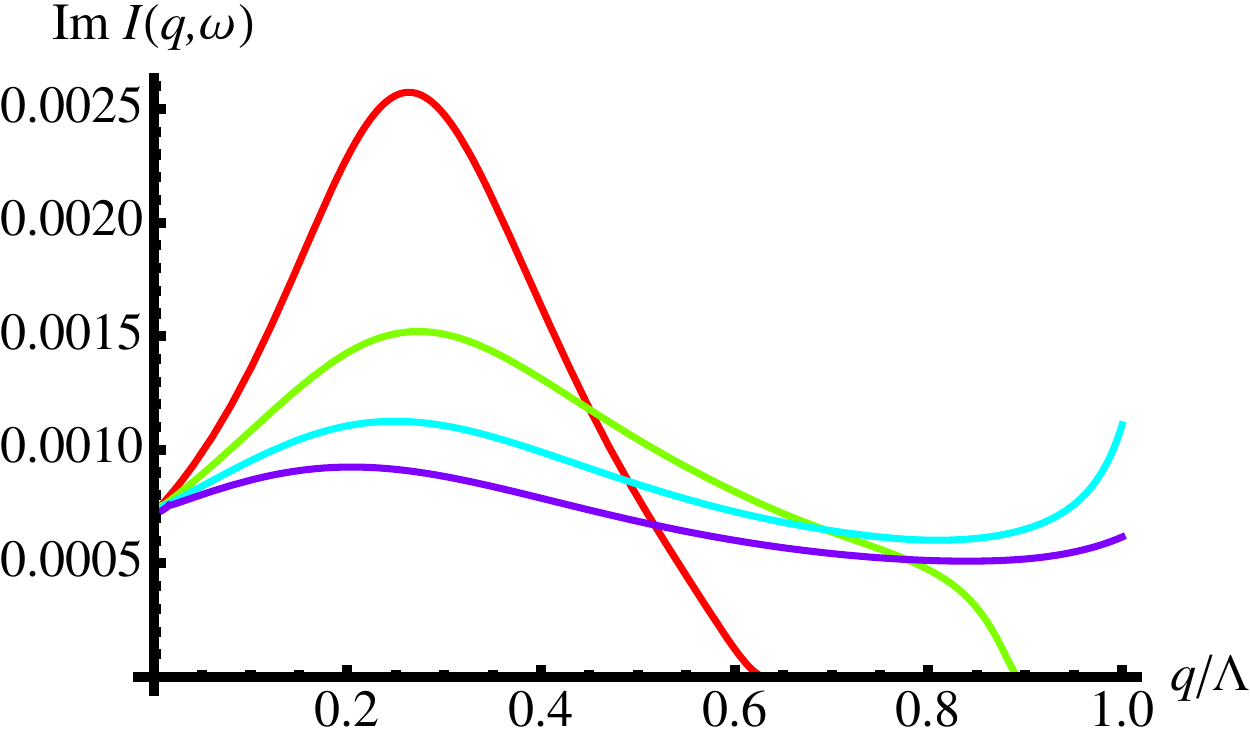}
\includegraphics[width=6cm]{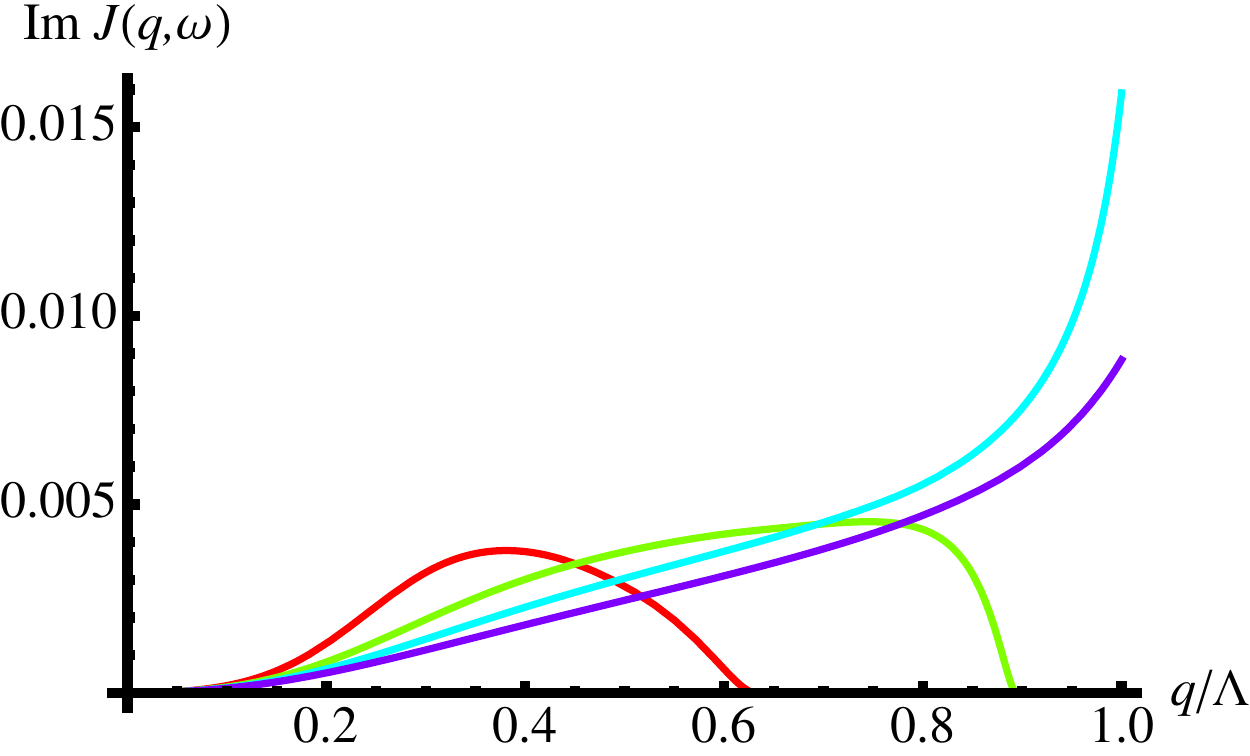}
\caption[fig]{\label{fig6}(Color online). Plot of the imaginary part of the propagators $\tilde I ( q , \omega )$ (left), and $\tilde J ( q , \omega )$ (right)  for $\lambda_{\rm e\text-fl} = 78$ in the absence of the attractive part of the interaction. Bottom: Cuts of the  same functions for for $\omega = 0.2 \omega_{\rm c} ,
 0.4 \omega_{\rm c} , 0.6 \omega_{\rm c} {\rm and} \, 0.8 \omega_{\rm c}$. Colors on the lower plots correspond to colors of the slices in the upper plots.}
\end{figure*}%

\subsubsection{Finite frequency}
In the quantum problem, the flexural bubble is interpreted as a two-phonon propagator, and it is
interesting to work out its frequency dependence. Consider $T=0$. Performing the analytical continuation
of the expressions in  Appendix \ref{a:phonon-bubble} to real time, we obtain in real frequency, the real part, as follows:

\begin{widetext}
\bea \label{realI0}
 {\rm Re}~ I_0(q, \omega_n \to - i \omega + \delta) &=& \frac{1}{256 \pi  \kappa ^{3/2} \sqrt{\rho } w} \nn \\
&& \times \bigg[ w \left(3 \log \Big|(s-w+1)
   (s+w+1)\Big| -\left(w^2+12\right) \log\Big|w^2-4\Big|+\left(w^2+9\right)
   \log \Big|w^2-1\Big|+9\right) \nn \\
   && ~~~+4 \log\bigg|\frac{s+w+1}{s-w+1}\bigg| +\left(6 w^2+4\right) \log\bigg|\frac{w+1}{w-1}\bigg|-2 \left(3 w^2+4\right) \log\bigg|\frac{w+2}{w-2}\bigg| \bigg]
.\eea
We  used the dimensionless  variables\be
w:=\frac{2 \sqrt{\rho} \omega}{\sqrt{\kappa} q^2} =  \frac{2 \omega}{\omega_{\rm fl}(q)} , \qquad  s:= \frac{4\Lambda^2}{q^2}
.\ee
The imaginary part reads

\be
 \label{imI0}
   {\rm Im} ~  I(p,i \omega_n \to \omega + i \delta) =
\frac{\Theta(|w|<1+s)}{256 \kappa ^{3/2} \sqrt{\rho} w}  \Big[ (3 | w| -4) \Theta
   (| w|-2)+  (4-|w|)(|w|-1)^2 \Theta (1<| w| <2) \Big]
.\ee
%
Hence it exhibits a two-threshold behavior. The lowest one ($w=1$)
arises from the minimum energy $\omega = 2 \sqrt{\kappa / \rho} (\frac{q}{2})^2 = \frac{1}{2} \omega_{\rm fl}(q)$
of a pair of flexural phonons with total momentum $q$, i.e.\ each with momentum $q/2$.
\end{widetext}
%
%
\subsection{Membrane coupled to electrons}\label{ss:mem+ele}
We now study the coupled system.

\begin{figure*}[t]
\includegraphics[height=6cm]{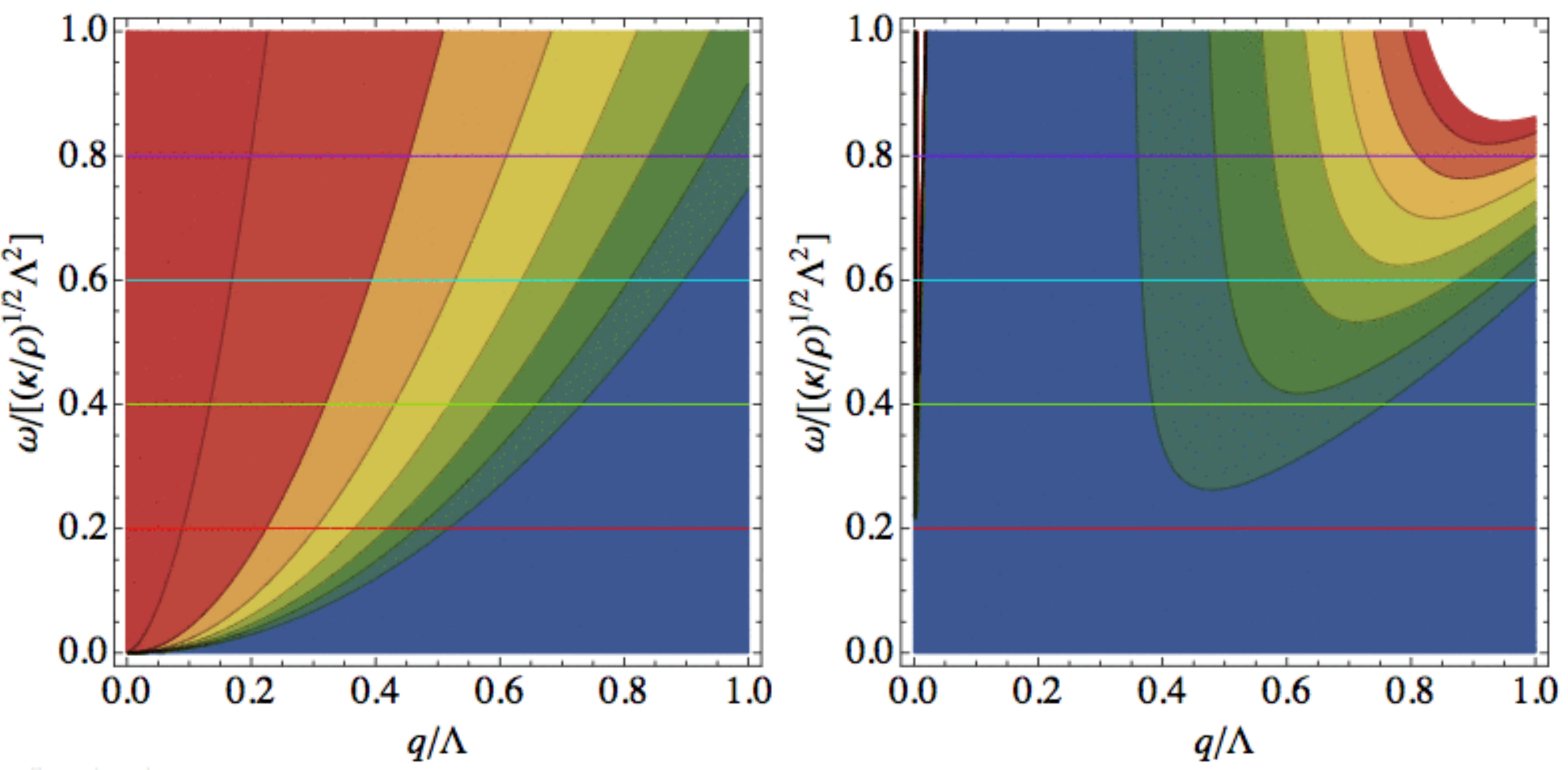}\\
\includegraphics[width=6cm]{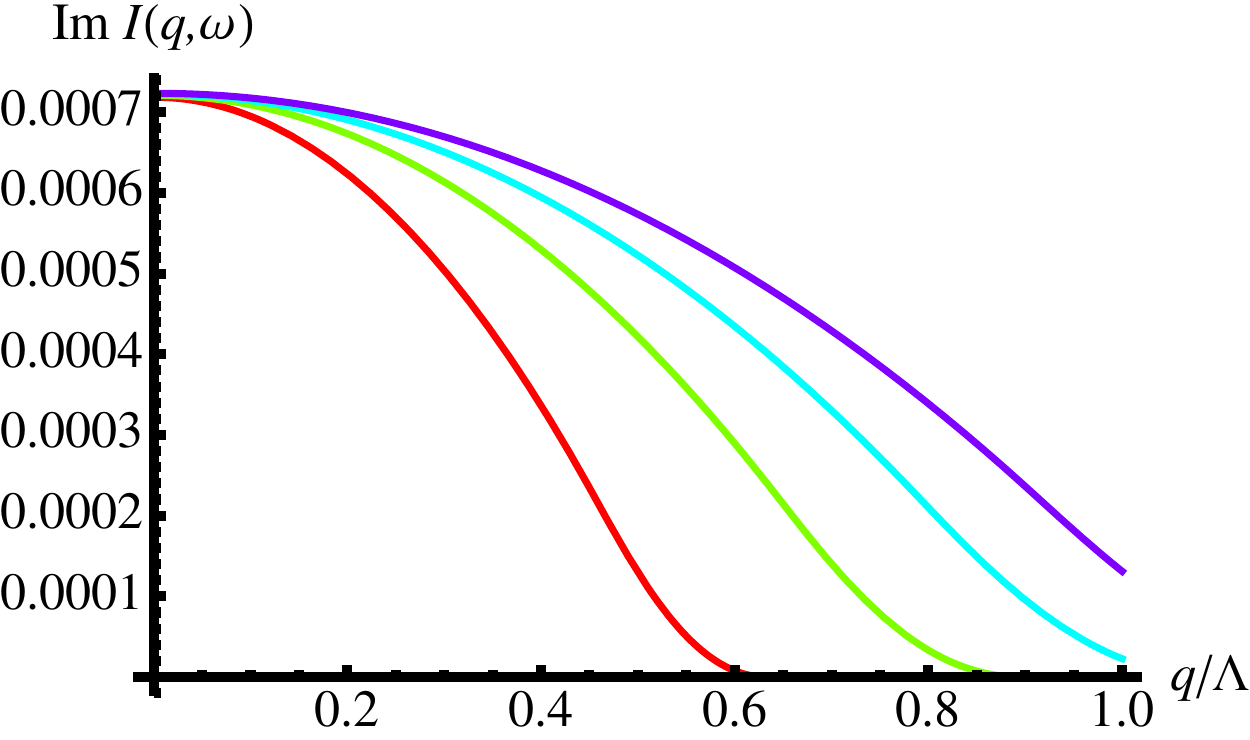}
\includegraphics[width=6cm]{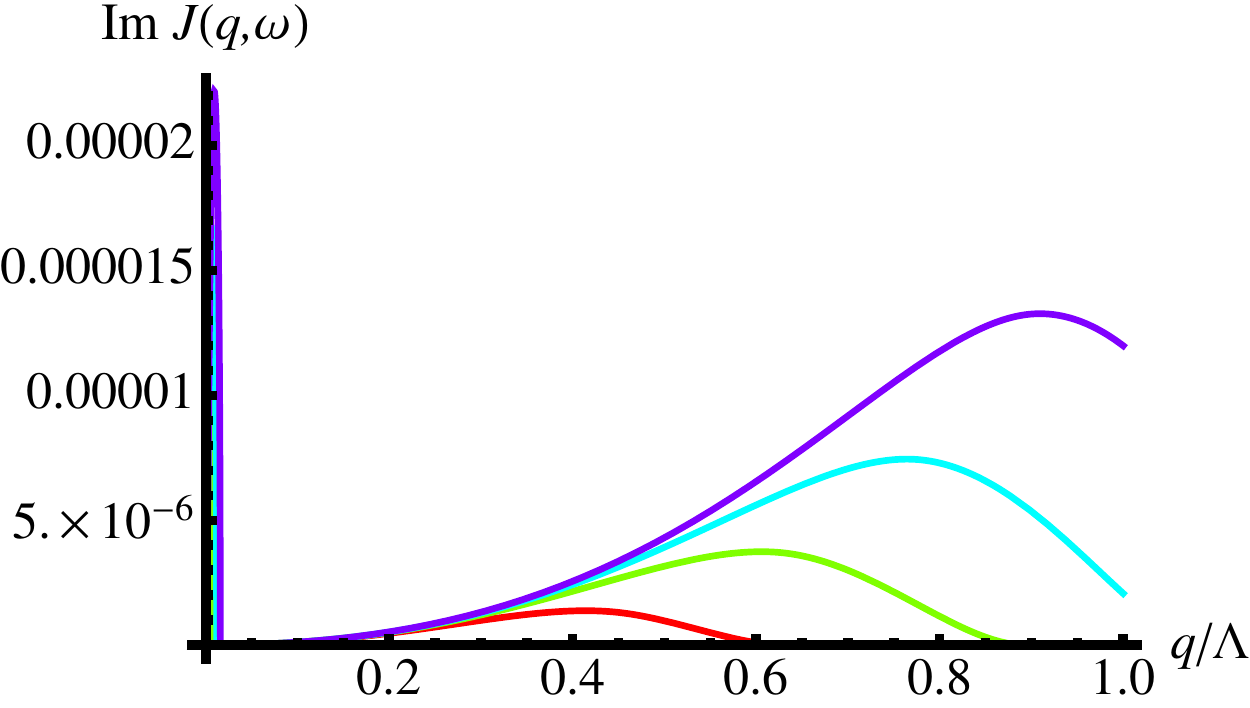}
\caption[fig]{\label{fig8}(Color online). Plot of the imaginary part of the propagators $\tilde I ( q , \omega )$ (left), and $\tilde J ( q , \omega )$ (right) for $\lambda_{\rm e\text-fl} = 2.5$ in the presence of the attractive part of the interaction. Bottom: Cuts of the  same functions for for $\omega = 0.2 \omega_{\rm c} ,
 0.4 \omega_{\rm c} , 0.6 \omega_{\rm c} {\rm and} \, 0.8 \omega_{\rm c}$. Colors on the lower plots correspond to colors of the slices in the upper plots.}
\end{figure*}%

\subsubsection{Qualitative discussion: pole in the two particle propagators}
\label{disc}

Schematically, the quartic interactions in our bare model are expressed in terms of the matrix ${\cal V}$,
\be \label{inter0}
\frac{1}{2 d} \left(\begin{array}{cc}
\frac{1}{2} {\rm P}^{\rm T} \partial h \partial h  , & \delta \rho
\end{array} \right)
 \left(\begin{array}{cc}
K_0 & - g  \\
- g  & V
\end{array} \right)
\left(\begin{array}{c}
\frac{1}{2} {\rm P}^{\rm T} \partial h \partial h  \\
\delta \rho
\end{array} \right),
\ee
where $\delta \rho=\rho-\rho_0$ are the deviations from the uniform electron density. One
legitimate question is whether the bare interaction matrix ${\cal V}$ is positive definite. In previous
work  \cite{G09,SGG11} the electron degrees of freedom were integrated over  within a Gaussian approximation before integrating over the
in-plane phonon modes. It is easy to reproduce these manipulations in our framework. Integrating
(\ref{inter0}) over $\delta \rho$ assuming a Gaussian distribution schematically leads to
\be
\frac{1}{2 d} \left(K_0 - \frac{g^2}{V} \right) \left(\frac{1}{2} {\rm P}^{\rm T} \partial h \partial h\right)^2.
\ee
i.e.\ a $q$-dependent Young modulus $K_0(q) = K_0 - g^2/V(q)$. If one inserts $V(q) = \frac{2 \pi e^2}{|q|}
 - \frac{g_0^2}{\lambda + 2 \mu}$ and $g=\frac{2 \mu}{2 \mu +
\lambda} g_0$ one recovers  the expression for the effective, $q$-dependent Young modulus $K_0(q)$ displayed in  \cite{G09,SGG11}; it becomes negative for
\bea
\frac{\mu + \lambda}{2 \mu + \lambda} q_0 < q < q_0 \quad , \quad q_0 = \frac{2 \pi e^2 (2 \mu +\lambda)}{g_0^2}
\eea
%
More generally, without integrating over the electrons, this signals negative modes for the interaction matrix ${\cal V}$, modes which are a mixture of the Gaussian
curvature, and the electron density.

The fact that the bare quartic interaction matrix has negative modes does not necessary imply that the system is
unstable, since one has to take into account  thermal and quantum fluctuations. First, ${\cal V}$ is replaced by
$\tilde {\cal V}$ which, in the large-$d$ limit, takes into account the bubbles (which contain the leading fluctuations). One has
\be
\det \tilde {\cal V} = \frac{ \det {\cal V} }{D}  , \qquad D =  \det(\1 + {\cal J}   {\cal V}),
\ee
where $D$ is the determinant defined in Eqs.\ (\ref{det1})-(\ref{coupled2}). For the decoupled system,
$D>0$. In this article,  we claim that upon increasing the coupling, the true instability  occurs not when $\det {\cal V}=0$, but at the critical mode $q_{\rm c}$ where
\be
D(q_{\rm c},\omega=0) = 0.
\ee
Since
\be
\det \tilde {\cal J} = \frac{\det {\cal J}}{D}
\ ,\ee
this is equivalent to the appearance of a pole in the matrix (\ref{susc1}) of the 2-particle propagators, i.e.\ of the 2-point functions for the composite fields
$\frac{1}{2} {\rm P}^{\rm T} \partial h \partial h$ and $\delta \rho$. A coupled soft mode  appears for these composite fields at zero frequency, signaling a phase transition. In section \ref{sec:sp}
we argue that the instability makes the composite fields acquire a non-zero expectation value in the ordered phase, at the wave vector $q_{\rm c}$, in
 analogy with a charge-density wave.

\begin{figure*}
\includegraphics[height=6cm]{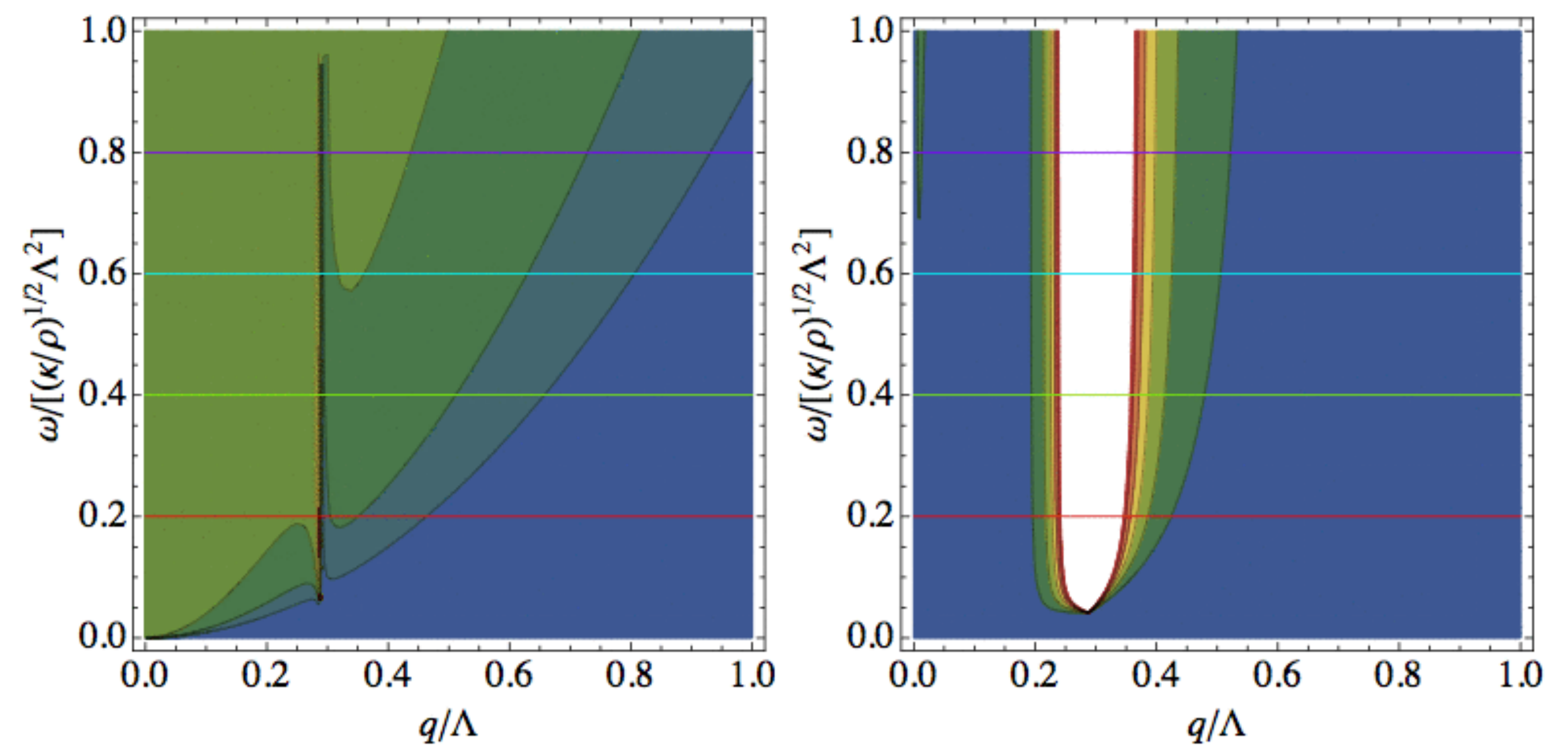}\\
\includegraphics[width=6cm]{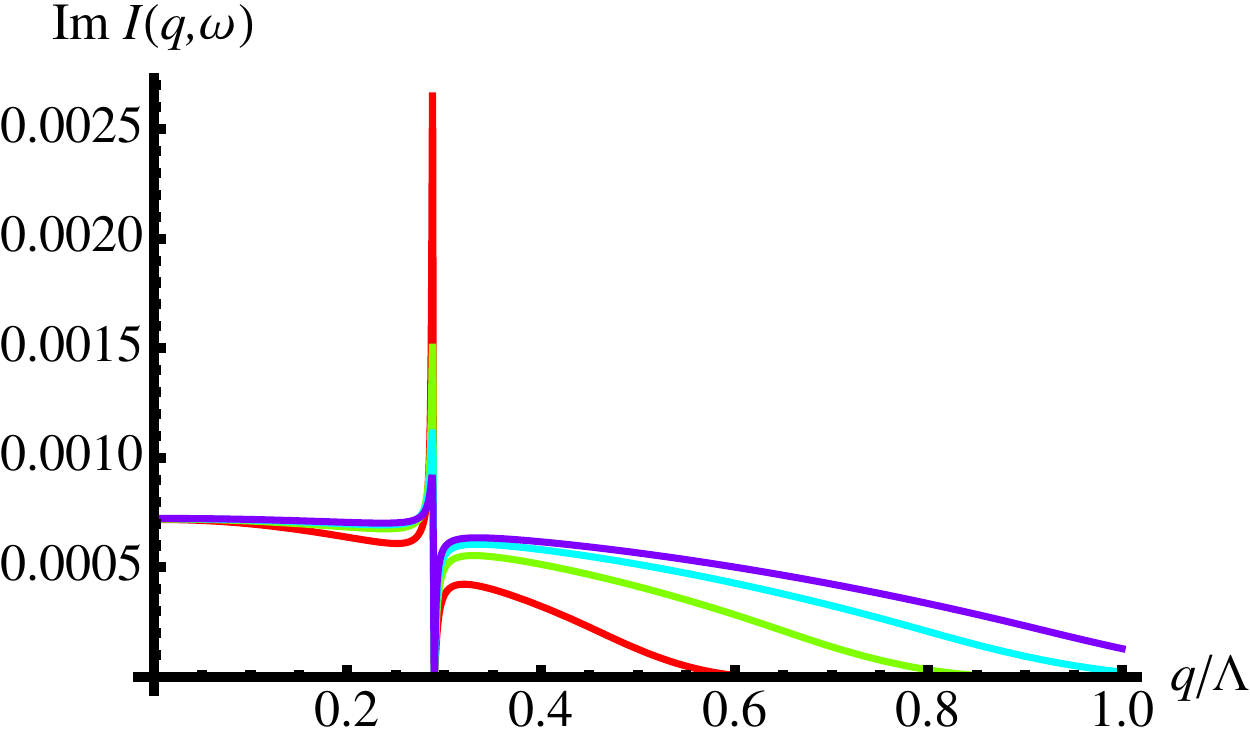}
\includegraphics[width=6cm]{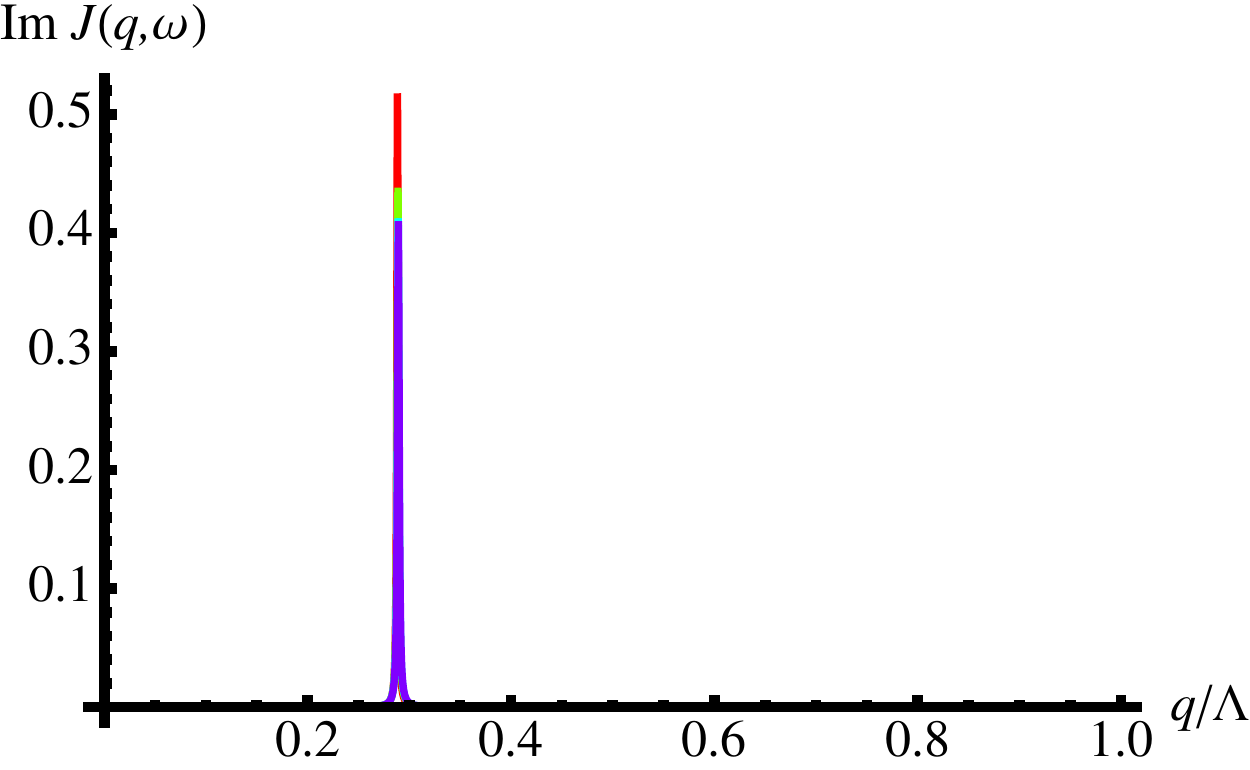}
\caption[fig]{\label{fig10}(Color online). Plot of the imaginary part of the propagators $\tilde I ( q , \omega )$ (left), and $\tilde J ( q , \omega )$ (right) for $\lambda_{\rm e\text-fl} = 6$ in the presence of the attractive part of the interaction. Bottom: Cuts of the  same functions for for $\omega = 0.2 \omega_{\rm c} ,
 0.4 \omega_{\rm c} , 0.6 \omega_{\rm c} {\rm and} \, 0.8 \omega_{\rm c}$. Colors on the lower plots correspond to colors of the slices in the upper plots.}
\end{figure*}%

\subsubsection{Results at finite frequency and collective excitations at $T=0$}

Let us start by studying the dependence in (real) frequency and momentum of the
dressed 2-particle propagators. In particular, we focus on their imaginary parts
${\rm Im} \tilde I ( q , \omega )$ and ${\rm Im}  \tilde J ( q , \omega )$ which
gives the weights of the collective 2 particle excitations in the phonon and
electronic sectors, respectively. These propagators are defined by the equation
\begin{align}
&\left( \begin{array}{cc} \frac{1}{2} \tilde I ( q , \omega ) & \Pi ( q , \omega ) \\ \Pi ( q , \omega ) & N_{\rm f} \tilde J ( q , \omega) \end{array} \right) = \left( \begin{array}{cc} \frac{1}{2} I_0 ( q , \omega ) &
 0 \\ 0 & N_{\rm f} J_0  ( q , \omega ) \end{array} \right) \nn\\
 &~~~~\times \left[ 
\1 { +} \left( \begin{array}{cc} K_0 & { -} g \\ { -}  g &V(q) \end{array} \right)
 \left( \begin{array}{cc} { \frac{1}{2}} I_0 ( q , \omega ) &0 \\ 0 & { N_{\rm f}} J_0 ( q , \omega ) \end{array} \right) \right]^{-1}
 \label{prop}
 \end{align}
The real and imaginary parts of the flexural bubble $I_0$ at $T=0$ are given in (\ref{realI0}) and (\ref{imI0}). The bubble for the Dirac fermions
has been calculated in several articles \cite{GGV94,NGPNG09} and its calculation is recalled in Appendix \ref{a:fermion-bubble}. Upon continuation to real time it reads:
\bea
&&  {\rm Re} ~  J_0 ( q , \omega_n \to - i \omega + \delta ) =  \theta(v_{\rm F} q - |\omega|) \frac{q^2}{16 \sqrt{v_{\rm F}^2 q^2 - \omega^2}} \nn\\ \\
&& {\rm Im} ~ J_0 ( q , \omega_n \to - i \omega + \delta ) = \nn\\
&& \qquad ~~~~~~~~  {\rm sgn}(\omega) \theta(|\omega|-v_{\rm F} q) \frac{q^2}{16 \sqrt{\omega^2 - v_{\rm F}^2 q^2}} .
\eea
We plot in  figures \ref{fig2}-\ref{fig10} the imaginary parts of the dressed 2-particle propagators
${\rm Im} \tilde I ( q , \omega )$ and ${\rm Im}  \tilde J ( q , \omega )$.

We first give the results when the electron-electron interaction is purely Coulomb, i.e.\ for $V(q)=V_0(q)$, disregarding the
attraction induced by the in-plane phonons in Eq.\ (\ref{newV}). Results in the absence of the coupling 
$\lambda_{\rm e\text-fl} = 0$, are plotted for reference in Fig.~\ref{fig2}. This figure and the ones below display the imaginary part of the propagators, which can be interpreted as the density of excitations, which, in the case of flexural phonon pairs, is weighted by a matrix element, involving the transverse projector.  These quantities could, in principle, be measured in inelastic scattering experiments. Note  that the scale in $q/\Lambda$ of Fig.~\ref{fig2} is much expanded as compared to the following figures, since the peak characteristic of pure fermionic excitations
takes place at a small momentum. The results at intermediate coupling, $\lambda_{\rm e\text-fl} = 10,$ but still below the phase transition, are shown in Fig.~\ref{fig4}. One sees the appearance of some structure in ${\rm Im} \tilde J$, at momenta well above the $\lambda_{\rm e\text-fl} = 0$ peak of Fig.~\ref{fig2}; the latter however survives (it is hard to see because of the different scales of $q/\Lambda$). Note that in the presence of a membrane-electron coupling, these plots show the {\it total density} of excitations projected either on the electronic degrees of freedom, or on the phonon degrees of freedom. One can imagine different experimental setups to measure either of them.
Finally, the results for a coupling just above  the phase transition, $\lambda_{\rm e\text-fl} = 78$, (see below) are shown in Fig.~\ref{fig6}. 
The results beyond the transition point should be interpreted with some care, since the calculation does not take into account the existence of a broken symmetry phase, discussed in the next sections. However, it is still informative since the high-energy excitations should remain unaffected by the low-energy changes induced by the phase transition.

Unless explicitly mentioned otherwise, the choice of parameters in plotting all the figures in this paper is the one in (\ref{param}), with $\lambda_{\rm anh}=0.63$ and $\alpha_e=2,$ which corresponds to unscreened graphene. (Screening is examined below). 

It appears clearly on these figures that the electron-hole pairs and the flexural phonons become more and more hybridized as the coupling increases, leading to collective excitations of mixed character.

We then give the results taking into account the attraction induced by the in-plane phonons in Eq.\ \ref{newV}. 
As discussed previously, the attractive interaction does not depend on momentum, and it overcomes the repulsive Coulomb interactions for sufficiently large momenta. This affects significantly the instability, which also occurs at a finite momentum. The value of the critical coupling constant is considerably reduced, and the transition is much facilitated and occurs for realistic values of the parameters. 
The results for a coupling $\lambda_{\rm e\text-fl} = 2.5$ near but below the phase transition are shown in Fig.~\ref{fig8}. The results for a coupling $\lambda_{\rm e\text-fl} = 6$ above the phase transition are shown in Fig.~\ref{fig10}. 
Again, these low-energy spectra at large $\lambda_{\rm e\text-fl}$ beyond the transition cannot be taken at face value since they do not include effects from the phase transition. Note however that the spectral weight is concentrated on the region of momenta where the unstable modes appear. 

\subsubsection{Results at zero frequency, phase transition for pure Coulomb interaction}

When the coupling increases beyond a critical value, $D$ vanishes and a pole appears in the 2-particle excitations.
The phase transition is defined by the appearance of a zero in
\begin{align}
\!\!\!\!0 &=\lim_{\omega \rightarrow 0} D ( q , \omega ) \nn\\
&= \lim_{\omega \rightarrow 0} {\rm \det}\! \left( \begin{array}{cc} 1 { + \frac{1}{2}} K_0 I_0 ( q , \omega ) &- g { N_{\rm f}  J_0 ( q, \omega )}  \\ - \rule{0mm}{2.5ex} {  \frac{1}{2} g I_0 ( q, \omega ) } &1  + V(q) N_{\rm f} J_0 ( q , \omega ) \end{array} \right)
\label{det}.
\end{align}
In this section we analyze this condition when the electron-electron interaction is purely Coulomb, i.e.\ for $V(q)=V_0(q)= \frac{2\pi e^2}{|q|}$, disregarding the
attraction induced by the in-plane phonons in Eq.\ (\ref{newV}). 

Let us start with $T=0$. Using the results (and definitions)  (\ref{bubblezero})ff. for the phonon-bubble at zero frequency,  \be \label{J0}
J_0(q,0)=\frac{|q|}{16 v_{\rm F}},\ee
and the dimensionless coupling constants of  Eq.\ (\ref{11}) we find
\bea
D(q,0) &=& \left(1+ \frac{\pi N_{\rm f}}{8} \alpha_{\rm e} \right) \left[1+ \lambda_{\rm anh} \frac{3}{128 \pi} f(s) \right]  \nn\\
&&
- \lambda_{\rm e\text-fl}^2 \frac{N_{\rm f}}{16} \frac{3}{64 \pi}  \frac{1}{\sqrt{s}} f(s).
\eea
We recall that $s=4 \Lambda^2/q^2$.
Since in the present case $\lambda_{\rm anh} \frac{3}{64 \pi} \approx 10^{-2}$ a
reasonable approximation is to neglect the corresponding term. We note that
$\frac{1}{\sqrt{s}} f(s)$ is maximal for $s_{\rm c}=18.5413$ and there equal to $0.205317$.
Hence the wave-vector where the effect of the coupling is maximal
is $q = q_{\rm c}^{\rm first}= 2 \Lambda/\sqrt{s_{\rm c}} =  0.464472 \Lambda$. This wave vector is not particularly small, but is  well within the Brillouin zone. This implies that
for
\be
\lambda_{\rm e\text-fl}^2 \geq 4.87052 \frac{64 \pi}{3}  \left (\frac{16}{N_{\rm f}} + 2 \pi \alpha_{\rm e} \right) ,
\ee
modes around $q=q_{\rm c}$ become unstable, while $q_{\rm c}$ is the first unstable mode. Taking $N_{\rm f}=4$ and $\alpha_{\rm e}=2,$ we obtain
the critical coupling as
\be
\lambda_{\rm e\text-fl,c} \approx 73.5,
\ee
while for non-interacting electrons one would find by setting $\alpha_e=0$:
\be
\lambda_{\rm e\text-fl,c} \approx 36.13.
\ee
Hence by screening the electron-electron interaction with a substrate  renders the
transition easier.

If one increases $\lambda_{\rm e\text-fl}$ beyond its critical value, a broader range of wave vectors becomes unstable.
Larger wavelengths become available for the ripple instability. For instance, for $\lambda_{\rm e\text-fl}=80,$
the minimum instable vector is $q^{\rm min}_{\rm c}=0.217 \Lambda$, while for $\lambda_{\rm e\text-fl}=100$, this value decreases to $q^{\rm min}_{\rm c}=0.09 \Lambda $.

\begin{figure*}
\includegraphics[width=6cm]{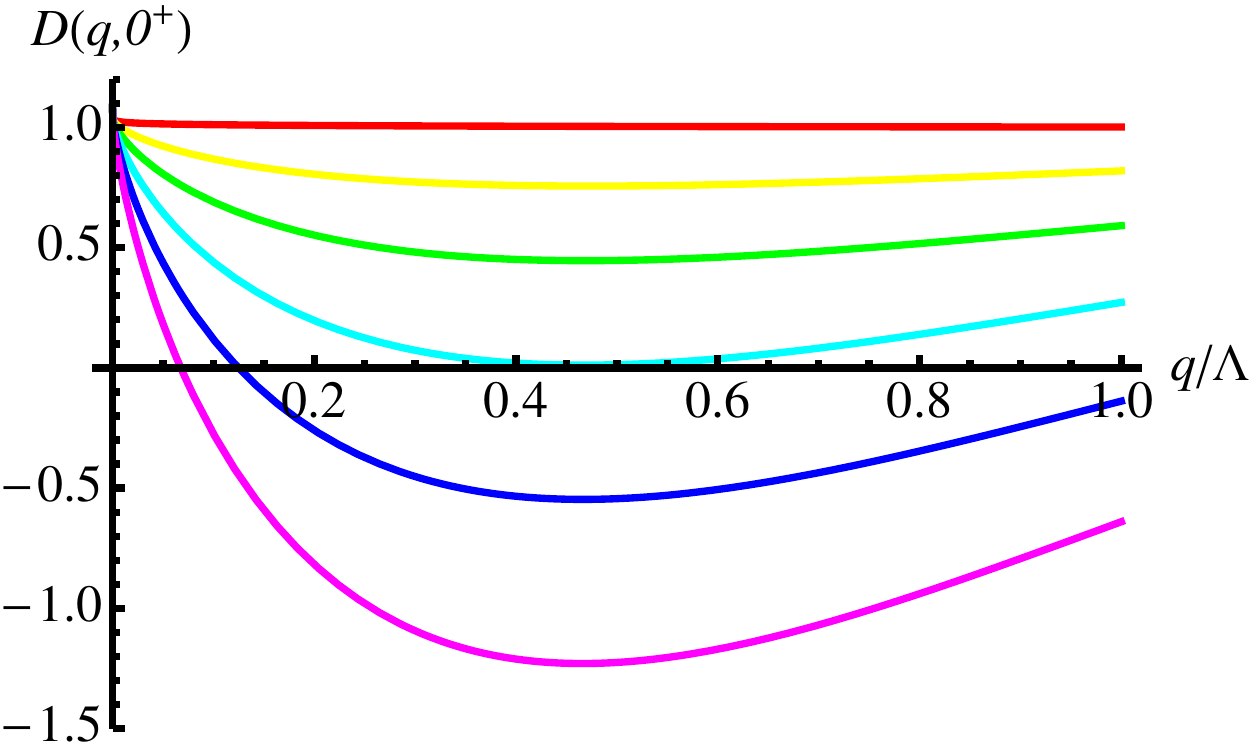}%
\includegraphics[width=6cm]{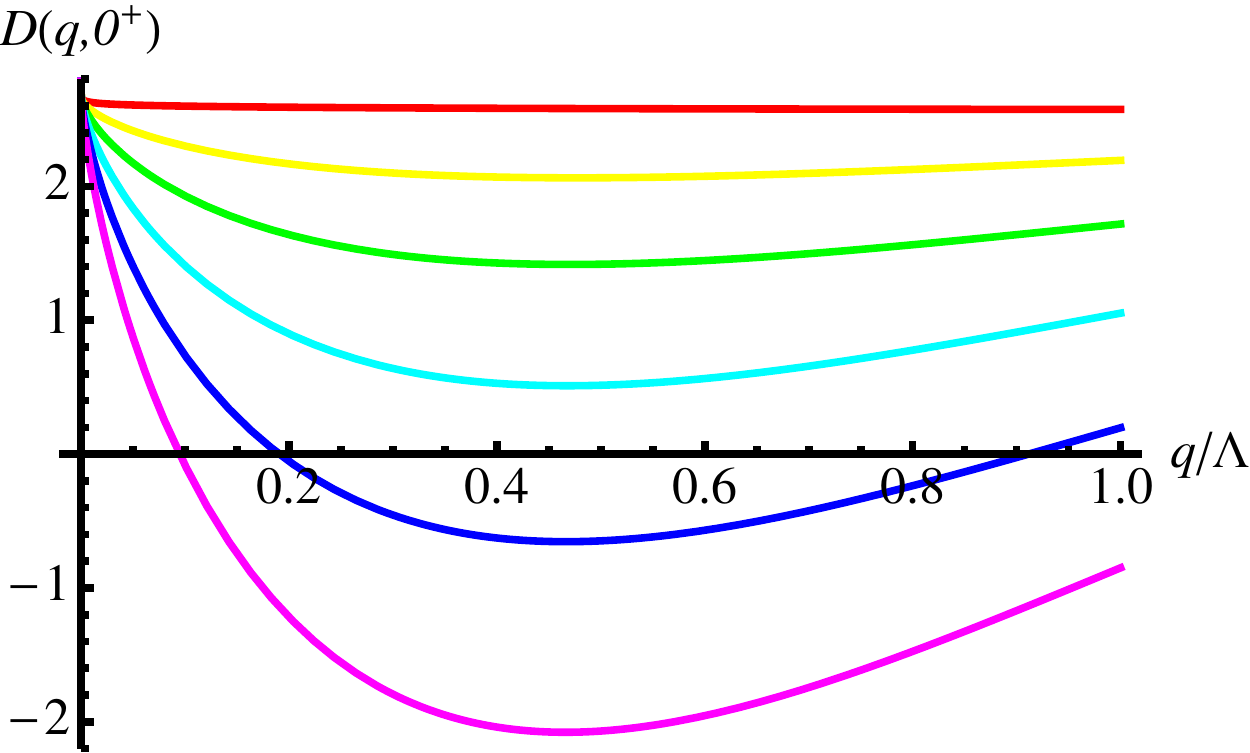}%
\includegraphics[width=6cm]{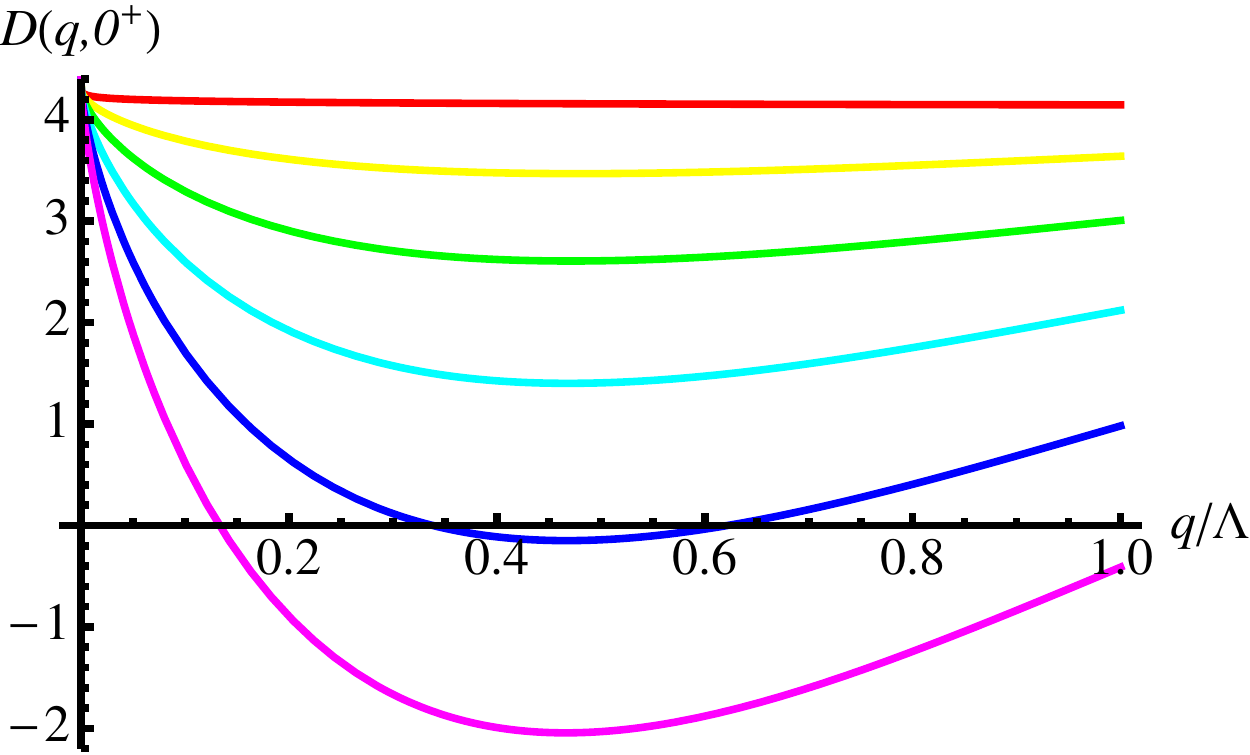}%
\caption[fig]{\label{fig12}(Color online). Function $D( q , 0 )$ as defined in Eq.~(\ref{det}), {\it in the absence of the attractive part of the interaction} for various dimensionless coupling $\lambda_{\rm e\text-fl}$ increasing from top (red) to bottom (violet)
Left: $\alpha_e = 0.5$, for $\lambda_{\rm e\text-fl} = 0,18,27,36,45,54$, 
Center: $\alpha_e=1$, for $\lambda_{\rm e\text-fl} = 0,26,39,52,65,78$,
Right: $\alpha_e=2$, for $\lambda_{\rm e\text-fl} = 0,30,45,60,75,90$.}
\end{figure*}

To confirm these results we plot in Fig. \ref{fig12} the evolution of $D(q,0)$ for various couplings and various effective electron charges
$\alpha_e$. The analysis of the eigenvectors of the matrix in (\ref{det}), i.e.\ $\1 + {\cal J}   {\cal V}$ (whose determinant is $D(q,0)$) at the wavevector $q_c$ where the transition occurs (with $D(q_c,0)=0$) describes the nature of the collective excitation which becomes unstable. A simple
numerical calculation using the above formulae, not detailed here, shows that this mode has a mixed electron-pair and flexural phonon pair character, with the amplitudes in either channel of the same order of magnitude.

From the discussion in Section \ref{sec:crossoverT}
 we see that thermal effects cannot be neglected at wave-vectors
of the order of $q_{\rm c}$. Hence the $T=0$ picture must be modified, whenever
\bea
&&T > T_\Lambda \frac{q_{\rm c}^2}{4 \Lambda^2} \ln\!\left(\frac{2 \Lambda}{q_{\rm c}}\right) \approx \frac{\ln 4}{16} T_\Lambda \approx 295\, {\rm K} ,\\
&&T_\Lambda = \sqrt{\frac{\kappa}{\rho}} \Lambda^2 = 0.31T_{1  {\rm eV}} = 3400\, {\rm K},
\eea
with $T_{1  {\rm eV}} = 11605\rm\, K$. However, thermal effects may modify it before that. To study the temperature dependence let us assume that
$I_0(q,0)$ is given by its classical limit, with $\frac{1}{2} K_0 I_0(q,0)=q_a^2/q^2$ and $q_a^2 = \frac{3}{32 \pi} \frac{K_0 T}{\kappa^2}$.
Then
\be \label{75}
D(q,0) = \left(1+ \frac{\pi N_{\rm f}}{8} \alpha_{\rm e} \right)  \left(1 + \frac{q_a^2}{q^2}\right) - \lambda_{\rm cl}^2 \frac{N_{\rm f}}{16} \frac{q_a^2}{\Lambda q}
\ee
in terms of the classical dimensionless coupling $\lambda_{\rm cl}=\lambda_{\rm e\text-fl}/\sqrt{\lambda_{\rm anh}}$.
The transition occurs when
\be
\lambda_{\rm cl}^2 > \frac{4 \Lambda}{q_a} \left(\frac{8}{N_{\rm f}} + \pi \alpha_{\rm e}\right)
\ee
for the wave-vector $q_{\rm c}=q_a$. This gives
\be
\lambda_{\rm cl,c} = \frac{6.54}{T[\rm eV]} ,
\ee
which is consistent with the above estimates. One should check whether the fermion bubble
remains the same until these temperatures, see Appendix \ref{a:fermion-bubble}.

\subsubsection{Results at zero frequency, phase transition in presence of an attraction}

\begin{figure*}
\includegraphics[width=6cm]{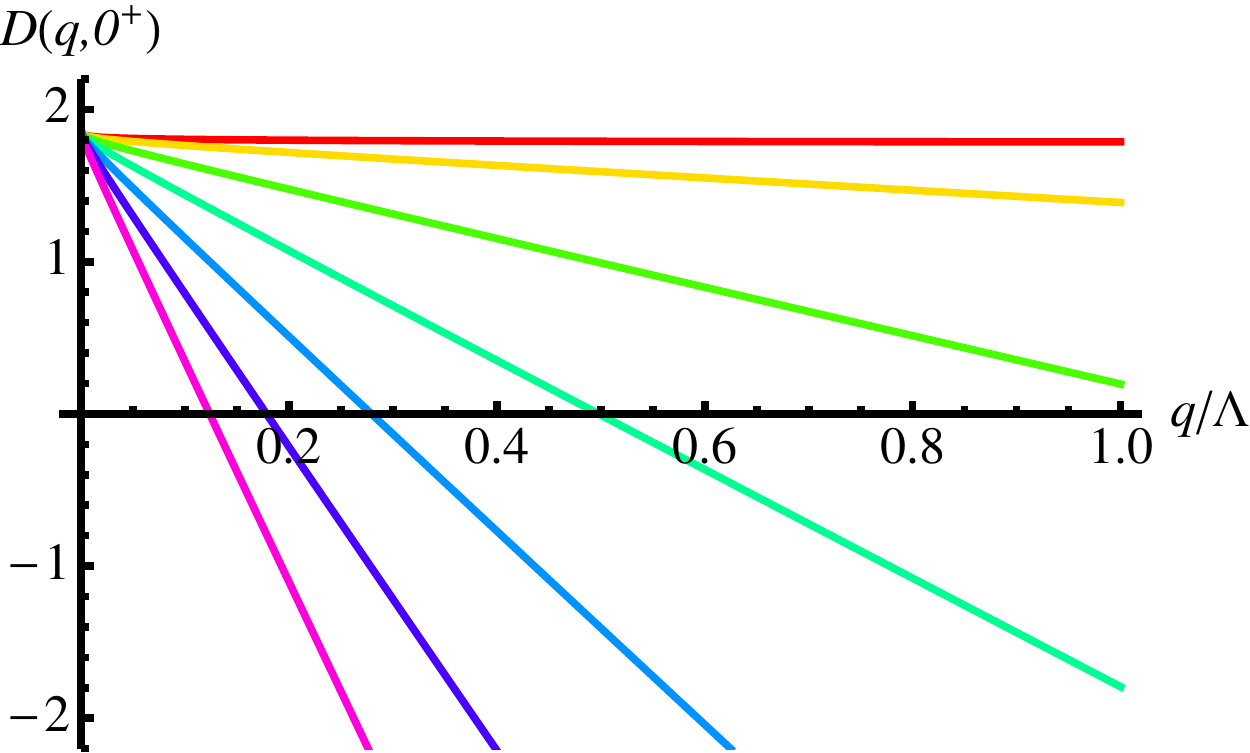}%
\includegraphics[width=6cm]{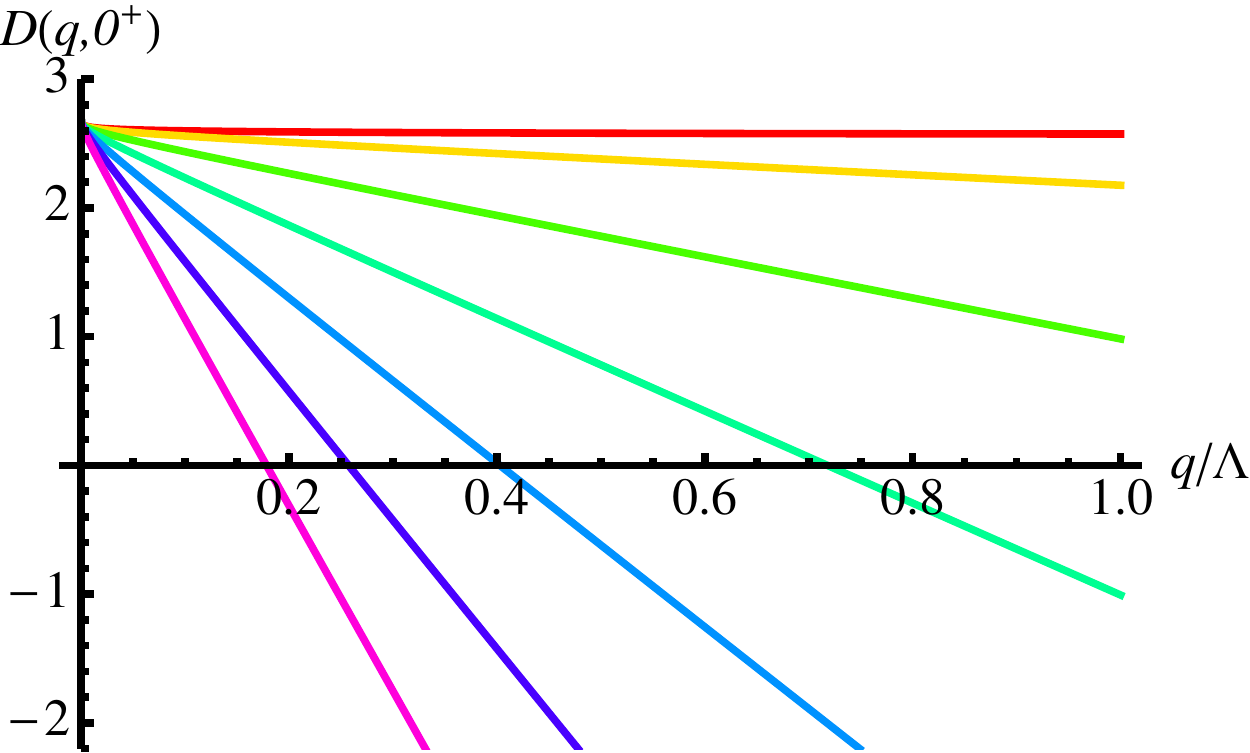}%
\includegraphics[width=6cm]{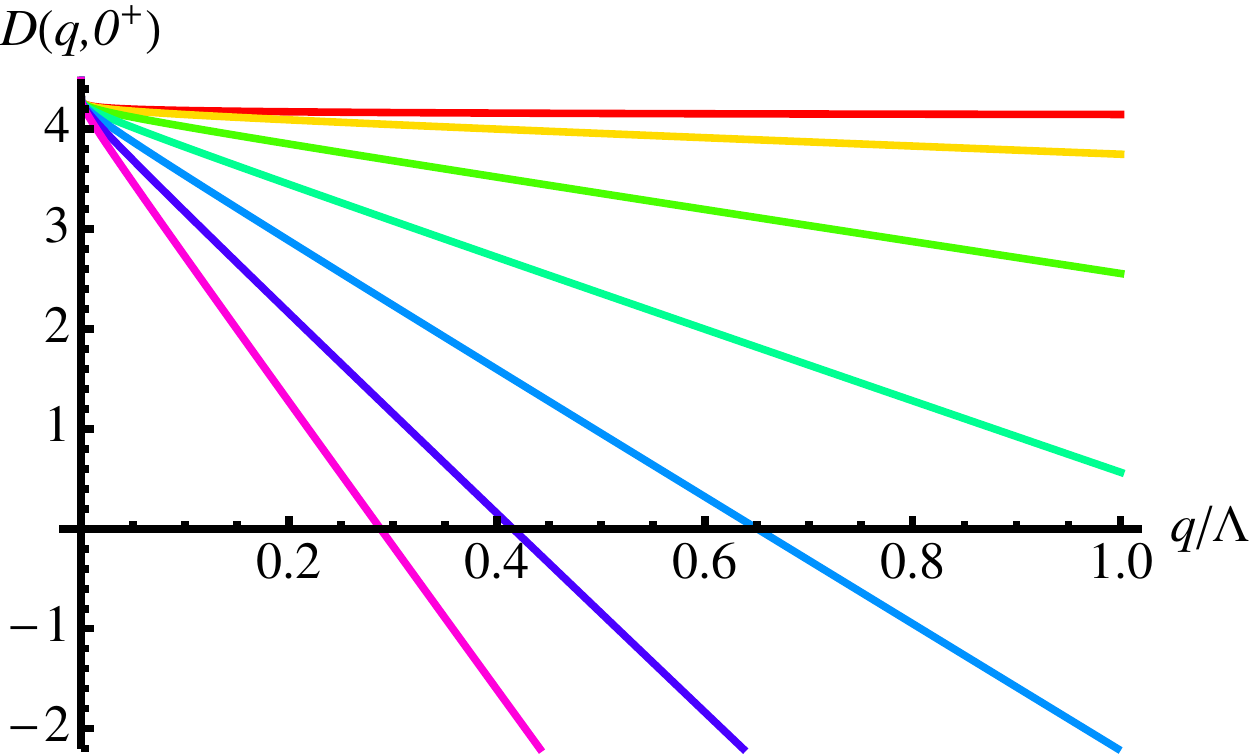}%
\caption[fig]{\label{fig13}(Color online). Function $D( q , 0 )$ as defined in Eq.~(\ref{det}), {\it in presence of the attractive part of the interaction} for dimensionless coupling $\lambda_{\rm e\text-fl} =  0 , 1 , 2 , 3 , 4,5,6 $, from top (red) to bottom (violet). Left: $\alpha_e = 0.5$. Center: $\alpha_e=1$. Right: $\alpha_e=2$.}
\end{figure*}

As we now discuss, the attractive interaction between electrons generated by the integration over the in-plane phonons, i.e.
\be \label{change} 
\frac{2 \pi e^2}{q} \to \frac{2 \pi e^2}{q} - \frac{2 \mu + \lambda}{4 \mu^2} g^2
,\ee
dramatically lowers the value of the coupling necessary to induce the phase transition. 
Equation (\ref{change}) can be rewritten as
\be
\frac{2 \pi e^2}{q} \to \frac{2 \pi e^2}{q} \left[1 - \frac{q}{2 \pi \Lambda \alpha_{\rm e}} \left(1+\frac{\lambda}{\mu}\right) \lambda_{\rm cl}^2\right].
\ee
With these replacements,  equation (\ref{75}) becomes
\bea
D(q,0) &=& \left[1+ \frac{\pi N_{\rm f}}{8} \alpha_{\rm e}  - \frac{N_{\rm f}}{8} \frac{1}{ \sqrt{s} } \left(1+\frac{\lambda}{\mu}\right) \lambda_{\rm cl}^2  \right] \nn\\
&& \times \left[1+ \lambda_{\rm anh} \frac{3}{128 \pi} f(s) \right]  \nn\\
&&
- \lambda_{\rm e\text-fl}^2 \frac{N_{\rm f}}{16} \frac{3}{64 \pi}  \frac{1}{\sqrt{s}} f(s).
\eea
If we again neglect the anharmonic corrections, we obtain
\bea
D(q,0) \approx 1&+& \frac{\pi N_{\rm f}}{8} \alpha_{\rm e}\nn\\
&  - &\frac{N_{\rm f}}{8 \sqrt{s} } \left[ \Big(1+\frac{\lambda}{\mu}\Big) \lambda_{\rm cl}^2
+ \lambda_{\rm e\text-fl}^2  \frac{3}{128 \pi} f(s) \right]\!.~~~~~~~
\eea
Further neglecting the last term, we find that the first instability occurs near the cutoff $s_{\rm c}=4$ when
$\lambda_{\rm cl}$ reaches the critical value
\be
\lambda_{\rm cl,c}^2 = \frac{2}{1+ \frac{\lambda}{\mu}} \left(\frac{8}{N_{\rm f}} + \pi \alpha_{\rm e}\right),
\ee
hence using $N_f=4$ and $\alpha_e=2$ at 
\be
\lambda_{\rm cl,c} \approx 3.8\ .
\ee
On the other hand, if we use $\lambda_{\rm anh}=0.63$, we find
\be
\lambda_{\rm e-fl} \approx 3\ .
\ee As announced, this is a much lower critical value
than the typical critical coupling values found above.
The first instability however occurs this time for $q_{\rm c} = \Lambda$.
This can be seen in Fig.\ \ref{fig13} where we have plotted how $D(q,0)$ evolves with the coupling constant for various values of the effective electron charge $\alpha_{\rm e}$. Again, larger values $\lambda_{\rm cl} > \lambda_{\rm cl,c}$ result in smaller wave vectors
becoming unstable. Thus, although the criterion for the transition $D(q_c,0)=0$ is different from the one used in previous articles,
we do find that this transition is facilitated by the electronic attraction mediated by the in-plane phonons, an effect which
has a counterpart as $K_0(q)$ becoming negative at  some wave-vector, as discussed in Section \ref{disc}.

\section{saddle point: beyond the instability}
\label{sec:sp}
\subsection{The saddle-point equations}
In this section we study the free energy of the model at $d=\infty$. As is well known from the
$O(N)$ model at large $N$ the saddle-point equations allow to determine whether a non-trivial
minimum exists,  signaling a non-trivial phase with ripples.

To this aim we introduce fluctuating
auxiliary fields $\sigma(x,\tau)$ and $\alpha(x,\tau)$ and consider the (e.g.\  Matsubara) action
\begin{eqnarray} \label{slarge}
 S' &=& S_0 + S'_{\rm int} \\
 S'_{\rm int} &=& \int_{x\tau} \sigma \left[ \frac{1}{2} {\rm P}^{\rm T}_{ij}(\partial)  \sum_{a=1}^{d} \partial_i h_a  \partial_j h_a \right]
+ \alpha \sum_{\gamma=1}^{d N_{\rm f}} \bar \Psi_\gamma \1 \Psi_\gamma  \nn\\
&&
\!\!\!\!\!\!  - \frac{d}{2} \int_{xx'\tau}
\left(\begin{array}{cc}
\sigma  , & \alpha
\end{array} \right)_{x\tau}
 \left(\begin{array}{cc}
K_0 & - g  \\
- g  & V
\end{array} \right)^{-1}_{xx'}
\left(\begin{array}{c}
\sigma \\
\alpha
\end{array} \right)_{x'\tau} \nn .
\end{eqnarray}
The interaction matrix can be non-local but is assumed to be static (i.e.\ frequency independent).
After integration over  auxiliary fields it reproduces the action (\ref{mats}). As we show
below, at the transition the fields $\sigma$ and $\alpha$ acquire static and space-dependent expectation
values, which we denote $\sigma_0(x) := \langle \sigma(x,\tau) \rangle$ and
$\alpha_0(x) := \langle \alpha(x,\tau) \rangle$. Since the
theory is Gaussian in the auxiliary fields, we have the exact relations
\be
\left(\begin{array}{c}
 \sigma_0(x) \\
\alpha_0(x)
\end{array} \right) =  \int_{x'} \left(\begin{array}{cc}
K_0 & - g  \\
- g  & V
\end{array} \right)_{xx'}
\left(\begin{array}{c}
\Phi_0(x') \\
\delta \rho_0(x')
\end{array} \right).
\ee 
We have defined the expectation values
\bea
&& \frac{1}{d} \left< \sum_{a=1}^d {\rm P}^{\rm T}_{ij}(\partial) \partial_i h^a(x,\tau) \partial_j h^a(x,\tau) \right> = \Phi_0(x)\qquad  \\
&& \frac{1}{d} \left< \sum_{\gamma=1}^{d N_{\rm f}} \bar \Psi_\gamma(x,\tau) \1 \Psi_\gamma(x,\tau)  \right> - \rho_0 = \delta \rho_0(x),
\eea
Hence the phase transition is equivalently characterized by
these composite fields, the Gaussian curvature and the electronic charge density,
acquiring expectation values, which are static and {\it non-uniform in space}.
Since these order parameters are defined only at a non-zero wave-vector, they
obviously vanish for the free action $S_0$. They also vanish in the small-$g$  phase. We show
below that the pole in the coupled propagator of the composite fields ${\rm P}^{\rm T} \partial h \partial h$ and $\bar \Psi \Psi$
 corresponds to an instability, allowing  $\sigma_0(x)$ and $\alpha_0(x)$ to
become non-zero.

We now derive the effective action for the fields $\sigma$ and $\alpha$. We allow for a breaking of  the $O(d)$ symmetry, i.e.\ the vector field $h_a(x,\tau)$ may acquire a non-zero expectation value and pick one direction in the transverse space, denoted $a=1$, with $\langle h_a(x,\tau) \rangle = \delta_{a1} h_1(x,\tau) \neq 0$. For the physical model $d=1$ this is the Ising symmetry related to the two possible orientations of the normal vector. There are various equivalent ways to implement that breaking,
either by integrating over the $d-1$ flexural modes except $h_1$, see   \cite{Zinn}, section 26, or decomposing
$h_a(x,\tau) = \langle h_a(x,\tau) \rangle  + \delta h_a(x,\tau)$ into an average and a fluctuating part.

Integrating over the fermions and the fluctuating part of the flexural modes, we find that the action reads, in the large-$d$ limit,
\begin{eqnarray}
&& \frac{S'}{d} = \frac{1}{2}\tr \ln\left( - \rho \partial_\tau^2 + \kappa \nabla^4  - [ {\rm P}^{\rm T}_{ij}(\partial)  \sigma(x,\tau) ] \partial_i \partial_j   \right) \nn\\
&&-  \frac{N_{\rm f}}{2}\tr \ln\Big( -  v_{\rm F} [ {\boldsymbol\sigma} \cdot (- i \nabla)] + \left[\alpha(x,\tau) - \mu - \partial_\tau \right]  \1  \Big) \nn
\\
&& - \frac{1}{2}  \int_{x,x',\tau}
\left(\begin{array}{cc}
\sigma  , & \alpha
\end{array} \right)_{x \tau}
 \left(\begin{array}{cc}
K_0 & - g  \\
- g  & V
\end{array} \right)_{xx'}^{-1}
\left(\begin{array}{c}
\sigma \\
\alpha
\end{array} \right)_{x' \tau} \nn\\
&&
 + \frac{1}{d}  \int_{x,\tau}  \left[ \frac{\kappa}{2} (\nabla^2 h_1)^2 + \frac{\rho}{2} (\partial_\tau h_1)^2 + \sigma \frac{1}{2} {\rm P}^{\rm T}_{ij}(\partial) \partial_i h_1  \partial_j h_1 \right].  \nn\\
 \label{actiond}
\end{eqnarray}
We have used that there are several ways to rewrite the term containing the transversal projector,
\bea
&& \int_{x\tau} \sigma(x,\tau) {\rm P}^{\rm T}_{ij}(\partial) \left[\partial_i h(x,\tau)  \partial_j h(x,\tau) \right] \nn\\
&&=
 \int_{x\tau}  \left[ {\rm P}^{\rm T}_{ij}(\partial)   \sigma(x,\tau) \right] \partial_i h(x,\tau)  \partial_j h(x,\tau) \nn  \\
 &&
 = \int_{p,k,\omega,\omega'} \sigma(p,\omega') {\rm P}^{\rm T}_{ij}(p) k_i k_j h(-k,-\omega) h(k-p,\omega-\omega') \nn\\
 && =
-  \int_{x\tau}  \left[ {\rm P}^{\rm T}_{ij}(\partial)   \sigma(x,\tau) \right] h(x,\tau)  \partial_i  \partial_j h(x,\tau)
\eea
If we suppose that $h_1 \sim \sqrt{d}$, which
 is the usual scaling for  $O(d)$ breaking, the action is uniformly
 proportional to $d$ and one can thus look for a saddle point.

We now derive the saddle-point equations. Since we look for a static solution,
our ansatz is in terms of time-independent fields. The variation w.r.t.\ $\sigma(x,\tau)$ yields \begin{widetext}
\begin{equation}\label{SP2}
\frac1d {\rm P}^{\rm T}_{ij} \partial_{i} h_1 (x)
 \partial_{j} h_1 (x) - \frac{1}{2 \beta} \sum_{\omega_n}
 \mathrm P^{\rm T}_{ij}(\partial_x)  \partial_{xi} \partial_{x j}
 \bigg[\rho \omega_n^2 +\kappa \nabla_y^4 -  [ {\rm P}^{\rm T}_{ij}(\partial_y)  \sigma(y) ] \partial_{iy}  \partial_{jy} \bigg]^{-1}_{xx} = \int_{x'} \Big(1,0\Big)
\left(\begin{array}{cc}K_{0} & -g \\
-g & V
\end{array} \right)_{\!xx'}^{\!-1} \left({\sigma(x') \atop  \alpha(x')}
\right) .
\end{equation}\end{widetext}
The variation w.r.t.\ $\alpha$ yields, setting the chemical potential $\mu\to 0$,~~~~~~~~~~~~~~\qquad 
\bea \label{SP3}
&& N_{\rm f} \frac{1}{\beta} \sum_{\omega'_n} (i \omega'_n) [ (- \alpha(y) +i \omega'_{n})^{2} + \nabla_y^{2}]^{-1}_{xx}
\nn\\
&&=  \int_{x'} \Big(0,1\Big)
\left(\begin{array}{cc}K_{0} & -g \\
-g & V
\end{array} \right)_{xx'} ^{-1} \left({\sigma(x') \atop  \alpha(x')}
\right) .
\eea
Finally, the variation w.r.t.\ $h_1$ yields
\be\label{SP1}
\kappa k^{4} h_1 (k) + \int_{p}
\left(k^{2}-\frac{(k\cdot p)^{2}}{p^{2}} \right)
 \sigma({ p) }h_1 (k-p) = 0\ .
\ee
Clearly there is always the trivial solution to these equations $\sigma(x)=\alpha(x)=h_1(x)=0,$
which corresponds to the weak-coupling phase. Consider now the action $\frac1d S'[\sigma,\alpha,h_1]$ in this phase, as a functional of
the fields. It is easy to see by expanding Eq.~(\ref{actiond}) in powers of $\sigma$ and $\alpha$ that
\bea \label{instab}
\frac{ S'[\sigma,\alpha,h_1=0]}{d} &=& - \frac{d}{2}
\left(\begin{array}{cc}
\sigma  , & \alpha
\end{array} \right)
({\cal J}  + {\cal V}^{-1} )
\left(\begin{array}{c}
\sigma \\
\alpha
\end{array} \right) \nn\\
&&+ {\cal O}((\sigma,\alpha)^3) ,
\eea
where ${\cal J}$ is the matrix of bubbles introduced in Eq.\ (\ref{J}). Now from the relations given there, one  finds that, in terms of the dressed interaction,
\be
{\cal J}  + {\cal V}^{-1} = {\cal V}^{-1} (\1 + {\cal V} {\cal J}) = \tilde {\cal V}^{-1} .
\ee
The important point is that if $D = \det (\1 + {\cal V} {\cal J})$ vanishes  at $q=q_c$,
then the quadratic part of the action in $(\sigma,\alpha)$ has a zero mode at $q=q_c$, and the solution
$\sigma(x)=\alpha(x)=0$ becomes unstable. The same instability can also be seen on the
above saddle-point equations expanded to linear order in $(\sigma,\alpha)$. Hence
the vanishing of the determinant, demonstrated in Section \ref{ss:mem+ele} implies a phase transition, and that one must
look for a non-trivial solution of the saddle-point equations.

Note, from (\ref{instab}), that we did not need to allow for a non-vanishing $h_1$ to find the instability.
Indeed, to quadratic order the $(\sigma,\alpha)$ and $h_1$ sectors decouple, since the leading
coupling is $O(\sigma h_1^2)$. Whether $h_1$ acquires or not an expectation value beyond the instability -- i.e.\ whether the rippling and breaking of Ising symmetry (here $O(d)$ symmetry)
occur simultaneously or not -- remains to be
investigated.

Searching for a solution of the above saddle-point equations in the rippled
phase is beyond the goal of this article. In Appendix \ref{solu1} however, we remark  that the magnitude 
of $\sigma$ fixes a scale for a possible $O(d)$ symmetry breaking.

\section{Conclusion}
\label{s:Conclusion}

In conclusion, we have studied in this article a model for graphene as an elastic membrane coupled to
Dirac electrons. By extending the model to $d$-component flexural phonons and $N_{\rm f} d$-component
Dirac fermions, we obtained a solvable limit for large $d$, while retaining a lot of the physics, e.g.\
screening of non-linearities by thermal and quantum fluctuations. 
We derived the Self Consistent Screening Approximation (SCSA) equations, 
which are extensions of the standard classical SCSA equations to
(i) the quantum membrane, and (ii) the coupled quantum membrane-electron problem.

By a careful study of the temperature dependence of the flexural bubble we obtained the first controlled
description of the quantum to classical, and harmonic to anharmonic crossover for the problem of the
membrane alone. 

We have analyzed, within the same approximation, the effect on the membrane of the electronic degrees of freedom. We find that the electron excitations, i.e.\ the electron-hole pairs, mix with the flexural modes, leading to collective excitations of hybrid character. For sufficiently large values of the electron-phonon coupling, new modes appear below the continuum of excitations made up of two flexural phonons. As the coupling is increased, the frequency of these modes goes to zero at a finite value of the momentum $q_c$. If the coupling is increased further, the frequency of the modes within a range of finite momenta becomes imaginary, signalling a phase transition and the appearance of a broken symmetry phase.

The instability appears first at momenta comparable with the high-momentum cutoff, of the order of the lattice spacing. As the electron-phonon coupling increases, the range of unstable modes shifts towards lower momenta. The character of these modes changes between mostly phonon-like to electron-like.

We have found that the attractive interaction between electrons mediated by in-plane phonons greatly facilitates the
transition which then occurs at lower and quite realistic values of the coupling. In addition, the transition is also
found to be facilitated by screening of the Coulomb interaction. 

Evidence for this instability was demonstrated in the $d=\infty$ limit.
It is different from previous approaches, because it does not involve the
 renormalization of the bending rigidity \cite{SGG11} and it does not
rely on the effective Young modulus becoming negative in some
window of wave vectors \cite{G09}.

It is tempting to associate this transition to the spontaneous and simultaneous formation of ripples
coupled to electronic puddles. To make this more precise
we have derived the saddle-point equations, exact at $d=\infty$, which allow us to
study the transition and in principle to describe the rippled phase. It confirms that
the instablility occurs at a finite wave-vector and mixes electronic and flexural degrees of freedom. The study of the
coupled non-linear saddle-point equations which describe the rippled phase is however complicated, and
left for the future. (It could be done either numerically or in some expansion, e.g.\ for large coupling.). 
We have not studied here the renormalization of $\kappa,\rho,g,v_{\rm F}$, which
can be added and occurs to next order in $1/d$. 
Although we do not expect renormalization  to  qualitatively change the mechanism proposed here,
it is likely to change the estimates for the transition.

The results presented here confirm that the coupling between flexural modes and electron-hole pairs 
significantly changes the structural properties of graphene. The main changes, and the instability for sufficiently large couplings, occur at a finite momentum. Hence, the results reported here should not be modified by the presence of a finite carrier concentration, provided that the square root of the density of carriers is small compared to the wave vector at which the instability takes place. On the other hand, the existence of a gap comparable to the bandwidth or the electronic cutoff will suppress the effects reported here. The two-dimensional material boron nitride  is structurally very similar to graphene, but it has a larger gap in the electronic spectrum. It would be interesting to analyze the properties of free standing boron nitride. Other two dimensional systems, like MoS$_2$ or MoW$_2$ are semiconductors with a small gap. Their tendency towards ripple formation   should be intermediate between that of graphene and of boron nitride.

An interesting extension would be to apply our approach in the presence of a substrate. Indeed it is known that 
graphene on many metallic substrates, where the Coulomb interaction is screened, has long-ranged height corrugations  \cite{Vetal08}. The study of these corrugations requires to add to our model the interaction between graphene and the substrate.

\acknowledgments

We are grateful to J.\ Gonzalez for stimulating discussions. 
FG acknowledges support from
the Spanish Ministry of Economy (MINECO) through
Grant No.\ FIS2011-23713, the European Research Council
Advanced Grant (contract 290846) and from the European
Commission under the Graphene Flagship contract
CNECT-ICT-604391.  The authors thank the KITP for hospitality within the program ``The Physics
of Graphene" (2012),
where this work was started. The work is partially supported
by the National Science Foundation under Grant No.\ NSF PHY11-25915.

\appendix

\section{Integration over in-plane phonons}
\label{a:in-plane-phonons}

We start from the elastic energy (\ref{elasten}) plus the coupling term (\ref{coupl}), together with their associated Matsubara actions.
For notational simplicity we will omit the index $a=1,...,d$, and set $\partial h_a \partial h_a \to \partial h \partial h$, i.e.\ in practice we consider only the physical case $d=1$, while the index can  easily be restored at the end. We note that the total coupling of the in-plane displacements to the flexural modes and electron density can be written, upon integration by part, as
\be
S_{\rm u\text-fl,e}= \int \rmd^2 x\, \rmd \tau\,  ~ u_m  [ - A_{ijm}(\partial) \partial_i h \partial_j h + g \partial_m \delta \rho ] .
\ee
Hence integrating over in-plane modes $u_i$ we find the total effective Matsubara action
for the flexural modes,
\begin{widetext}
\be
S_{\rm eff,fl} = \frac{1}{8} \int \rmd^2 x\, \rmd \tau\, [ \lambda (\partial_i h \partial_i h)^2 + 2 \mu (\partial_i h \partial_j h)^2 ] - \frac{1}{2} \int_{q,\omega}
(\partial_i h \partial_j h)_{q,\omega} (\partial_k h \partial_l h) _{-q,-\omega} A_{ijm}(q) A_{klp}(q)  \langle u_m(q,\omega) u_p(-q,\omega)  \rangle_0.
\ee
Here and above we denote $A_{ijm}(q) = \frac{\lambda}{2} \delta_{ij} q_m + \frac{\mu}{2} (q_i \delta_{jm} + q_j \delta_{im})$ and we
use the notation $\int_\omega \equiv \frac{1}{\beta} \sum_{\omega_n}$. Inserting the quadratic bare in-plane phonon  propagator gives
\be
\langle u_m(q) u_p(-q)  \rangle_0 = \frac{P_{mp}^L(q)}{\rho \omega^2 + (\lambda+2 \mu) q^2} + \frac{P_{mp}^T(q)}{\rho \omega^2 + \mu q^2}.
\ee
We find, after a tedious calculation,
\bea \label{integ1}
 S_{\rm eff,fl} &=& \int_{q,\omega} \frac{(\lambda+\mu) \mu q^2}{2 (\rho \omega^2 + \mu q^2) (\rho \omega^2 + (\lambda + 2 \mu) q^2)} \Big[ \mu q^2 \Big|({\rm P}^{\rm T} \partial h \partial h)_{q,\omega}\Big|^2
+ \rho \omega^2  ({\rm P}^{\rm T} \partial h \partial h)_{q,\omega}  (\partial h \partial h)_{-q,-\omega} \Big] \nn \\
&& + \frac{(\lambda + 2 \mu) \rho \omega^2}{8(\rho \omega^2 + (\lambda + 2 \mu) q^2)} \Big|(\partial h \partial h)_{q,\omega}\Big|^2
+ \frac{\mu \rho \omega^2}{2(\rho \omega^2 + \mu q^2)}  \Big[ (\partial_1 h \partial_2 h)_{q,\omega} (\partial_1 h \partial_2 h)_{-q,-\omega} -
(\partial_1 h \partial_1 h)_{q,\omega} (\partial_2 h \partial_2 h)_{-q,-\omega} \Big].\nn\\
\eea
We have used the notations $({\rm P}^{\rm T} \partial h \partial h)(x,\tau) = {\rm P}^{\rm T}_{ij}(\partial) \partial_i h(x,\tau) \partial_j h(x,\tau)$ and
$(\partial h \partial h)(x,\tau) = \partial_i h(x,\tau) \partial_i h(x,\tau)$ for the bilinears in the gradient of the height field, and their Fourier transforms.
We have used that $\int_{q,\omega} (\partial_1 h \partial_2 h)_{q,\omega} (\partial_1 h \partial_2 h)_{-q,-\omega}
= \int_{q,\omega}  (\partial_1 h \partial_1 h)_{q,\omega} (\partial_2 h \partial_2 h)_{-q,-\omega}$ to rewrite some terms. An equivalent more compact form is given by
\bea \label{integ2}
 S_{\rm eff,fl} &=& \int_{q,\omega} \frac{4 \mu (\lambda+\mu) q^2 + \rho \omega^2 (\lambda+2 \mu)}{8 ( (\lambda+2 \mu) q^2 + \rho \omega^2)} |H^T(q,\omega)|^2 \nn \\
 && + \frac{1}{2} \rho \omega^2 \bigg[  \frac{(\lambda+2 \mu) |H^L(q,\omega)|^2  + 2 \lambda H^L(q,\omega) H^T(-q,-\omega) }{4 ( (\lambda+2 \mu) q^2 + \rho \omega^2)}  + \frac{\mu |H^M(q,\omega)|^2}{\mu q^2 + \rho \omega^2} \bigg]
,\eea
where we have used the general decomposition of the matrix $H_{ij} = \partial_i h \partial_j h$,
\bea
&& H_{ij}(q,\omega) = {\rm P}^{\rm T}_{ij}(q) H^T(q,\omega) + {\rm P}^{\rm L}_{ij}(q) H^L(q,\omega) + {\rm P}^{\rm M}_{ij}(q) H^L(q,\omega).
\eea
Here ${\rm P}^{\rm M}_{ij}(q) := (q_i q^T_j + q^T_i q_j)/q^2$, with $q^T_i=\epsilon_{ij} q_j$, is not a projector but satisfies $({\rm P}^{\rm M})^2=1$ and is orthogonal to $\rm P^T$ and $\rm P^L$. We further define\bea
&& H^T(x,\tau) = {\rm P}^{\rm T}_{ij}(\partial) \partial_i h(x,\tau) \partial_j h(x,\tau), \\
&& H^L(x,\tau) = {\rm P}^{\rm L}_{ij}(\partial) \partial_i h(x,\tau) \partial_j h(x,\tau), \\
&& H^M(x,\tau) = \frac{1}{2} {\rm P}^{\rm M}_{ij}(\partial) \partial_i h(x,\tau) \partial_j h(x,\tau).
\eea We note that (\ref{integ1}) and (\ref{integ2}) lead to the usual result for $\omega=0$, i.e.\ in the classical (high $T$) limit,
$S_{\rm eff,fl} = \frac{K_0}{8} |({\rm P}^{\rm T} \partial h \partial h)_{q,\omega}|^2$, with $K_0=4 \mu (\lambda+ \mu)/(\lambda+ 2 \mu)$.
The novelty is the appearance of a coupling to the {\it longitudinal part} of the tensor $\partial_i h \partial_j h$ which arises from an
incomplete screening due to retardation effects. (This coupling is proportional to $\omega^2$).

Integration over in-plane modes also generates a cross-term
\be
\delta S_{\rm eff,fl\text-e} =  \int_{q,\omega} \frac{g}{(\lambda + 2 \mu) q^2 + \rho \omega^2} q_m A_{ijm}(q) (\partial _i h \partial_j h)_{q,\omega} \delta \rho(-q,-\omega).
\ee
\end{widetext}
It has to be added to the direct coupling (\ref{coupl}), \be
S_{\rm fl,e,direct}= - g_0  \int \rmd^2 x\, \rmd \tau\, \frac{1}{2} (\partial _i h \partial_i h) \delta \rho\ ,
\ee
and produces in total
\bea
S_{\rm fl,e} &=& - g_0  \int_{q,\omega} \frac{2 \mu q^2 (\delta_{ij}- \hat q_i \hat q_j) + \rho \omega^2 \delta_{ij}}{(\lambda+2 \mu)q^2 + \rho \omega^2}  \nn\\
&&~~~~~~~~~\times \frac{1}{2} (\partial _i h \partial_j h)_{q,\omega} \delta \rho(-q,-\omega).
~~~~~~\eea
In the limit where one neglects the $\omega$ dependence (e.g. in the classical limit,  as described in the text) it reduces to
\be
S_{\rm fl,e}= - g_0 \frac{2 \mu}{\lambda + 2 \mu}  \int \rmd^2 x\, \rmd \tau\, \frac{1}{2} ({\rm P}^{\rm T}_{ij}(\partial) \partial _i h \partial_j h) \delta \rho .
\ee
In addition integrating over the in-plane phonons generates a short-ranged attraction between electrons,
\be
\delta S = - \frac{1}{2} g^2 \int_{q,\omega} |\rho_{\rm el}(q,\omega)|^2 \frac{q^2}{(\lambda + 2 \mu) q^2 + \rho \omega^2}
.\ee
In the classical limit, or neglecting the frequency dependence, this gives the result (\ref{newV}) quoted in the text.

%
Finally, for completeness we should mention that there is also a fluctuation determinant, which gives an additional contribution to the Matsubara action,
\be
\frac{1}{2}\tr \ln\! \Big( \rho \omega^2 + (\lambda+2 \mu) q^2 \Big) + \frac{1}{2}\tr \ln ( \rho \omega^2 + \mu q^2 ),
\ee
a field-independent temperature dependent constant (which contributes to the specific heat)
but which does not play an  important role in our discussion in the text.

\begin{widetext}
\section{Flexural bubble}
\label{a:phonon-bubble}
Consider the flexural bubble
\be   \label{Ip}
I_0(p,\omega)  = \int_{k} \frac{1}{\beta}\sum_{\omega_{n}}\frac{ \left[k^{2}-
\frac{(k\cdot p)^{2}}{p^{2}} \right]^{2} }{[\kappa ( k+\frac{p}{2})^{4} +\rho
(\omega_{n}+\omega)^{2}] [{\kappa ( k-\frac{p}{2})^{4} +\rho
(\omega_{n})^{2}}]}
,\ee
where the summation is over the Matsubara frequencies $\omega_n=2 \pi n/  \beta$, $n\in \mathbb Z$.

First, in the high-temperature limit,  zero Matsubara frequencies dominate, and (\ref{Ip}) reduces to
the classical result
\be\label{B2}
I_0(p,\omega_m) = \delta_{m,0} I_0(p) \quad , \quad  I_0(p) = \frac{T}{\kappa^2}  \int_{k} \frac{ \left[k^{2}-
\frac{(k\cdot p)^{2}}{p^{2}} \right]^{2} }{( k+\frac{p}{2})^{4} ( k-\frac{p}{2})^{4}} = \frac{3}{16 \pi} \frac{T}{\kappa^2 p^2}\, ,
\ee
which is a convergent integral.
At finite temperature, where quantum effects are important, one must perform the
summation over the Matsubara frequencies $\omega_n=2 \pi n/  \beta$.  Using that $\omega=2\pi j/\beta$, with $j \in \mathbb Z$, and the symmetry $k\to -k$,  one
obtains
\be
 I_0(p,\omega)  = - \int_k  \left[k^{2}-
\frac{(k\cdot p)^{2}}{p^{2}} \right]^{2} \frac{16 \left[\kappa  (k\cdot p) \left(4
   k^2+p^2\right)-\rho  \omega ^2\right] \coth
   \left(\frac{\beta   \sqrt{\kappa } (2 k + p)^2}{8
   \sqrt{\rho }}\right)}{\sqrt{\kappa } \sqrt{\rho
   } (2 k+p)^2
   \left[4 \kappa  (k\cdot p)^2+\rho  \omega
   ^2\right] \left[\kappa  \left(4
   k^2+p^2\right)^2+4 \rho  \omega ^2\right]}
.\ee
It simplifies, for $\omega=0,$ into
\bea
 I_0(p,\omega=0)  &=& -   \frac{1}{(2 \pi)^2 \kappa ^{3/2} \sqrt{\rho }}  \int_0^\Lambda \rmd k \int_0^{2 \pi} \rmd \theta
\frac{4 k^4 \sin ^3(\theta ) \tan (\theta )
   \coth \left(\frac{\beta  \sqrt{\kappa }
   \left(4 k^2+4 k p \cos (\theta )+p^2\right)}{8 \sqrt{\rho}
   }\right)}{ p
   \left(4 k^2+p^2\right) \left(4
   k^2+4 k p \cos (\theta )+p^2\right)}\nn\\
   &=&  - \int_0^{\Lambda/p} \rmd k \int_{-1}^{1} \rmd z\,\frac{2 k^4 \left(1-z^2\right)^{3/2} \coth \left(\frac{\beta  \sqrt{\kappa }
   p^2 (4 k (k+z)+1)}{8 \sqrt{\rho }}\right)}{\pi ^2 \kappa ^{3/2} \sqrt{\rho } \left(4
   k^2+1\right) z (4 k (k+z)+1)}
\label{B4}
,\eea
which one may further symmetrize in $\theta \to \pi+\theta$. Although it looks superficially UV divergent as $O(\Lambda)$, after
symmetrization the UV divergence is only logarithmic: As we will see below, the coefficient of the logarithmic divergence is independent of
temperature.

In the quantum limit $T=0$ we can set $\coth(...) \to 1$ and we obtain, after symmetrization $k \to - k$,
\be  \label{qq}
 I_0(p,\omega)\Big|_{{T=0}}  = \int_k  \left[k^{2}-
\frac{(k\cdot p)^{2}}{p^{2}} \right]^{2}
\frac{16  \left(4 k^2+p^2\right)}{\sqrt{\kappa }
   \sqrt{\rho } \left(4 k^2-4 (k\cdot p)+p^2\right)
   \left(4 \left(k^2+(k\cdot p)\right)+p^2\right)
   \left(\kappa  \left(4 k^2+p^2\right)^2+4 \rho
   \omega ^2\right)}
.\ee
Using the same variable transforms as in (\ref{B4}), we can write it after performing the angular integral as
\be
I_0(p,\omega)\Big|_{{T=0}}  =-\int_0^{\Lambda/p} \rmd k
\frac{k p^4 \left(64 k^6-48 k^4-12
   k^2+1 -\left| 1-4 k^2\right| ^3\right)}{32 \pi  \sqrt{\kappa } \sqrt{\rho } \left(\kappa  \left(4
   k^2+1\right)^2 p^4+4 \rho  \omega ^2\right)}\ .
\ee
This integral is IR convergent and logarithmically UV divergent,
\begin{equation}\label{a19b}
 \left. \frac{\Lambda \partial }{\partial \Lambda } I_0(p,\omega)\right|_{{T=0}} = \frac{1}{(2\pi)^{2}}
 \int_{0}^{2\pi}\rmd \theta \, \frac{\sin ^4(\theta )}{4 \kappa ^{3/2}
 \sqrt{\rho }} =\frac{3}{64 \pi  \kappa ^{3/2} \sqrt{\rho }}
.\end{equation}
The integral can be calculated analytically. With $s:=(2 k/p)^2$ one has
\bea
I_0(p,\omega)\big|_{{T=0}} &=& \int_{0}^{\frac{ 4\Lambda^{2}}{p^{2}}} \rmd s\, \frac{p^4 \left[(s-1)^2 |1-s|-s^3+3 s^2+3 s-1\right]}{256 \pi  \sqrt{\kappa  \rho
   } \left[\kappa  p^4 (s+1)^2+4 \rho  \omega ^2\right]} \nn \\
   & =&-\frac{p^2 \cot ^{-1}\left(\frac{2 \sqrt{\rho } \omega
   }{\sqrt{\kappa } \left(4 \Lambda ^2+p^2\right)}\right)}{64 \pi  \kappa
   \rho  \omega }
+\frac{ \left(\kappa  p^4-3 \rho
   \omega ^2\right) \cot ^{-1}\left(\frac{\sqrt{\rho } \omega }{\sqrt{\kappa }
   p^2}\right)}{32 \pi  \kappa ^{2} p^2 \rho  \omega
   } -\frac{ \left(\kappa  p^4-6 \rho  \omega ^2\right) \cot ^{-1}\left(\frac{2  \sqrt{\rho } \omega }{\sqrt{\kappa } p^2}\right)}{64 \pi  \kappa ^{2} p^2
   \rho  \omega } \nn \\
&&
+\frac{\left(9 \kappa  p^4-4 \rho  \omega ^2\right) \log
   \left(\frac{\kappa  p^4+4 \rho  \omega ^2}{4 \left(\kappa  p^4+\rho  \omega
   ^2\right)}\right)-3 \kappa  p^4 \left(\log \left(\frac{4 \left(\kappa
   p^4+\rho  \omega ^2\right)}{16 \kappa  \Lambda ^4+\kappa  p^4+8 \kappa  \Lambda
   ^2 p^2+4 \rho  \omega ^2}\right)-3\right)}{256 \pi  \kappa ^2 p^4 \sqrt{\kappa \rho} }\ .
\eea
At $\omega=0$ its value is
\be \label{B9}
I_0(p,0)\big|_{{T=0}}=\frac{6 \log \left(
   \frac{4 \Lambda ^2+p^2}{16 p^2} \right)+\frac{8 p^2}{4 \Lambda
   ^2+p^2}+9}{256 \pi  \kappa ^{3/2} \sqrt{\rho }}
,\ee
which leads to (\ref{bubblezero})ff.~in the main text.

Let us now study the crossover as a function of temperature. One can write
\be I_0(p,T,\omega=0)=\frac{6 \log \left(
   \frac{4 \Lambda ^2+p^2}{16 p^2} \right)+\frac{8 p^2}{4 \Lambda
   ^2+p^2}+9}{256 \pi  \kappa ^{3/2} \sqrt{\rho }} +\frac{3}{16 \pi} \frac{T}{\kappa^2
p^2}\, g\!\left(     \frac{  \sqrt{\kappa
} p^2}{8T \sqrt{\rho }} , \frac{\Lambda}{p} \right)
,\ee
where from (\ref{B4}) we obtain the crossover function
\bea
&& g(x,y) : = x \int\limits_{0}^{1}\rmd z\int_0^{y}\limits\rmd k\,
\frac{256 k^4 \left(1-z^2\right)^{3/2} }{3 \pi  \left(4 k^2+1\right) z
   \left(\left(4 k^2+1\right)^2-16 k^2 z^2\right)} \nn\\
&& ~~~~~~~~~~~~~~~~~~~~~~~~~\times  [(4 k (k{+}z)+1)
   \coth (x (4 k (k{-}z)+1))+(4 k (z{-}k)-1) \coth (x (4 k
   (k{+}z)+1))-8 k z].  \label{gx}
~~~~~~~~~\eea
This expression is suitable  for numerical evaluation. Remarkably, the $k$ integration is now
UV convergent, thanks to the substraction of the $T=0$ result, hence there is a well-defined
limit\bea
g(x) := \lim_{\Lambda \to \infty} g\left(x, \frac{\Lambda}{p}\right) = g(x,\infty).
\eea
This function $g(x)$ is given by the integral (\ref{gx}) where the upper integration bound on $k$ is
set to $\infty$. It is not easy to calculate analytically, hence we evaluated it numerically. It is plotted
on figure \ref{f:gfig}. It satisfies $g(0)=1$.
\begin{figure}\
\Fig{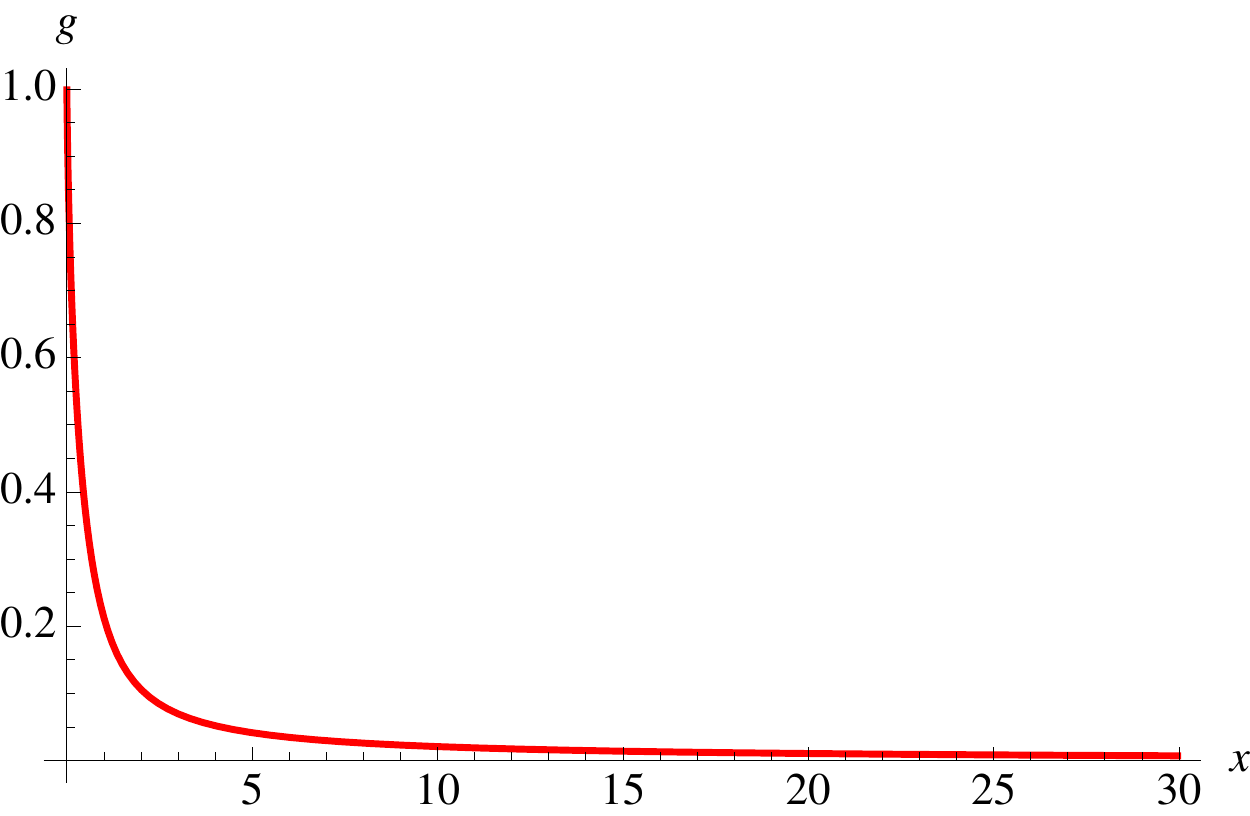}~~~\Fig{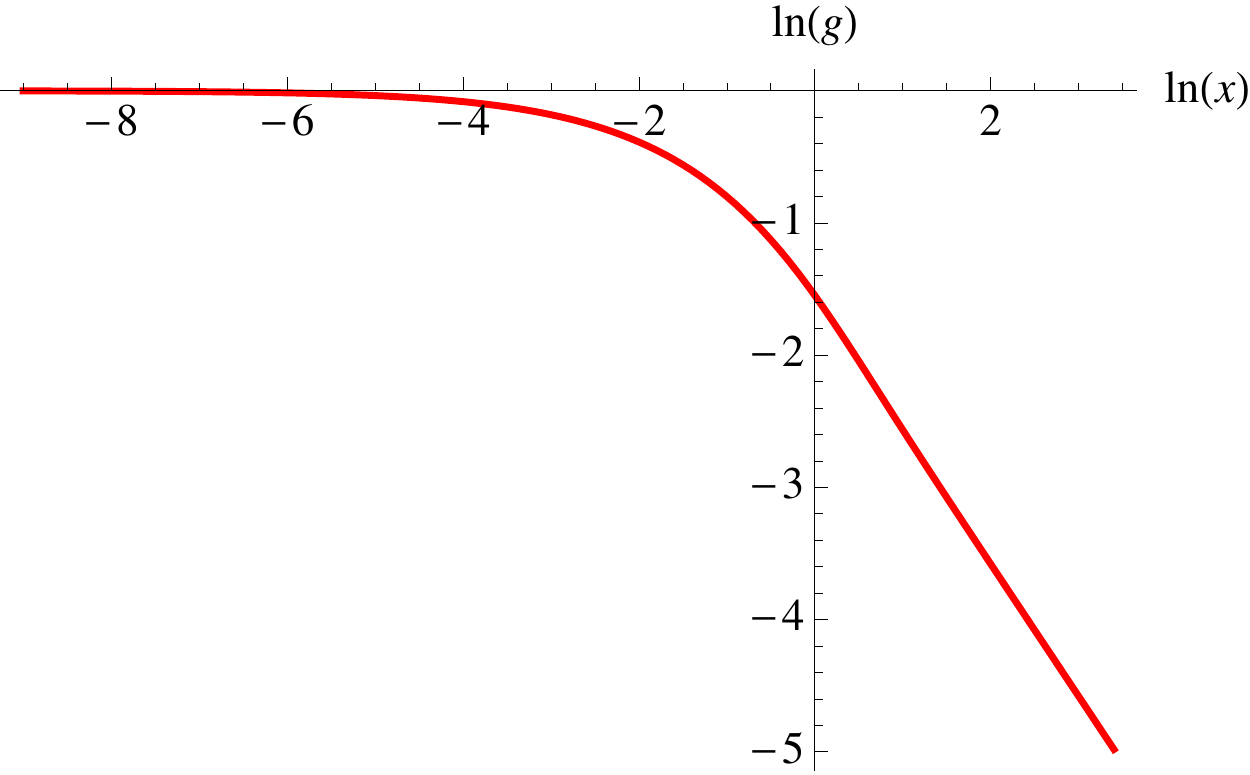}
\caption{$g(x)$ defined in Eq.\ (\ref{B11}).}
\label{f:gfig}
\end{figure}
It describes the thermal crossover for $p \ll \Lambda$.  More precisely,
\be I_0(p,T,\omega=0) = \frac{3 \log \left(
   \frac{\Lambda ^2}{4 p^2} \right)+\frac{9}{2}}{128 \pi  \kappa ^{3/2} \sqrt{\rho }} +\frac{3}{16 \pi} \frac{T}{\kappa^2
p^2}\, g\!\left(     \frac{\omega_{\rm fl}(p)}{8 T} \right) + O\Big(\frac{p^2}{\Lambda^2}\Big) \quad , \quad
\omega_{\rm fl}(p)=p^2 \sqrt{\kappa/\rho} \label{crossover1}
\ ,\ee
are the first two terms in the expansion in $p/\Lambda$ {\it at fixed $T$}. Since  this thermal crossover
occurs for $p^2 \sim T \sqrt{\rho/\kappa}$, the formula (\ref{crossover1}) is  useful
only for $T \ll T_\Lambda = \Lambda^2 \sqrt{\kappa/\rho}= \omega_{\rm fl}(\Lambda)$, i.e.\ the Debye temperature for the flexural phonons.

To obtain the small-$T$ behavior we need to expand $g(x)$ at large $x$. For that we note that $4 k(k{+}z)+1>1$
in the whole integration domain, hence $\coth (x (4 k(k{+}z)+1))$ can be replaced
by $1$ at exponential accuracy (i.e.\ up to $e^{-2 x}$) in the the integral (\ref{gx}).
By contrast, the term $4 k(k{-}z)+1$ in the other argument vanishes for $(z,k)=(1,1/2)$. Expanding
around that point and rescaling by defining new variables $z=1 - \frac{v}{x}$, $k=\frac{1}{2} + \frac{q}{\sqrt{x}}$,
we find at large $x$:
\bea
&& g(x) \simeq_{x \to  \infty}  \frac{C}{x} \quad , \quad C= \int_0^{\infty} dv \int_{-\infty}^{\infty} dq \frac{8 \sqrt{2} v^{3/2} (\coth(2 v + 4 q^2)-1)}{3 \pi (v + 2 q^2)} \approx 0.205617
\ .\eea
This yields the low-temperature behavior
\be I_0(p,T,\omega=0) = \frac{1}{\kappa^{3/2} \sqrt{\rho}} \bigg[  \frac{3}{128 \pi} \left( \log \left(
   \frac{\Lambda ^2}{4 p^2} \right)+\frac{3}{2}\right) +\frac{3 C}{2 \pi} \frac{T^2}{\omega_{\rm fl}(p)^2}  + O\left(\frac{p^2}{\Lambda^2},\frac{T^3}{\omega_{\rm fl}(p)^3}\right)  \bigg]
.\ee
In the opposite limit of $\omega_{\rm fl}(p) \ll T \ll T_\Lambda$, one finds the leading correction to
the classical result,
\be I_0(p,T,\omega=0) \simeq  \frac{3}{16 \pi} \frac{T}{\kappa^2 p^2}
+ \frac{3}{128 \pi} \frac{1}{\kappa^{3/2} \sqrt{\rho}}  \,\ln\!\left(\frac{\omega_{\rm fl}(p)}{8T}\right)
,\ee
using that $g(x) \approx_{x \to 0}  1 + x (\ln x + c) $.
Note this is equivalent to the second term in  Eq.\ (\ref{crossover1}).

For completeness we give the very-high temperature expansion, $T \gg T_\Lambda$,
\bea
I_0(p,0)\Big|_{\beta\to 0} &=&\frac{48 \Lambda ^4-p^4}{16 \beta  \pi
   \kappa ^2 p^2 (4 \Lambda
   ^2+p^2)^2}+\frac{\beta ^3 \Lambda
   ^6}{23040 \pi  \rho ^2}-\frac{\beta ^5
   \kappa  \Lambda ^6 \left(48 \Lambda ^4+5
   p^4+40 \Lambda ^2 p^2\right)}{38707200
   \left(\pi  \rho ^3\right)}\nn\\&&+\frac{\beta ^7 \kappa
   ^2 \Lambda ^6 \left(34560 \Lambda ^8+315 p^8+5880
   \Lambda ^2 p^6+34608 \Lambda ^4 p^4+62720 \Lambda
   ^6 p^2\right)}{1040449536000 \pi  \rho
   ^4}+O(\beta ^9)  .
\label{B11}\eea
This is not useful for graphene  since $T_\Lambda = O(3000K)$.

Let us now consider the analytical continuation to real time via $i \omega_n \to \omega + i \delta$. The $\omega$-dependent factor in (\ref{qq}) yields
the continuation
\be
 {\rm Im}\;
\frac{1}{\kappa (4 k^2+p^2)^2+4 \rho  \omega_n ^2} \;\to  \; \pi\,{\rm sgn}(\omega)\, \delta\!\left(\kappa (4 k^2+p^2)^2-4 \rho  \omega ^2\right)\ .
\ee
Calculating the remaining angular integral we obtain\bea
 {\rm Im}\, I_0(p,i \omega_n \to \omega + i \delta) &=&
\frac{3 \rho  |\omega| -2   p^2 \sqrt{{\rho }{\kappa }}}{256
   \kappa ^{3/2} \rho ^{3/2} \omega }
\, \Theta\!\left(\sqrt{ \frac{\rho }{\kappa} } |\omega| > p^2\right)  \nn\\
&& + \frac{\sqrt{\rho } \left(2 |\omega|  \sqrt{\frac{\rho }{\kappa
   }}-p^2\right)^2 \left(2 p^2-|\omega|  \sqrt{\frac{\rho }{\kappa
   }}\right)}{256 \kappa ^{5/2} p^4 \omega  \left(\frac{\rho }{\kappa
   }\right)^{3/2}}\, \Theta\!\left(p^2 > \sqrt{\frac{\rho }{\kappa}} |\omega| > p^2/2\right).
\eea
This result is presented in the main text in a dimensionless form.


\section{Fermion bubble}
\label{a:fermion-bubble}

We recall  for completeness the calculation of $J_0(p, \omega),$ which is minus the fermion bubble,
using free propagators, \begin{eqnarray}\label{a4}
J_0(p,\omega) &:=& - \int\frac{\rmd^{2}k}{(2\pi)^{2}}
\frac{1}{\beta }\sum_{\omega'_{n}} \tr \left[ \frac{\left(\begin{array}{cc}
i\omega'_{n}+i\omega & -k-p \\
-k^{* }-p^{*}&i\omega'_{n}+i\omega
\end{array} \right)}{(i \omega'_{n}+i\omega)^{2}-|k+p|^{2}}\frac{\left(\begin{array}{cc}
i\omega'_{n} & -k \\
-k^{* }&  i\omega'_{n}
\end{array} \right)}{(i \omega'_{n})^{2}-|k|^{2}} \right]\nn\\
&=& - 2\int\frac{\rmd^{2}k}{(2\pi)^{2}}\frac{1}{\beta }
\sum_{\omega'_{n}}
\frac{ k^{2} - \frac{p^2}{4} + (i \omega'_{n}) (i \omega'_{n}+i\omega ) }
{[(i \omega'_{n}+i\omega)^{2}-|k+\frac{p}{2}|^2][(i \omega'_{n})^{2}-|k-\frac{p}{2}|^{2}]}.
\end{eqnarray}
Here $\omega\equiv \omega_m$ stands for a {\em bosonic} Matsubara frequency while $\omega_n'=\pi(2 n+1)/\beta$ is a {\em fermionic} one.
We have set $v_{\rm F}=1,$ to be restored later.
Summing over the $\omega'_n$ we obtain
\bea
J_0(p,\omega) &=& - 2 \int\frac{\rmd^{2}k}{(2\pi)^{2}} \frac{(k_-^2 - k_+^2) (k_+^2 + k^2-\frac{p^2}{4}) + \omega^2 (k^2-\frac{p^2}{4}-k_+^2)}{k_+ [(k_+ + k_-)^2 + \omega^2] [(k_+ - k_-)^2 + \omega^2]} \tanh\left(\frac{\beta k_+}{2}\right) \quad , \quad k_\pm = \left|k \pm \frac{p}{2}\right| \nn\\
&=&\int\frac{\rmd^{2}k}{(2\pi)^{2}} \frac{ 2 k \cdot p \left(4
   k^2+2 k \cdot p+\omega ^2\right)+p^2 \omega ^2}{k_+ [\omega ^2
   \left(4 k^2+p^2+\omega ^2\right)+4 (k \cdot p)^2]} \tanh \left(\frac{\beta  k_+}{2}\right) .
\eea
This can be symmetrized over $p \to -p$. In the limit $T=0$, this reduces to {\red}\be
J_0(p,\omega) =  \int\frac{\rmd^{2}k}{(2\pi)^{2}} \frac{(k_++k_-)(k_+ k_- - k^2 + \frac{p^2}{4})}{k_+ k_- [(k_++k_-)^2 + \omega^2]} \ .
\ee
Evaluation of this integral can be done, using distance geometry, \begin{eqnarray}
\int\frac{\rmd^2 k}{(2\pi)^2} f(k_+,k_-)  &=&   \frac1{\pi^2} \int \frac{ \rmd k_+\, \rmd
k_-\,  k_- k_+ f(k_+,k_-)\Theta(|k_+-k_-|>p) }{\sqrt{\left(p-k_-+k_+\right)
   \left(k_--k_++p\right) \left(k_-+k_+-p\right)
   \left(k_-+k_++p\right)}}\nn\\
& =&  \frac{1}{4\pi^2}\int_{0}^\infty \rmd x\int_{-p/2}^{p/2}\rmd y\, \frac{(p+x)^2-4 y^2}{\sqrt{x (2 p+x) \left(p^2-4 y^2\right)}} f(p+x+y,p+x-y).
\end{eqnarray}\end{widetext}
This gives, restoring the $v_{\rm F}$ factor,
\be
J_0(p,\omega) = \frac{p^2}{16 \sqrt{v_{\rm F}^2 p^2 + \omega^2} }
\ .
\ee
A similar calculation for arbitrary $\beta$ yields  \cite{Gazit2} 
\be
J_0(p,0) = \frac{ T}{\pi v_{\rm F}^2} \int_0^1 \rmd x\, \ln\left(2 \cosh\left(\frac{v_{\rm F} \beta p}{2} \sqrt{x(1-x)}\right) \right) .
\ee
This leads to the classical limit for $v_{\rm F} \beta p \ll 1$,
\be
J_0(p,0) \simeq \frac{ \ln 2}{\pi v_{\rm F}^2} T ,
\ee
and a sharp crossover to $J_0(q,0)= p/(16 v_{\rm F})$ at $v_{\rm F} \beta p \approx 3$.

%
%
%

\section{$O(d)$ symmetry breaking}
\label{solu1}

Although we found in the text that for realistic couplings the instability arises for intermediate
wave vectors $q_c$, it is still interesting to investigate how an
almost uniform order parameter $\sigma(x)\approx\sigma$ and $\alpha(x)\approx\alpha$
could induce a breaking of the $O(d)$ symmetry at a finite $q$ for $h_1(x)$.

Smearing out $\sigma(p)$ isotropically around $p=0$,
one can replace $
\int \sigma(p) P^{\rm T}_{ij} (p) \partial_{i}
 \partial_{j} \to \left(1-\frac1d\right) \int \sigma(p) \partial_{i}
 \partial_{j}.$ Now the saddle-point equation (\ref{SP1}) reduces to \be
\kappa k^4 h_1(k) + \frac{k^{2}}2 \sigma h_1(k) = 0.
\ee
This equation has two solutions, either $h (k)=0$, or the non-trivial solution:
\begin{eqnarray}\label{sol}
h_1 (k)= {\sf h} \delta^{2}
(k-k_{0})\ , \\
\sigma= - 2 \kappa k_0^2 \ .
\end{eqnarray}
This shows that the magnitude of $\sigma(p \approx 0)$ sets a scale for the
$O(d)$ symmetry breaking. More investigations are needed to see if a
closed solution to the full set of saddle-point equations can be constructed 
along these lines.


\begin{thebibliography}{37}%
\makeatletter
\providecommand \@ifxundefined [1]{%
 \@ifx{#1\undefined}
}%
\providecommand \@ifnum [1]{%
 \ifnum #1\expandafter \@firstoftwo
 \else \expandafter \@secondoftwo
 \fi
}%
\providecommand \@ifx [1]{%
 \ifx #1\expandafter \@firstoftwo
 \else \expandafter \@secondoftwo
 \fi
}%
\providecommand \natexlab [1]{#1}%
\providecommand \enquote  [1]{``#1''}%
\providecommand \bibnamefont  [1]{#1}%
\providecommand \bibfnamefont [1]{#1}%
\providecommand  \citenamefont [1]{#1}%
\providecommand \href@noop [0]{\@secondoftwo}%
\providecommand \href [0]{\begingroup \@sanitize@url \@href}%
\providecommand \@href[1]{\@@startlink{#1}\@@href}%
\providecommand \@@href[1]{\endgroup#1\@@endlink}%
\providecommand \@sanitize@url [0]{\catcode `\\12\catcode `\$12\catcode
  `\&12\catcode `\#12\catcode `\^12\catcode `\_12\catcode `\%12\relax}%
\providecommand \@@startlink[1]{}%
\providecommand \@@endlink[0]{}%
\providecommand \url  [0]{\begingroup\@sanitize@url \@url }%
\providecommand \@url [1]{\endgroup\@href {#1}{\urlprefix }}%
\providecommand \urlprefix  [0]{URL }%
\providecommand \Eprint [0]{\href }%
\providecommand \doibase [0]{http://dx.doi.org/}%
\providecommand \selectlanguage [0]{\@gobble}%
\providecommand \bibinfo  [0]{\@secondoftwo}%
\providecommand \bibfield  [0]{\@secondoftwo}%
\providecommand \translation [1]{[#1]}%
\providecommand \BibitemOpen [0]{}%
\providecommand \bibitemStop [0]{}%
\providecommand \bibitemNoStop [0]{.\EOS\space}%
\providecommand \EOS [0]{\spacefactor3000\relax}%
\providecommand \BibitemShut  [1]{\csname bibitem#1\endcsname}%
\let\auto@bib@innerbib\@empty
\bibitem [{ \citenamefont {Novoselov}\ \emph {et~al.}(2004) \citenamefont
  {Novoselov},  \citenamefont {Geim},  \citenamefont {Morozov},  \citenamefont
  {Jiang},  \citenamefont {Zhang},  \citenamefont {Dubonos},  \citenamefont
  {Grigorieva},\ and\  \citenamefont {Firsov}}]{Netal04}%
  \BibitemOpen
  \bibfield  {author} {\bibinfo {author} {\bibfnamefont {K.~S.}\ \bibnamefont
  {Novoselov}}, \bibinfo {author} {\bibfnamefont {A.~K.}\ \bibnamefont {Geim}},
  \bibinfo {author} {\bibfnamefont {S.~V.}\ \bibnamefont {Morozov}}, \bibinfo
  {author} {\bibfnamefont {D.}~\bibnamefont {Jiang}}, \bibinfo {author}
  {\bibfnamefont {Y.}~\bibnamefont {Zhang}}, \bibinfo {author} {\bibfnamefont
  {S.~V.}\ \bibnamefont {Dubonos}}, \bibinfo {author} {\bibfnamefont {I.~V.}\
  \bibnamefont {Grigorieva}}, \ and\ \bibinfo {author} {\bibfnamefont {A.~A.}\
  \bibnamefont {Firsov}},\ }\href@noop {} {\bibfield  {journal} {\bibinfo
  {journal} {Science}\ }\textbf {\bibinfo {volume} {306}},\ \bibinfo {pages}
  {666} (\bibinfo {year} {2004})}\BibitemShut {NoStop}%
\bibitem [{ \citenamefont {Novoselov}\ \emph {et~al.}(2005) \citenamefont
  {Novoselov},  \citenamefont {Jiang},  \citenamefont {Schedin},  \citenamefont
  {Booth},  \citenamefont {Khotkevich},  \citenamefont {Morozov},\ and\
   \citenamefont {Geim}}]{Netal05}%
  \BibitemOpen
  \bibfield  {author} {\bibinfo {author} {\bibfnamefont {K.~S.}\ \bibnamefont
  {Novoselov}}, \bibinfo {author} {\bibfnamefont {D.}~\bibnamefont {Jiang}},
  \bibinfo {author} {\bibfnamefont {F.}~\bibnamefont {Schedin}}, \bibinfo
  {author} {\bibfnamefont {T.~J.}\ \bibnamefont {Booth}}, \bibinfo {author}
  {\bibfnamefont {V.~V.}\ \bibnamefont {Khotkevich}}, \bibinfo {author}
  {\bibfnamefont {S.~V.}\ \bibnamefont {Morozov}}, \ and\ \bibinfo {author}
  {\bibfnamefont {A.~K.}\ \bibnamefont {Geim}},\ }\href@noop {} {\bibfield
  {journal} {\bibinfo  {journal} {Proc. Natl. Acad. Sci. U.S.A.}\ }\textbf
  {\bibinfo {volume} {102}},\ \bibinfo {pages} {10451} (\bibinfo {year}
  {2005})}\BibitemShut {NoStop}%
\bibitem [{ \citenamefont {Castro~Neto}\ \emph {et~al.}(2009) \citenamefont
  {Castro~Neto},  \citenamefont {Guinea},  \citenamefont {Peres},  \citenamefont
  {Novoselov},\ and\  \citenamefont {Geim}}]{NGPNG09}%
  \BibitemOpen
  \bibfield  {author} {\bibinfo {author} {\bibfnamefont {A.~H.}\ \bibnamefont
  {Castro~Neto}}, \bibinfo {author} {\bibfnamefont {F.}~\bibnamefont {Guinea}},
  \bibinfo {author} {\bibfnamefont {N.~M.~R.}\ \bibnamefont {Peres}}, \bibinfo
  {author} {\bibfnamefont {K.~S.}\ \bibnamefont {Novoselov}}, \ and\ \bibinfo
  {author} {\bibfnamefont {A.~K.}\ \bibnamefont {Geim}},\ }\href {\doibase
  10.1103/RevModPhys.81.109} {\bibfield  {journal} {\bibinfo  {journal} {Rev.
  Mod. Phys.}\ }\textbf {\bibinfo {volume} {81}},\ \bibinfo {pages} {109}
  (\bibinfo {year} {2009})}\BibitemShut {NoStop}%
\bibitem [{ \citenamefont {Lee}\ \emph {et~al.}(2008) \citenamefont {Lee},
   \citenamefont {Wei},  \citenamefont {Kysar},\ and\  \citenamefont
  {Hone}}]{LWKH08}%
  \BibitemOpen
  \bibfield  {author} {\bibinfo {author} {\bibfnamefont {C.}~\bibnamefont
  {Lee}}, \bibinfo {author} {\bibfnamefont {X.}~\bibnamefont {Wei}}, \bibinfo
  {author} {\bibfnamefont {J.~W.}\ \bibnamefont {Kysar}}, \ and\ \bibinfo
  {author} {\bibfnamefont {J.}~\bibnamefont {Hone}},\ }\href {\doibase
  10.1126/science.1157996} {\bibfield  {journal} {\bibinfo  {journal}
  {Science}\ }\textbf {\bibinfo {volume} {321}},\ \bibinfo {pages} {5887}
  (\bibinfo {year} {2008})}\BibitemShut {NoStop}%
\bibitem [{ \citenamefont {Stolyarova}\ \emph {et~al.}(2008) \citenamefont
  {Stolyarova},  \citenamefont {Rim},  \citenamefont {Ryu},  \citenamefont
  {Maultzsch},  \citenamefont {Kim},  \citenamefont {Brus},  \citenamefont
  {Heinz},  \citenamefont {Hybertsen},\ and\  \citenamefont {Flynn}}]{Setal08}%
  \BibitemOpen
  \bibfield  {author} {\bibinfo {author} {\bibfnamefont {E.}~\bibnamefont
  {Stolyarova}}, \bibinfo {author} {\bibfnamefont {K.~T.}\ \bibnamefont {Rim}},
  \bibinfo {author} {\bibfnamefont {S.}~\bibnamefont {Ryu}}, \bibinfo {author}
  {\bibfnamefont {J.}~\bibnamefont {Maultzsch}}, \bibinfo {author}
  {\bibfnamefont {P.}~\bibnamefont {Kim}}, \bibinfo {author} {\bibfnamefont
  {L.~E.}\ \bibnamefont {Brus}}, \bibinfo {author} {\bibfnamefont {T.~F.}\
  \bibnamefont {Heinz}}, \bibinfo {author} {\bibfnamefont {M.~S.}\ \bibnamefont
  {Hybertsen}}, \ and\ \bibinfo {author} {\bibfnamefont {G.~W.}\ \bibnamefont
  {Flynn}},\ }\href@noop {} {\bibfield  {journal} {\bibinfo  {journal} {Proc.
  Nat. Ac. Sci. (USA)}\ }\textbf {\bibinfo {volume} {104}},\ \bibinfo {pages}
  {9209} (\bibinfo {year} {2008})}\BibitemShut {NoStop}%
\bibitem [{ \citenamefont {Geringer}\ \emph {et~al.}(2009) \citenamefont
  {Geringer},  \citenamefont {Liebmann},  \citenamefont {Echtermeyer},
   \citenamefont {Runte},  \citenamefont {Schmidt},  \citenamefont {R\"uckamp},
   \citenamefont {Lemme},\ and\  \citenamefont {Morgenstern}}]{Getal09}%
  \BibitemOpen
  \bibfield  {author} {\bibinfo {author} {\bibfnamefont {V.}~\bibnamefont
  {Geringer}}, \bibinfo {author} {\bibfnamefont {M.}~\bibnamefont {Liebmann}},
  \bibinfo {author} {\bibfnamefont {T.}~\bibnamefont {Echtermeyer}}, \bibinfo
  {author} {\bibfnamefont {S.}~\bibnamefont {Runte}}, \bibinfo {author}
  {\bibfnamefont {M.}~\bibnamefont {Schmidt}}, \bibinfo {author} {\bibfnamefont
  {R.}~\bibnamefont {R\"uckamp}}, \bibinfo {author} {\bibfnamefont {M.~C.}\
  \bibnamefont {Lemme}}, \ and\ \bibinfo {author} {\bibfnamefont
  {M.}~\bibnamefont {Morgenstern}},\ }\href {\doibase
  10.1103/PhysRevLett.102.076102} {\bibfield  {journal} {\bibinfo  {journal}
  {Phys. Rev. Lett.}\ }\textbf {\bibinfo {volume} {102}},\ \bibinfo {pages}
  {076102} (\bibinfo {year} {2009})}\BibitemShut {NoStop}%
\bibitem [{ \citenamefont {Viola~Kusminskiy}\ \emph {et~al.}(2011) \citenamefont
  {Viola~Kusminskiy},  \citenamefont {Campbell},  \citenamefont {Castro~Neto},\
  and\  \citenamefont {Guinea}}]{KCNG11}%
  \BibitemOpen
  \bibfield  {author} {\bibinfo {author} {\bibfnamefont {S.}~\bibnamefont
  {Viola~Kusminskiy}}, \bibinfo {author} {\bibfnamefont {D.~K.}\ \bibnamefont
  {Campbell}}, \bibinfo {author} {\bibfnamefont {A.~H.}\ \bibnamefont
  {Castro~Neto}}, \ and\ \bibinfo {author} {\bibfnamefont {F.}~\bibnamefont
  {Guinea}},\ }\href {\doibase 10.1103/PhysRevB.83.165405} {\bibfield
  {journal} {\bibinfo  {journal} {Phys. Rev. B}\ }\textbf {\bibinfo {volume}
  {83}},\ \bibinfo {pages} {165405} (\bibinfo {year} {2011})}\BibitemShut
  {NoStop}%
\bibitem [{ \citenamefont {de~Parga}\ \emph {et~al.}(2008) \citenamefont
  {de~Parga},  \citenamefont {Calleja},  \citenamefont {Borca},  \citenamefont
  {Passeggi},  \citenamefont {Hinarejos},  \citenamefont {Guinea},\ and\
   \citenamefont {Miranda}}]{Vetal08}%
  \BibitemOpen
  \bibfield  {author} {\bibinfo {author} {\bibfnamefont {A.~L.~V.}\
  \bibnamefont {de~Parga}}, \bibinfo {author} {\bibfnamefont {F.}~\bibnamefont
  {Calleja}}, \bibinfo {author} {\bibfnamefont {B.}~\bibnamefont {Borca}},
  \bibinfo {author} {\bibfnamefont {M.~C.}\ \bibnamefont {Passeggi}}, \bibinfo
  {author} {\bibfnamefont {J.~J.}\ \bibnamefont {Hinarejos}}, \bibinfo {author}
  {\bibfnamefont {F.}~\bibnamefont {Guinea}}, \ and\ \bibinfo {author}
  {\bibfnamefont {R.}~\bibnamefont {Miranda}},\ }\href@noop {} {\bibfield
  {journal} {\bibinfo  {journal} {Phys. Rev. Lett.}\ }\textbf {\bibinfo
  {volume} {100}},\ \bibinfo {pages} {056807} (\bibinfo {year}
  {2008})}\BibitemShut {NoStop}%
\bibitem [{ \citenamefont {Meyer}\ \emph {et~al.}(2007) \citenamefont {Meyer},
   \citenamefont {Geim},  \citenamefont {Katsnelson},  \citenamefont {Novoselov},
   \citenamefont {Booth},\ and\  \citenamefont {Roth}}]{Metal07}%
  \BibitemOpen
  \bibfield  {author} {\bibinfo {author} {\bibfnamefont {J.~C.}\ \bibnamefont
  {Meyer}}, \bibinfo {author} {\bibfnamefont {A.~K.}\ \bibnamefont {Geim}},
  \bibinfo {author} {\bibfnamefont {M.~I.}\ \bibnamefont {Katsnelson}},
  \bibinfo {author} {\bibfnamefont {K.~S.}\ \bibnamefont {Novoselov}}, \bibinfo
  {author} {\bibfnamefont {T.~J.}\ \bibnamefont {Booth}}, \ and\ \bibinfo
  {author} {\bibfnamefont {S.}~\bibnamefont {Roth}},\ }\href@noop {} {\bibfield
   {journal} {\bibinfo  {journal} {Nature}\ }\textbf {\bibinfo {volume}
  {446}},\ \bibinfo {pages} {60} (\bibinfo {year} {2007})}\BibitemShut
  {NoStop}%
\bibitem [{ \citenamefont {Fasolino}\ \emph {et~al.}(2007) \citenamefont
  {Fasolino},  \citenamefont {Los},\ and\  \citenamefont {Katsnelson}}]{FLK07}%
  \BibitemOpen
  \bibfield  {author} {\bibinfo {author} {\bibfnamefont {A.}~\bibnamefont
  {Fasolino}}, \bibinfo {author} {\bibfnamefont {J.~H.}\ \bibnamefont {Los}}, \
  and\ \bibinfo {author} {\bibfnamefont {M.~I.}\ \bibnamefont {Katsnelson}},\
  }\href@noop {} {\bibfield  {journal} {\bibinfo  {journal} {Nature Mater.}\
  }\textbf {\bibinfo {volume} {6}},\ \bibinfo {pages} {858} (\bibinfo {year}
  {2007})}\BibitemShut {NoStop}%
\bibitem [{ \citenamefont {Horovitz}\ and\  \citenamefont
  {Doussal}(2002)}]{HorovitzDoussal2002}%
  \BibitemOpen
  \bibfield  {author} {\bibinfo {author} {\bibfnamefont {B.}~\bibnamefont
  {Horovitz}}\ and\ \bibinfo {author} {\bibfnamefont {P.~L.}\ \bibnamefont
  {Doussal}},\ }\href {\doibase 10.1103/PhysRevB.65.125323} {\bibfield
  {journal} {\bibinfo  {journal} {Phys. Rev. B}\ }\textbf {\bibinfo {volume}
  {65}},\ \bibinfo {pages} {125323} (\bibinfo {year} {2002})}\BibitemShut
  {NoStop}%
\bibitem [{ \citenamefont {Guinea}\ \emph {et~al.}(2008) \citenamefont {Guinea},
   \citenamefont {Horovitz},\ and\  \citenamefont {Le~Doussal}}]{GHL08}%
  \BibitemOpen
  \bibfield  {author} {\bibinfo {author} {\bibfnamefont {F.}~\bibnamefont
  {Guinea}}, \bibinfo {author} {\bibfnamefont {B.}~\bibnamefont {Horovitz}}, \
  and\ \bibinfo {author} {\bibfnamefont {P.}~\bibnamefont {Le~Doussal}},\
  }\href {\doibase 10.1103/PhysRevB.77.205421} {\bibfield  {journal} {\bibinfo
  {journal} {Phys. Rev. B}\ }\textbf {\bibinfo {volume} {77}},\ \bibinfo
  {pages} {205421} (\bibinfo {year} {2008})}\BibitemShut {NoStop}%
\bibitem [{ \citenamefont {Guinea}\ \emph {et~al.}(2009) \citenamefont {Guinea},
   \citenamefont {Horovitz},\ and\  \citenamefont {{Le Doussal}}}]{GHL09}%
  \BibitemOpen
  \bibfield  {author} {\bibinfo {author} {\bibfnamefont {F.}~\bibnamefont
  {Guinea}}, \bibinfo {author} {\bibfnamefont {B.}~\bibnamefont {Horovitz}}, \
  and\ \bibinfo {author} {\bibfnamefont {P.}~\bibnamefont {{Le Doussal}}},\
  }\href@noop {} {\bibfield  {journal} {\bibinfo  {journal} {Solid State
  Communications}\ }\textbf {\bibinfo {volume} {149}},\ \bibinfo {pages} {1140}
  (\bibinfo {year} {2009})}\BibitemShut {NoStop}%
\bibitem [{ \citenamefont {Mariani}\ and\  \citenamefont {von
  Oppen}(2008)}]{MO08}%
  \BibitemOpen
  \bibfield  {author} {\bibinfo {author} {\bibfnamefont {E.}~\bibnamefont
  {Mariani}}\ and\ \bibinfo {author} {\bibfnamefont {F.}~\bibnamefont {von
  Oppen}},\ }\href {\doibase 10.1103/PhysRevLett.100.076801} {\bibfield
  {journal} {\bibinfo  {journal} {Phys. Rev. Lett.}\ }\textbf {\bibinfo
  {volume} {100}},\ \bibinfo {pages} {076801} (\bibinfo {year}
  {2008})}\BibitemShut {NoStop}%
\bibitem [{ \citenamefont {Castro}\ \emph {et~al.}(2010) \citenamefont {Castro},
   \citenamefont {Ochoa},  \citenamefont {Katsnelson},  \citenamefont {Gorbachev},
   \citenamefont {Elias},  \citenamefont {Novoselov},  \citenamefont {Geim},\ and\
   \citenamefont {Guinea}}]{Cetal10}%
  \BibitemOpen
  \bibfield  {author} {\bibinfo {author} {\bibfnamefont {E.~V.}\ \bibnamefont
  {Castro}}, \bibinfo {author} {\bibfnamefont {H.}~\bibnamefont {Ochoa}},
  \bibinfo {author} {\bibfnamefont {M.~I.}\ \bibnamefont {Katsnelson}},
  \bibinfo {author} {\bibfnamefont {R.~V.}\ \bibnamefont {Gorbachev}}, \bibinfo
  {author} {\bibfnamefont {D.~C.}\ \bibnamefont {Elias}}, \bibinfo {author}
  {\bibfnamefont {K.~S.}\ \bibnamefont {Novoselov}}, \bibinfo {author}
  {\bibfnamefont {A.~K.}\ \bibnamefont {Geim}}, \ and\ \bibinfo {author}
  {\bibfnamefont {F.}~\bibnamefont {Guinea}},\ }\href {\doibase
  10.1103/PhysRevLett.105.266601} {\bibfield  {journal} {\bibinfo  {journal}
  {Phys. Rev. Lett.}\ }\textbf {\bibinfo {volume} {105}},\ \bibinfo {pages}
  {266601} (\bibinfo {year} {2010})}\BibitemShut {NoStop}%
\bibitem [{ \citenamefont {Mariani}\ and\  \citenamefont {von
  Oppen}(2010)}]{MO10}%
  \BibitemOpen
  \bibfield  {author} {\bibinfo {author} {\bibfnamefont {E.}~\bibnamefont
  {Mariani}}\ and\ \bibinfo {author} {\bibfnamefont {F.}~\bibnamefont {von
  Oppen}},\ }\href {\doibase 10.1103/PhysRevB.82.195403} {\bibfield  {journal}
  {\bibinfo  {journal} {Phys. Rev. B}\ }\textbf {\bibinfo {volume} {82}},\
  \bibinfo {pages} {195403} (\bibinfo {year} {2010})}\BibitemShut {NoStop}%
\bibitem [{ \citenamefont {Gazit}(2009{\natexlab{a}})}]{G09}%
  \BibitemOpen
  \bibfield  {author} {\bibinfo {author} {\bibfnamefont {D.}~\bibnamefont
  {Gazit}},\ }\href {\doibase 10.1103/PhysRevB.80.161406} {\bibfield  {journal}
  {\bibinfo  {journal} {Phys. Rev. B}\ }\textbf {\bibinfo {volume} {80}},\
  \bibinfo {pages} {161406} (\bibinfo {year} {2009}{\natexlab{a}})}\BibitemShut
  {NoStop}%
\bibitem [{ \citenamefont {San-Jose}\ \emph {et~al.}(2011) \citenamefont
  {San-Jose},  \citenamefont {Gonz\'alez},\ and\  \citenamefont
  {Guinea}}]{SGG11}%
  \BibitemOpen
  \bibfield  {author} {\bibinfo {author} {\bibfnamefont {P.}~\bibnamefont
  {San-Jose}}, \bibinfo {author} {\bibfnamefont {J.}~\bibnamefont
  {Gonz\'alez}}, \ and\ \bibinfo {author} {\bibfnamefont {F.}~\bibnamefont
  {Guinea}},\ }\href {\doibase 10.1103/PhysRevLett.106.045502} {\bibfield
  {journal} {\bibinfo  {journal} {Phys. Rev. Lett.}\ }\textbf {\bibinfo
  {volume} {106}},\ \bibinfo {pages} {045502} (\bibinfo {year}
  {2011})}\BibitemShut {NoStop}%
\bibitem [{ \citenamefont {Gibertini}\ \emph {et~al.}(2012) \citenamefont
  {Gibertini},  \citenamefont {Tomadin},  \citenamefont {Guinea},  \citenamefont
  {Katsnelson},\ and\  \citenamefont {Polini}}]{GTGKP12}%
  \BibitemOpen
  \bibfield  {author} {\bibinfo {author} {\bibfnamefont {M.}~\bibnamefont
  {Gibertini}}, \bibinfo {author} {\bibfnamefont {A.}~\bibnamefont {Tomadin}},
  \bibinfo {author} {\bibfnamefont {F.}~\bibnamefont {Guinea}}, \bibinfo
  {author} {\bibfnamefont {M.~I.}\ \bibnamefont {Katsnelson}}, \ and\ \bibinfo
  {author} {\bibfnamefont {M.}~\bibnamefont {Polini}},\ }\href {\doibase
  10.1103/PhysRevB.85.201405} {\bibfield  {journal} {\bibinfo  {journal} {Phys.
  Rev. B}\ }\textbf {\bibinfo {volume} {85}},\ \bibinfo {pages} {201405}
  (\bibinfo {year} {2012})}\BibitemShut {NoStop}%
\bibitem [{ \citenamefont {Nelson}\ and\  \citenamefont {Peliti}(1987)}]{NP87}%
  \BibitemOpen
  \bibfield  {author} {\bibinfo {author} {\bibfnamefont {D.~R.}\ \bibnamefont
  {Nelson}}\ and\ \bibinfo {author} {\bibfnamefont {L.}~\bibnamefont
  {Peliti}},\ }\href@noop {} {\bibfield  {journal} {\bibinfo  {journal} {J.
  Phys. France}\ }\textbf {\bibinfo {volume} {48}},\ \bibinfo {pages} {1085}
  (\bibinfo {year} {1987})}\BibitemShut {NoStop}%
\bibitem [{ \citenamefont {Aronovitz}\ and\  \citenamefont
  {Lubensky}(1988)}]{AL88}%
  \BibitemOpen
  \bibfield  {author} {\bibinfo {author} {\bibfnamefont {J.~A.}\ \bibnamefont
  {Aronovitz}}\ and\ \bibinfo {author} {\bibfnamefont {T.~C.}\ \bibnamefont
  {Lubensky}},\ }\href {\doibase 10.1103/PhysRevLett.60.2634} {\bibfield
  {journal} {\bibinfo  {journal} {Phys. Rev. Lett.}\ }\textbf {\bibinfo
  {volume} {60}},\ \bibinfo {pages} {2634} (\bibinfo {year}
  {1988})}\BibitemShut {NoStop}%
\bibitem [{ \citenamefont {David}\ and\  \citenamefont
  {Guitter}(1988)}]{DavidGuitter1988}%
  \BibitemOpen
  \bibfield  {author} {\bibinfo {author} {\bibfnamefont {F.}~\bibnamefont
  {David}}\ and\ \bibinfo {author} {\bibfnamefont {E.}~\bibnamefont
  {Guitter}},\ }\href@noop {} {\bibfield  {journal} {\bibinfo  {journal}
  {Europhys. Lett.}\ }\textbf {\bibinfo {volume} {5}},\ \bibinfo {pages} {709}
  (\bibinfo {year} {1988})}\BibitemShut {NoStop}%
\bibitem [{ \citenamefont {Le~Doussal}\ and\  \citenamefont
  {Radzihovsky}(1992)}]{LR92}%
  \BibitemOpen
  \bibfield  {author} {\bibinfo {author} {\bibfnamefont {P.}~\bibnamefont
  {Le~Doussal}}\ and\ \bibinfo {author} {\bibfnamefont {L.}~\bibnamefont
  {Radzihovsky}},\ }\href {\doibase 10.1103/PhysRevLett.69.1209} {\bibfield
  {journal} {\bibinfo  {journal} {Phys. Rev. Lett.}\ }\textbf {\bibinfo
  {volume} {69}},\ \bibinfo {pages} {1209} (\bibinfo {year}
  {1992})}\BibitemShut {NoStop}%
\bibitem [{ \citenamefont {Zakharchenko}\ \emph {et~al.}(2010) \citenamefont
  {Zakharchenko},  \citenamefont {Rold\'an},  \citenamefont {Fasolino},\ and\
   \citenamefont {Katsnelson}}]{ZRFK10}%
  \BibitemOpen
  \bibfield  {author} {\bibinfo {author} {\bibfnamefont {K.~V.}\ \bibnamefont
  {Zakharchenko}}, \bibinfo {author} {\bibfnamefont {R.}~\bibnamefont
  {Rold\'an}}, \bibinfo {author} {\bibfnamefont {A.}~\bibnamefont {Fasolino}},
  \ and\ \bibinfo {author} {\bibfnamefont {M.~I.}\ \bibnamefont {Katsnelson}},\
  }\href {\doibase 10.1103/PhysRevB.82.125435} {\bibfield  {journal} {\bibinfo
  {journal} {Phys. Rev. B}\ }\textbf {\bibinfo {volume} {82}},\ \bibinfo
  {pages} {125435} (\bibinfo {year} {2010})}\BibitemShut {NoStop}%
\bibitem [{ \citenamefont {Gonzalez}\ \emph {et~al.}(1994) \citenamefont
  {Gonzalez},  \citenamefont {Guinea},\ and\  \citenamefont
  {Vozmediano}}]{GGV94}%
  \BibitemOpen
  \bibfield  {author} {\bibinfo {author} {\bibfnamefont {J.}~\bibnamefont
  {Gonzalez}}, \bibinfo {author} {\bibfnamefont {F.}~\bibnamefont {Guinea}}, \
  and\ \bibinfo {author} {\bibfnamefont {M.~A.}\ \bibnamefont {Vozmediano}},\
  }\href@noop {} {\bibfield  {journal} {\bibinfo  {journal} {Nuclear Physics
  B}\ }\textbf {\bibinfo {volume} {424}},\ \bibinfo {pages} {595} (\bibinfo
  {year} {1994})}\BibitemShut {NoStop}%
\bibitem [{ \citenamefont {Elias}\ \emph {et~al.}(2011) \citenamefont {Elias},
   \citenamefont {Gorbachev},  \citenamefont {Mayorov},  \citenamefont {Morozov},
   \citenamefont {Zhukov},  \citenamefont {Blake},  \citenamefont {Ponomarenko},
   \citenamefont {Grigorieva},  \citenamefont {Novoselov},  \citenamefont
  {Guinea},\ and\  \citenamefont {Geim}}]{Eetal11}%
  \BibitemOpen
  \bibfield  {author} {\bibinfo {author} {\bibfnamefont {D.~C.}\ \bibnamefont
  {Elias}}, \bibinfo {author} {\bibfnamefont {R.~V.}\ \bibnamefont
  {Gorbachev}}, \bibinfo {author} {\bibfnamefont {A.~S.}\ \bibnamefont
  {Mayorov}}, \bibinfo {author} {\bibfnamefont {S.~V.}\ \bibnamefont
  {Morozov}}, \bibinfo {author} {\bibfnamefont {A.~A.}\ \bibnamefont {Zhukov}},
  \bibinfo {author} {\bibfnamefont {P.}~\bibnamefont {Blake}}, \bibinfo
  {author} {\bibfnamefont {L.~A.}\ \bibnamefont {Ponomarenko}}, \bibinfo
  {author} {\bibfnamefont {I.~V.}\ \bibnamefont {Grigorieva}}, \bibinfo
  {author} {\bibfnamefont {K.~S.}\ \bibnamefont {Novoselov}}, \bibinfo {author}
  {\bibfnamefont {F.}~\bibnamefont {Guinea}}, \ and\ \bibinfo {author}
  {\bibfnamefont {A.~K.}\ \bibnamefont {Geim}},\ }\href@noop {} {\bibfield
  {journal} {\bibinfo  {journal} {Nature Physics}\ }\textbf {\bibinfo {volume}
  {7}},\ \bibinfo {pages} {701} (\bibinfo {year} {2011})}\BibitemShut {NoStop}%
\bibitem [{ \citenamefont {Vozmediano}\ \emph {et~al.}(2010) \citenamefont
  {Vozmediano},  \citenamefont {Katsnelson},\ and\  \citenamefont
  {Guinea}}]{VKG10}%
  \BibitemOpen
  \bibfield  {author} {\bibinfo {author} {\bibfnamefont {M.}~\bibnamefont
  {Vozmediano}}, \bibinfo {author} {\bibfnamefont {M.}~\bibnamefont
  {Katsnelson}}, \ and\ \bibinfo {author} {\bibfnamefont {F.}~\bibnamefont
  {Guinea}},\ }\href {\doibase 10.1016/j.physrep.2010.07.003} {\bibfield
  {journal} {\bibinfo  {journal} {Physics Reports}\ }\textbf {\bibinfo {volume}
  {496}},\ \bibinfo {pages} {109 } (\bibinfo {year} {2010})}\BibitemShut
  {NoStop}%
\bibitem [{ \citenamefont {Ono}\ and\  \citenamefont {Sugihara}(1966)}]{OS66}%
  \BibitemOpen
  \bibfield  {author} {\bibinfo {author} {\bibfnamefont {S.}~\bibnamefont
  {Ono}}\ and\ \bibinfo {author} {\bibfnamefont {K.}~\bibnamefont {Sugihara}},\
  }\href@noop {} {\bibfield  {journal} {\bibinfo  {journal} {J. Phys. Soc.
  Jpn}\ }\textbf {\bibinfo {volume} {21}},\ \bibinfo {pages} {861} (\bibinfo
  {year} {1966})}\BibitemShut {NoStop}%
\bibitem [{ \citenamefont {Suzuura}\ and\  \citenamefont {Ando}(2002)}]{SA02}%
  \BibitemOpen
  \bibfield  {author} {\bibinfo {author} {\bibfnamefont {H.}~\bibnamefont
  {Suzuura}}\ and\ \bibinfo {author} {\bibfnamefont {T.}~\bibnamefont {Ando}},\
  }\href {\doibase 10.1103/PhysRevB.65.235412} {\bibfield  {journal} {\bibinfo
  {journal} {Phys. Rev. B}\ }\textbf {\bibinfo {volume} {65}},\ \bibinfo
  {pages} {235412} (\bibinfo {year} {2002})}\BibitemShut {NoStop}%
\bibitem [{ \citenamefont {Choi}\ \emph {et~al.}(2010) \citenamefont {Choi},
   \citenamefont {Jhi},\ and\  \citenamefont {Son}}]{CSS10}%
  \BibitemOpen
  \bibfield  {author} {\bibinfo {author} {\bibfnamefont {S.-M.}\ \bibnamefont
  {Choi}}, \bibinfo {author} {\bibfnamefont {S.-H.}\ \bibnamefont {Jhi}}, \
  and\ \bibinfo {author} {\bibfnamefont {Y.-W.}\ \bibnamefont {Son}},\ }\href
  {\doibase 10.1103/PhysRevB.81.081407} {\bibfield  {journal} {\bibinfo
  {journal} {Phys. Rev. B}\ }\textbf {\bibinfo {volume} {81}},\ \bibinfo
  {pages} {081407} (\bibinfo {year} {2010})}\BibitemShut {NoStop}%
\bibitem [{ \citenamefont {Nelson}\ \emph {et~al.}(1989) \citenamefont {Nelson},
   \citenamefont {Piran},\ and\  \citenamefont
  {Eds.}}]{JerusalemWinterSchool1989}%
  \BibitemOpen
  \bibfield  {author} {\bibinfo {author} {\bibfnamefont {D.}~\bibnamefont
  {Nelson}}, \bibinfo {author} {\bibfnamefont {T.}~\bibnamefont {Piran}}, \
  and\ \bibinfo {author} {\bibfnamefont {S.~W.}\ \bibnamefont {Eds.}},\
  }\href@noop {} {\emph {\bibinfo {title} {Statistical Mechanics of Membranes
  and Surfaces}}},\ Proceedings of the Fifth Jerusalem Winter School for
  Theoretical Physics\ (\bibinfo  {publisher} {World Scientific},\ \bibinfo
  {address} {Singapore},\ \bibinfo {year} {1989})\BibitemShut {NoStop}%
\bibitem {WieseHabil}%
  K.J.\ Wiese, in {\em Phase Transitions and Critical Phenomena}, C.~Domb and J.L.~Lebowitz, eds., Acadamic Press
  London, 1999.
\bibitem [{ \citenamefont {Wunsch}\ \emph {et~al.}(2006) \citenamefont {Wunsch},
   \citenamefont {Stauber},  \citenamefont {Sols},\ and\  \citenamefont
  {Guinea}}]{WSSG06}%
  \BibitemOpen
  \bibfield  {author} {\bibinfo {author} {\bibfnamefont {B.}~\bibnamefont
  {Wunsch}}, \bibinfo {author} {\bibfnamefont {T.}~\bibnamefont {Stauber}},
  \bibinfo {author} {\bibfnamefont {F.}~\bibnamefont {Sols}}, \ and\ \bibinfo
  {author} {\bibfnamefont {F.}~\bibnamefont {Guinea}},\ }\href
  {http://stacks.iop.org/1367-2630/8/i=12/a=318} {\bibfield  {journal}
  {\bibinfo  {journal} {New Journal of Physics}\ }\textbf {\bibinfo {volume}
  {8}},\ \bibinfo {pages} {318} (\bibinfo {year} {2006})}\BibitemShut {NoStop}%
\bibitem [{ \citenamefont {Brey}\ and\  \citenamefont {Palacios}(2008)}]{BP08}%
  \BibitemOpen
  \bibfield  {author} {\bibinfo {author} {\bibfnamefont {L.}~\bibnamefont
  {Brey}}\ and\ \bibinfo {author} {\bibfnamefont {J.~J.}\ \bibnamefont
  {Palacios}},\ }\href {\doibase 10.1103/PhysRevB.77.041403} {\bibfield
  {journal} {\bibinfo  {journal} {Phys. Rev. B}\ }\textbf {\bibinfo {volume}
  {77}},\ \bibinfo {pages} {041403} (\bibinfo {year} {2008})}\BibitemShut
  {NoStop}%
\bibitem [{ \citenamefont {Gazit}(2009{\natexlab{b}})}]{Gazit2009}%
  \BibitemOpen
  \bibfield  {author} {\bibinfo {author} {\bibfnamefont {D.}~\bibnamefont
  {Gazit}},\ }\href {\doibase 10.1103/PhysRevB.79.113411} {\bibfield  {journal}
  {\bibinfo  {journal} {Phys. Rev. B}\ }\textbf {\bibinfo {volume} {79}},\
  \bibinfo {pages} {113411} (\bibinfo {year} {2009}{\natexlab{b}})}\BibitemShut
  {NoStop}%
\bibitem [{ \citenamefont {Gazit}(2009{\natexlab{c}})}]{G09b}%
  \BibitemOpen
  \bibfield  {author} {\bibinfo {author} {\bibfnamefont {D.}~\bibnamefont
  {Gazit}},\ }\href {\doibase 10.1103/PhysRevE.80.041117} {\bibfield  {journal}
  {\bibinfo  {journal} {Phys. Rev. E}\ }\textbf {\bibinfo {volume} {80}},\
  \bibinfo {pages} {041117} (\bibinfo {year} {2009}{\natexlab{c}})}\BibitemShut
  {NoStop}%
\bibitem [{ \citenamefont {Zinn-Justin}(1989)}]{Zinn}%
  \BibitemOpen
  \bibfield  {author} {\bibinfo {author} {\bibfnamefont {J.}~\bibnamefont
  {Zinn-Justin}},\ }\href@noop {} {\emph {\bibinfo {title} {Quantum Field
  Theory and Critical Phenomena}}}\ (\bibinfo  {publisher} {Oxford University
  Press},\ \bibinfo {address} {Oxford},\ \bibinfo {year} {1989})\BibitemShut
  {NoStop}%
\bibitem{Gazit2}
R. Dillenschneider, Phys. Rev. B {\bf 78} 115417 (2008)
  
  
\end{thebibliography}
\end{document}